\documentclass[11pt,british]{article}
\usepackage[utf8]{inputenc}
\usepackage{mathtools}
\usepackage{amsmath}
\usepackage{amsthm}
\usepackage{amssymb}
\usepackage{microtype}
\usepackage{antmath}
\usepackage{babel}
\usepackage{stmaryrd}
\usepackage{slashed}
\usepackage{cancel}
\usepackage{centernot}
\usepackage{bbm}

\makeatletter
\usepackage{tikz}
\usepackage{tikz-cd}
\usetikzlibrary{backgrounds}
\usepackage[framemethod=tikz]{mdframed}
\AtBeginEnvironment{mdframed}{%
\tikzset{every picture/.style={}}%
}
\mdfsetup{roundcorner=.5ex}

\theoremstyle{definition}
\newtheorem*{defn*}{Definition}

\usepackage{jheppub}                   
\usepackage[linecolor=blue,backgroundcolor=blue!25,bordercolor=blue,textsize=scriptsize]{todonotes}

\makeatletter
\gdef\@fpheader{\ }                    
\makeatother

\usepackage{setspace}
\usepackage{amsfonts} 		
\SetTracking{encoding={*}, shape=sc}{0} 	
\usepackage{color} 		
\definecolor{darkblue}{rgb}{0.0,0.0,0.3} 	
\allowdisplaybreaks		
\date{\today} 		
\numberwithin{equation}{section}	

\makeatletter
\g@addto@macro\bfseries{\boldmath}
\makeatother

\let\originalleft\left
\let\originalright\right
\renewcommand{\left}{\mathopen{}\mathclose\bgroup\originalleft}
\renewcommand{\right}{\aftergroup\egroup\originalright}

\newcommand{\D}{\mathcal{D}}

\newcommand{\lb}{\mathcal{L}}

\newcommand{\dil}{\hat\varphi}

\newcommand{\cliff}[1]{\cancel{#1}}
\usepackage{centernot}

\title{Generalising G$_\text{2}$ geometry: involutivity, moment maps and moduli}

\author[a]{Anthony Ashmore,}
\emailAdd{aashmore@sas.upenn.edu}
\author[b]{Charles Strickland-Constable,}
\emailAdd{c.strickland-constable@herts.ac.uk}
\author[c]{David Tennyson}
\emailAdd{d.tennyson16@imperial.ac.uk}
\author[c]{\\and Daniel Waldram}
\emailAdd{d.waldram@imperial.ac.uk}

\affiliation[a]{Department of Physics and Astronomy, University of Pennsylvania,\\
Philadelphia, PA 19104, USA}
\affiliation[b]{School of Physics, Astronomy and Mathematics, University of Hertfordshire, \\College Lane, Hatfield, AL10 9AB, UK}
\affiliation[c]{Department of Physics,
	Imperial College London, \\
	Prince Consort Road, London, SW7 2AZ, UK} 

\subheader{\hfill\textrm{Imperial/TP/19/DW/2}}

\abstract{
We analyse the geometry of generic Minkowski $\mathcal{N}=1$, $D=4$ flux compactifications in string theory, the default backgrounds for string model building. In M-theory they are the natural string theoretic extensions of $\mathrm{G}_2$ holonomy manifolds. In type II theories, they extend the notion of Calabi--Yau geometry and include the class of flux backgrounds based on generalised complex structures first considered by Graña et al.~(GMPT). Using $\mathrm{E}_{7(7)}\times\mathbb{R}^+$ generalised geometry we show that these compactifications are characterised by an $\mathrm{SU}(7)\subset\mathrm{E}_{7(7)}$ structure defining an involutive subbundle of the generalised tangent space, and with a vanishing moment map, corresponding to the action of the diffeomorphism and gauge symmetries of the theory. The Kähler potential on the space of structures defines a natural extension of Hitchin's $\mathrm{G}_2$ functional. Using this framework we are able to count, for the first time, the massless scalar moduli of GMPT solutions in terms of generalised geometry cohomology groups. It also provides an intriguing new perspective on the existence of $\Gx2$ manifolds, suggesting possible connections to Geometrical Invariant Theory and stability.}

\begin{document}
\maketitle

\section{Introduction}

Supersymmetric string backgrounds play a central role in our understanding of string phenomenology and the AdS/CFT correspondence. Without flux, low-energy supersymmetry implies the internal manifold has special holonomy. For example M-theory compactified on a $\Gx2$ holonomy manifold leads to an $\mathcal{N}=1$ effective theory in four dimensions~\cite{Papadopoulos:1995da,Acharya:2000ps}. Studying the geometric properties of $\Gx2$ manifolds allows one to probe various properties of the compactified theory, such as the moduli spaces of massless fields, particle spectra and couplings~\cite{Acharya:1998pm,Acharya:2000gb,Witten:2001uq,Atiyah:2001qf,Acharya:2001gy,Beasley:2002db,Berglund:2002hw,Acharya:2002kv,Atiyah:2001qf,DallAgata:2005zlf,Grigorian:2008tc,Acharya:2008hi,Braun:2018vhk}. Despite several recent developments such as new constructions of large families of examples on twisted connected sums~\cite{corti2015}, $\Gx2$ manifolds remain far less well understood than, for example, their Calabi--Yau counterparts~\cite{Donaldson-g2}.  

String theory admits a much larger class of generic $\mathcal{N}=1$, $D=4$ backgrounds once one allows non-trivial flux. In M-theory, this moves one away from $\Gx2$ holonomy and greatly complicates conventional geometrical descriptions~\cite{Kaste:2003zd,Behrndt:2003zg,DallAgata:2003txk,Lukas:2004ip}. In type II theories, truly $\mathcal{N}=1$ backgrounds necessarily have non-zero flux, and are generically not of Calabi--Yau type (for a review see~\cite{Grana06}). This raises several natural questions: How does one extract the properties of the low-energy theory from the geometry of this much larger class? Do they have an ``nice'' geometrical description in analogy to that of special holonomy spaces? What tools do we have to find the number of massless moduli or construct examples? Does incorporating them in a larger class shed any light on the nature of $\Gx2$ manifolds?

In this paper we will address these questions using the formalism of $\ExR{7(7)}$ generalised geometry~\cite{Hull07,PW08,CSW11,CSW14}. It was shown in~\cite{PW05,CSW14b,CS16} that such backgrounds define a generalised $G$-structure, where the relevant group is $G=\SU7\subset\ExR{7(7)}$. Furthermore, supersymmetry for the ten- or eleven-dimensional solution was shown to be equivalent to the generalised $\SU7$ structure being torsion-free. This analysis provides a description of generic $\mathcal{N}=1$, $D=4$ backgrounds via an invariant spinor and a generalised connection, similar to giving an $\SU3$-invariant spinor and the Levi-Civita connection for a conventional Calabi--Yau background. Conventional $G$-structures can also be described by giving a set of nowhere-vanishing, $G$-invariant  tensors together with differential conditions that constrain the intrinsic torsion of the structure, for example in the $\Gx2$ case the three-form $\varphi$ satisfying $\dd\varphi=\dd\star\varphi=0$. The same is true for generalised structures and this is the analysis we will present in this paper. This viewpoint gives an elegant geometric reinterpretation of the supersymmetry conditions for generic flux backgrounds, providing a general formalism for understanding moduli, and giving general expressions for the perturbative superpotential and K\"ahler potential of the four-dimensional effective theories. The geometric structures present for generic $\mathcal{N}=2$ compactifications have been discussed elsewhere~\cite{GLW06,GLSW09,GO12,AW15,GN16} -- this paper can be seen as the $\mathcal{N}=1$ companion to those works.

In M-theory, the $\SU7$ generalised structures can generically be viewed as a sort of complexification of a conventional $\Gx2$ structure combining the three-form $\varphi$ with the M-theory gauge potential $A$. This is analogous to the way the symplectic structure and NSNS two-form $B$ combine to give a complexified Kähler form in the A-model topological string. Because of this complexification, the space of $\SU7$ structures $\ZS{\SU7}$ admits a Kähler metric. Furthermore, as we will see, the Kähler potential $\mathcal{K}$ on $\ZS{\SU7}$ is a generalisation of Hitchin's $\Gx2$ functional~\cite{Hitchin00a}. The supersymmetry conditions on the $\SU7$ structure are of two types. First there is a involutivity condition: in direct analogy to a conventional complex structure, the $\SU7$ structure defines a subbundle of the complexified generalised tangent space that must be involutive under the generalised Lie derivative. We call this an ``exceptional complex structure''. In the special case of $\Gx2$ structures, it imposes $\dd\varphi=0$. The second supersymmetry condition is the vanishing of a moment map, defined for the action of generalised diffeomorphisms (that is, conventional diffeomorphisms plus form-field gauge transformations) and in the $\Gx2$ case imposes the condition $\dd\star\varphi=0$. 

Interestingly, this reformulation puts the analysis of generic supersymmetric backgrounds, and $\Gx2$ structures in particular, in the same setting as many classic problems in differential geometry: we have a complex condition (the involutivity of the subbundle) together with an infinite-dimensional moment map. This set up appears for example in Atiyah and Bott's work on flat connections on Riemann surfaces~\cite{AB83}, in the supersymmetric hermitian Yang--Mills equations of Donaldson--Uhlenbeck--Yau~\cite{Donaldson85,UY86,UY89} and the equations of Kähler--Einstein geometry~\cite{Fujiki90,Donaldson97,KEproof}. In each case, the existence of solutions to the moment map equations can be reformulated in terms of an algebraic notion of stability, using the ideas of Geometrical Invariant Theory (GIT). As we will discuss, at least formally, something similar happens here. The generalised Hitchin functional $\mathcal{K}$ plays the role of the norm functional of GIT, and in addition, the square of the moment map (the Yau functional in the case of Kähler--Einstein metrics) is equal to the generalised Ricci scalar. In the context of $\Gx2$ structures, this complexified picture suggests a possible notion of stability on the space of closed $\Gx2$ structures, the stable orbits being the ones that extremise the generalised extension of Hitchin's $\Gx2$ functional and satisfy $\dd\star\varphi=0$. 

In the type II context, the original $\Orth{d,d}$ version of generalised geometry of Hitchin and Gualtieri~\cite{Hitchin02,Gualtieri04} ``geometrises'' the NSNS sector of the supergravity fluxes, giving a unified description of the metric and $B$ field. The connection to supergravity has already been used to characterise $D=4$, $\mathcal{N}=1$ backgrounds in the seminal work of Graña et al.~(GMPT)~\cite{GMPT04b,GMPT05}. The RR fluxes can be included but are not geometrised in the same way as the NSNS sector and so the conditions do not have a simple interpretation in terms of integrability. This has made it difficult to analyse the general properties of these backgrounds, such as their moduli spaces. Despite this, these methods have been very useful in finding new solutions~\cite{GMPT07,Andriot08,Apruzzi:2013yva,AFPRT14,AFPT15,Passias:2018zlm} and investigating the AdS/CFT correspondence~\cite{MPZ06,BFMMPZ08,GGPSW10}. By going to $\ExR{7(7)}$ generalised geometry, all the fluxes become geometrical and as we have mentioned the $\mathcal{N}=1$ conditions are equivalent to a torsion-free generalised structure~\cite{CSW14b}. Crucially, as we show, this allows us to treat the moduli of flux backgrounds in a unified manner, as well as giving simple expressions for the corresponding low-energy Kähler potential and superpotential. We can then re-derive known results in the $\Gx2$ case as well as in a more general class we denote ``type-0''. Furthermore, we can extend previous results for GMPT backgrounds, in particular the work of Tomasiello~\cite{Tomasiello07}, to find, for the first time, the exact moduli of GMPT backgrounds. 

Throughout our analysis we restrict ourselves to warped flux backgrounds with a four-dimensional Minkowski factor. An important caveat is that all our backgrounds are hence subject to the ``no-go'' theorems that preclude fluxes precisely when the internal space is compact~\cite{Candelas:1984yd,Candelas:1986di,deWit:1986mwo,MN01,GKP02,GMW03,GMPW04}. Thus the spaces we discuss should be understood either as non-compact or, if one is interested in model building, as spaces with boundaries where the sources (branes and orientifolds) have been removed. At various points in the derivations we make use of integrations by parts that will be valid provided we adopt suitable boundary conditions at infinity and/or at the sources. It would be interesting to include the effects of localised sources in the analysis, in particular to see how they might modify the naive moduli space calculations.

The layout of the paper is as follows. In section~\ref{sec:review} we review the notion of $G$-structures and the role of involutivity and moment maps in defining conventional complex structures and generalised complex structures in six dimensions. These give simple models for the general analysis of $\SU7$ structures we then give in section~\ref{sec:Gen N=1 Structures}. Section~\ref{sec:examples} shows explicitly how $\Gx2$ manifolds and the solutions of GMPT fit into the general analysis. Section~\ref{sec:WK-GIT} first shows how the involutivity and moment map conditions can be viewed as $F$- and $D$-term supersymmetry conditions in a rewriting of the supergravity as an effective $D=4$, $\mathcal{N}=1$ theory. It then connects our analysis to the Geometrical Invariant Theory picture and the $\Gx2$ functional of Hitchin. In particular, we see that Hitchin's extremisation is equivalent to finding the stationary points of the norm functional, and we go on to outline the naive connection to stability. Section~\ref{sec:moduli} addresses the general moduli problem, and calculates the moduli of generic ``type-0'' structures (including $\Gx2$) and the full set of moduli of GMPT solutions. We conclude with some discussion. 

\section{Review of integrability, involutivity and moment maps for \texorpdfstring{$G$}{G}-structures}
\label{sec:review}

We begin with a review of two examples of familiar geometric structures that appear when describing supersymmetric backgrounds: conventional complex structures in six dimensions and their generalised geometry extensions first introduced by Hitchin and Gualtieri~\cite{Hitchin02,Gualtieri04b}. In each case, involutivity of an appropriate vector bundle under a bracket is equivalent to the integrability of the structure.\footnote{Note that we use ``integrable'' and ``torsion-free'' interchangeably. For a conventional $G$-structure, integrable is a stronger condition: torsion-free implies the $G$-structure is flat to first-order, while integrable implies the $G$-structure is locally equivalent to the flat model. See \cite{Bryant03} for some remarks on this nomenclature.} We will then also discuss how the extra differential conditions that promote these structures to integrable $\SL{3,\mathbb{C}}$ and generalised Calabi--Yau structures come from a moment map for the action of diffeomorphisms and, in the latter case, gauge symmetries. These two examples will provide the model for how we analyse generic four-dimensional $\mathcal{N}=1 $ flux backgrounds. 

\subsection{Complex structures}\label{sec:complex}

Let $M$ be a six-dimensional manifold with tangent bundle $T$. Recall that an almost complex structure on $M$ is a conventional $G$-structure with $G=\GL{3,\mathbb{C}}$. It is defined by a nowhere-vanishing tensor $I\in\Gamma(\End T)$, with $I^2 = -\id$, that allows one to decompose the complexified tangent bundle into subbundles
\begin{equation}
T\otimes\mathbb{C}\coloneqq T_{\mathbb{C}}=L_{1}\oplus L_{-1},
\end{equation}
where sections of $L_{1}$ have charge $+\ii$ under the action of $I$, and $\bar{L}_1 \simeq L_{-1}$. Typically, $L_{1}$ is written $T^{1,0}$ but we will use this notation to highlight the similarities to the work in the later sections. Consider two vectors $V,W\in\Gamma(L_{1})$. A standard way to define an integrable structure is to require that the Lie bracket of two $(1,0)$-vector fields gives another $(1,0)$-vector field. In other words, $L_{1}$ is involutive under the Lie bracket
\begin{equation}
[V,W]\in\Gamma(L_{1}) \qquad \forall\;V,W\in \Gamma(L_1).
\end{equation}
Using $I$ to project onto $L_1$ it is then straightforward to show that involutivity of the bracket is equivalent to the vanishing of the Nijenhuis tensor, or equivalently, in the language of $G$-structures, the vanishing of the intrinsic torsion.

Every almost complex structure $I$ defines a unique ``canonical'' line bundle $\mathcal{U}_I\subset \ext^3T_\bbC^*$ satisfying
\begin{equation}
\imath_V \Omega = 0 \quad \forall\; V\in \Gamma( L_{-1}), \qquad \Omega \wedge \bar{\Omega} \neq 0, 
\end{equation}
where $\Omega$ is a local section of $\mathcal{U}_I$. If this bundle is trivial, one  can introduce a refinement of the almost complex structure by considering $G=\SL{3,\mathbb{C}}$ structures. Each such structure is defined by a nowhere-vanishing section $\Omega\in\Gamma(\mathcal{U}_I)$ so that any two such structures defining the same complex structure differ by nowhere-vanishing complex function $f$
\begin{equation}
    \Omega' = f \Omega . 
\end{equation}
Note that, as $\SL{3,\mathbb{C}}\subset\GL{3,\mathbb{C}}$, given a suitable complex three-form $\Omega$ (one stabilised by $\SL{3,\mathbb{C}}$)  one can construct an almost complex structure $I$, as described by Hitchin~\cite{Hitchin00}. It is natural then to ask the question, if we have a torsion-free $\GL{3,\mathbb{C}}$ structure (a complex structure), what extra condition do we need to impose to have a torsion-free $\SL{3,\mathbb{C}}$ structure? From the intrinsic torsion in each case, it is straightforward to see that the $\GL{3,\mathbb{C}}$ structure is torsion-free if
\begin{equation}
\label{eq:dOmega}
\dd\Omega=A\wedge\Omega,
\end{equation}
for some $(0,1)$-form $A$, while for a torsion-free $\SL{3,\mathbb{C}}$ structure we should have
\begin{equation}
\label{eq:dOmega0}
\dd\Omega=0.
\end{equation}
Thus $A$ encodes the extra intrinsic torsion components of the $\SL{3,\mathbb{C}}$ structure.  

This additional integrability condition can be viewed as the vanishing of a moment map. One first notes that the space of $\SL{3,\mathbb{C}}$ structures admits a natural pseudo-Kähler metric~\cite{Hitchin00}. 
At a point $p\in M$, a choice of $\Omega$ is equivalent to picking a point in the coset
\begin{equation}
\Omega|_{p}\in Q_{\SL{3,\mathbb{C}}} = \frac{\GL{6,\mathbb{R}}}{\SL{3,\mathbb{C}}}.
\end{equation}
The choice of $\SL{3,\mathbb{C}}$ structure on $M$ thus corresponds to a section of the fibre bundle
\begin{equation}
Q_{\SL{3,\mathbb{C}}}\to \mathcal{Q}_{\SL{3,\mathbb{C}}}\to M,
\end{equation}
that is, we can identify
\begin{equation}
    \text{space of $\SL{3,\bbC}$ structures, $\ZS{\SL{3,\bbC}}$}
    \simeq \Gamma(\mathcal{Q}_{\SL{3,\mathbb{C}}}) . 
\end{equation}
This infinite-dimensional space then inherits a pseudo-Kähler structure from the pseudo-Kähler structure\footnote{This metric has signature $(18,2)$~\cite{Hitchin00}.} on the coset space $Q_{\SL{3,\mathbb{C}}}$, with a K\"ahler potential given by
\begin{equation}
\label{eq:Kahlerpot}
\mathcal{K}=\ii\int_{M}\Omega\wedge\bar{\Omega} , 
\end{equation}
where $\Omega$ can be viewed as a complex coordinate on the space of structures (or more precisely as a holomorphic embedding $\Omega\colon\ZS{\SL{3,\bbC}}\hookrightarrow \Gamma(\ext^3T^*_\bbC)$). One can also restrict to the subspace of structures that define an (integrable) complex structure, so that $L_1$ is involutive,
\begin{equation}
    \ZI{\SL{3,\bbC}} 
    = \{ \Omega\in \ZS{\SL{3,\bbC}} \; | \; \text{$I$ is integrable} \} .
\end{equation}
Given that the integrability condition~\eqref{eq:dOmega} is holomorphic -- it is independent of $\bar{\Omega}$ -- this space inherits a pseudo-Kähler metric from $\ZS{\SL{3,\bbC}}$ with the same Kähler potential.

Diffeomorphisms act on $\ZI{\SL{3,\bbC}}$ since the integrability conditions on $I$ are diffeomorphism invariant. Infinitesimally they define a vector field $\rho_V\in\Gamma(T\ZI{\SL{3,\bbC}})$ such that 
\begin{equation}
\imath_{\rho_V}\delta\Omega= \mathcal{L}_V \Omega ,
\end{equation}
where $\delta$ is the exterior (functional) derivative on $\ZI{\SL{3,\bbC}}$ and $V\in\Gamma(T)$ generates the diffeomorphism. Clearly the Kähler potential~\eqref{eq:Kahlerpot} is diffeomorphism invariant. Furthermore, since $\mathcal{L}_V \Omega$ is independent of $\bar{\Omega}$, we see that diffeomorphisms also preserve the complex structure on $\ZI{\SL{3,\bbC}}$. Together this implies they preserve the Kähler form.\footnote{Note that there may be further subtleties if the integrability condition defines a null subspace within $\ZS{\SL{3,\bbC}}$ or if the group action defining the moment map is null. We comment on this for the case of $\SU7$ structures in section \ref{sec:GIT}.}
Explicitly this is given by
\begin{equation}
\label{eq:Kahler-form}
\varpi=\ii\,\partial'\bar{\partial}'\mathcal{K},
\end{equation}
where we have decomposed $\delta=\partial'+\bar{\partial}'$ into holomorphic and antiholomorphic derivatives. 
For an arbitrary vector $\alpha\in \Gamma(T\ZI{\SL{3,\bbC}})$ we then have 
\begin{equation}
\begin{split}\imath_{\rho_{V}}\imath_{\alpha}\varpi & =-\int_{M}\bigl(\imath_{\alpha}\delta\Omega\wedge\mathcal{L}_{V}\bar{\Omega}-\mathcal{L}_{V}\Omega\wedge\imath_{\alpha}\delta\bar{\Omega}\bigr)
=\int_{M}\bigl(\mathcal{L}_{V}\imath_{\alpha}\delta\Omega\wedge\bar{\Omega}+\mathcal{L}_{V}\Omega\wedge\imath_{\alpha}\delta\bar{\Omega}\bigr)\\
& =\int_{M}\imath_{\alpha}\bigl(\mathcal{L}_{V}\delta\Omega\wedge\bar{\Omega}+\mathcal{L}_{V}\Omega\wedge\delta\bar{\Omega}\bigr)
=\imath_{\alpha}\delta\mu(v),
\end{split}
\end{equation}
where 
\begin{equation}
\mu(V)=\int_{M}\mathcal{L}_{V}\Omega\wedge\bar{\Omega}.\label{eq:moment_map}
\end{equation}
defines a moment map $\mu\colon\ZI{\SL{3,\bbC}}\to\diff^*$, where $\diff$ is the Lie algebra of diffeomorphisms. It is straightforward to check that $\mu$ is equivariant.

Given the integrability condition~\eqref{eq:dOmega}, we can integrate by parts, to write
\begin{equation}
\begin{split}
\mu(V) & =\int_{M}\Bigl(-\imath_{V}\Omega\wedge\bar{A}\wedge\bar{\Omega}
+\imath_{V}(A\wedge\Omega)\wedge\bar{\Omega}\Bigr)\\
& =\int_{M}(\imath_{V}A-\imath_{V}\bar{A})\,\Omega\wedge\bar{\Omega} ,
\end{split}
\end{equation}
where we have used $A\wedge\bar{\Omega}=\bar{A}\wedge\Omega=0$. The moment map vanishes for all $V\in\Gamma(T)$ if and only if 
\begin{equation}
A=\bar{A}=0.
\end{equation}
In other words, we see that the vanishing of the moment map imposes the final condition~\eqref{eq:dOmega0} that promotes a complex structure to a torsion-free $\SL{3,\mathbb{C}}$ structure.

Since two $\SL{3,\mathbb{C}}$ structures that are related by a diffeomorphism are equivalent, the moduli space of $\SL{3,\mathbb{C}}$ structures is naturally a quotient, defined as
\begin{equation}
\mathcal{M}_{\Omega}=\{\Omega\in\ZI{\SL{3,\bbC}} \; | \; \mu=0 \}\quotient \Diff.
\end{equation}
As we have seen, the Kähler geometry on the space of structures $\ZI{\SL{3,\bbC}}$ is preserved by the action of the diffeomorphism group, thus we can view the moduli space either as a symplectic quotient by $\Diff$ or as a quotient by the complexified group 
\begin{equation}
\label{eq:Omega-mod}
\mathcal{M}_{\Omega} = \ZI{\SL{3,\bbC}} \qquotient \Diff 
  \simeq \ZI{\SL{3,\bbC}} \quotient \Diff_\mathbb{C} .
\end{equation}
Note that the complexification of the diffeomorphism group $\Diff_\mathbb{C}$ is not really well defined. What is really meant is the complexification of the orbits, that is, if the vector field $\rho_V\in\Gamma(T\ZI{\SL{3,\bbC}})$ generates the action of diffeomorphisms on the spaces of structures, we can complexify this to also include the orbits generated by $\mathcal{I}\rho_V$, where $\mathcal{I}$ is the complex structure on $\ZI{\SL{3,\bbC}}$. Since $\Omega$ is a holomorphic function on $\ZI{\SL{3,\bbC}}$ we have 
\begin{equation}
\label{eq:complex-diff-Omega}
    \imath_{\mathcal{I}\rho_V} \delta\Omega 
       = - \imath_{\rho_V} (\mathcal{I}\delta\Omega)
       = \ii\,\imath_{\rho_V} \delta\Omega
       = \ii \,\mathcal{L}_V \Omega 
       = \mathcal{L}_{IV}\Omega + 2\ii (\imath_VA) \Omega, 
\end{equation}
where in the last expression we have used~\eqref{eq:dOmega} and the fact that $\imath_{IV}\Omega=\ii\,\imath_V\Omega$ and $\imath_{IV}A=-\ii\,\imath_{V}A$. Thus in~\eqref{eq:Omega-mod}, up to diffeomorphisms, for each fixed complex structure, the action of $\Diff_\bbC$ simply rescales $\Omega$ until~\eqref{eq:dOmega0} is satisfied and the moment map vanishes. 

\subsection{Generalised complex structures}\label{sec:gen complex structures}

Let us now review the analogous story for the generalised complex structures (GCS) of Hitchin and Gualtieri~\cite{Hitchin02,Gualtieri04b}.  We will see again that involutivity and a moment map characterise the integrable structures and lead to a local description of the moduli space as a Kähler quotient.

Consider a six-dimensional manifold $M$ with a generalised tangent bundle $\EO=T\oplus T^{*}$. This admits a natural $\Orth{6,6}$ structure given by the inner product
\begin{equation}
\left<x+\xi,y+\eta\right> = \eta(x)+\xi(y).
\end{equation}
As was noted in \cite{CSW11b}, the relevant structure group for supergravity is actually $\Orth{6,6}\times \mathbb{R}^{+}$ to account for the dilaton. We take all generalised vectors to be weight zero under the $\mathbb{R}^{+}$ action. Given a generalised vector $V=x+\xi\in\Gamma(E)$, there is a natural generalised Lie derivative $L_V$ such that, acting on a generalised vector $W=y+\eta$, 
\begin{equation}\label{eq:o66_GLie}
L_{x+\xi}\left(y+\eta\right) = [x,y] + \mathcal{L}_{x}\eta - \imath_y\dd\xi .
\end{equation}
This generates conventional diffeomorphisms and one-form gauge transformations, parametrised by $x$ and $\xi$ respectively. Its antisymmetrisation $\llbracket V,W \rrbracket \coloneqq \frac{1}{2}\left(L_V W-L_WV\right)$ is the Courant bracket. $\EO$ generates a Clifford algebra $\Cliff{6,6}$ via the inner product above, which has a natural representation on sections $\Psi$ of the spinor bundle $S\coloneqq\ext^{\bullet}T^{*}$ via 
\begin{equation}
\label{eq:Clifford}
\cliff{V}\Psi = \imath_{x}\Psi +\xi\wedge \Psi.
\end{equation}
The slash notation signifies the Clifford action and can be viewed as contraction with the $\Orth{6,6}$ gamma matrices $\Gamma^{M}$. There is an invariant antisymmetric pairing $(\Psi,\Sigma)$ on spinors given by the Mukai pairing~\eqref{eq:mukai}, with the property that 
\begin{equation}
\label{eq:Clifford-adjoint}
    (\Psi,\cliff{V}\Sigma) = (-\cliff{V}\Psi,\Sigma). 
\end{equation}
As a representation of $\mathrm{Spin}(6,6)\times\mathbb{R}^{+}$ the spinor bundle is reducible as one can define the analogue of Majorana--Weyl spinors\footnote{It was important that we take the structure group to be $\Orth{6,6}\times\mathbb{R}^{+}$, or its double cover $\Spin{6,6}\times\mathbb{R}^{+}$, here since polyforms do not form a representation of $\Spin{6,6}$ alone. It also implies the antisymmetric pairing gives a top-form rather than a scalar. Without the $\bbR^+$ factor, one would have to take $S\simeq \ext^{\bullet}T^{*}\otimes (\det T)^{1/2}$. }
\begin{equation}
S^{+}=\ext^{\text{even}}T^{*}, \qquad S^{-} = \ext^{\text{odd}}T^{*}.
\end{equation}
The exterior derivative gives a map $\dd\colon S^\pm\to S^\mp$ such that the action of the generalised Lie derivative can be written as
\begin{equation}
\label{eq:Lgen-spinor}
    L_V \Psi = \dd (\cliff{V}\Psi) + \cliff{V} \dd\Psi ,
\end{equation}
for any $\Psi\in\Gamma(S)$.

In analogy to a conventional almost complex structure, a generalised almost complex structure $\mathcal{J}$ is a endomorphism $\mathcal{J}\colon \Gamma(\EO)\rightarrow \Gamma(\EO)$ such that
\begin{equation}
    \mathcal{J}^{2}=-\id, \qquad 
    \left<\mathcal{J}V,\mathcal{J}V\right> = \left<V,V\right> \quad \forall \;V\in\Gamma(E) .
\end{equation}
As a generalised tensor, $\mathcal{J}$ is nowhere vanishing so defines reduction of the structure group of $\EO$ from $\Orth{6,6}\times\bbR^{+}$ to $\Uni{3,3}\times\bbR^{+}$. It gives a decomposition of the complexified generalised tangent bundle
\begin{equation}
\EO_{\mathbb{C}} = L_{1}\oplus L_{-1},
\end{equation}
where $L_{\pm1}$ has charge $\pm\ii$ under $\mathcal{J}$. Note that $L_1$ is maximally isotropic: $\langle L_{1},L_{1}\rangle=0$. This defines an isomorphism $L_{1}^{*}\simeq \bar{L}_{1}=L_{-1}$.
A generalised almost complex structure is integrable if $L_{1}$ is involutive with respect to the generalised Lie derivative 
\begin{equation}
L_VW\in \Gamma(L_{1}) \qquad \forall\;V,W\in \Gamma(L_{1}),
\end{equation}
which also implies $L_V W=\llbracket V,W \rrbracket$. Using the notion of generalised intrinsic torsion introduced in~\cite{CSW14b}, one can show that this involution condition is equivalent to the vanishing of the generalised intrinsic torsion of the $\Uni{3,3}\times\bbR^+$ structure defined by $\mathcal{J}$.

Each generalised almost complex structure defines a unique pure spinor line bundle $\mathcal{U}_{\mathcal{J}}\subset S$ satisfying
\begin{equation}
\cliff{V} \Phi = 0 \quad \forall \;V\in\Gamma(L_1), \qquad (\Phi,\bar{\Phi})\neq 0, 
\end{equation}
where $\Phi$ is a local section of $\mathcal{U}_{\mathcal{J}}$ and $(\cdot,\cdot)$ is the Mukai pairing defined in~\eqref{eq:mukai}. If the  pure spinor line bundle is trivial, one can choose a global nowhere-vanishing section. This defines an $\SU{3,3}$ or generalised Calabi--Yau (GCY) structure~\cite{Hitchin02}.\footnote{Given a GCY structure, one can recover the generalised almost complex structure by identifying $L_1$ as the null space of $\Phi$.} Two such structures defining the same GCY structure differ by nowhere vanishing complex function $f$
\begin{equation}
    \Phi' = f \Phi . 
\end{equation}
From the generalised intrinsic torsion
it is straightforward to see that the corresponding generalised complex structure is integrable if
\begin{equation}
\label{eq:dPhi}
\dd\Phi = \cliff{A} \Phi,
\end{equation}
where $A\in \Gamma(L_{-1})$ acts on $\Phi$ via the Clifford action. The generalised Calabi--Yau structure is integrable if
\begin{equation}
\dd\Phi=0,
\end{equation}
and hence $A$ encodes the extra components of the intrinsic torsion of the $\SU{3,3}$ structure. 

As in the previous example of a complex structure, one can view the additional integrability condition as the vanishing of a moment map. One first notes that the space of $\SU{3,3}$ structures on $M$ admits a natural pseudo-K\"ahler metric~\cite{Hitchin02,Hitchin00a} -- the construction follows that of the almost complex structure case. At a point $p\in M$, a choice of $\Phi$ is equivalent to picking a point in the coset
\begin{equation}
\Phi|_{p}\in Q_{\SU{3,3}} = \frac{\Orth{6,6}\times\bbR^+}{\SU{3,3}},
\end{equation}
so that an $\SU{3,3}$ structure on $M$ corresponds to a section of the fibre bundle
\begin{equation}
Q_{\SU{3,3}}\to \mathcal{Q}_{\SU{3,3}}\to M.
\end{equation}
We can then identify
\begin{equation}
    \text{space of $\SU{3,3}$ structures, $\ZS{\SU{3,3}}$}
    \simeq \Gamma(\mathcal{Q}_{\SU{3,3}}) . 
\end{equation}
This infinite-dimensional space inherits a pseudo-Kähler structure from the pseudo-Kähler structure\footnote{This metric has signature $(30,2)$~\cite{Hitchin02,GLW06}.} on the coset space $Q_{\SU{3,3}}$, with a Kähler potential given by
\begin{equation}
\mathcal{K}=\ii\int_{M}(\Phi,\bar{\Phi}). 
\end{equation}
Again $\Phi$ can be viewed as a complex coordinate on the space of structures (or more precisely as a holomorphic embedding $\Phi\colon\ZS{\SU{3,3}}\hookrightarrow \Gamma(S_\bbC)$) and one can also restrict to the subspace of structures that define an (integrable) generalised complex structure, so that $L_1$ is involutive,
\begin{equation}
    \ZI{\SU{3,3}} 
    = \{ \Phi\in \ZS{\SU{3,3}} \; | \; \text{$\mathcal{J}$ is integrable} \} .
\end{equation}
The condition~\eqref{eq:dPhi} is holomorphic and so $\ZI{\SU{3,3}}$ inherits a pseudo-Kähler metric from $\ZS{\SU{3,3}}$, with the same Kähler potential.

The group of generalised diffeomorphisms $\GDiff$, that is diffeomorphisms and gauge transformations, acts on $\ZI{\SU{3,3}}$ and preserves the Kähler structure. The action is generated by vector fields $\rho_V\in\Gamma(T\ZI{\SU{3,3}})$ defined via the generalised Lie derivative
\begin{equation}
    \imath_{\rho_V}\delta\Phi = L_V \Phi .
\end{equation}
Given the Kähler form as defined in \eqref{eq:Kahler-form} and an arbitrary vector $\alpha\in\Gamma(T\ZI{\SU{3,3}})$, one finds
\begin{equation}
\begin{split}
\imath_{\rho_{V}}\imath_{\alpha}\varpi &= -\int_{M}( \imath_{\alpha}\delta\Phi,\Dorf_{V}\bar{\Phi} ) - ( \Dorf_{V}\Phi,\imath_{\alpha}\delta\bar{\Phi} ) 
= \int_{M}( \Dorf_{V}\imath_{\alpha}\delta\Phi,\bar{\Phi} ) 
+ (\Dorf_{V}\Phi,\imath_{\alpha}\delta\bar{\Phi} ) \\
&=\imath_{\alpha}\delta\int_{M}( \Dorf_{V}\Phi,\bar{\Phi} ) 
= \imath_{\alpha}\delta\mu(V)
\end{split}
\end{equation}
where 
\begin{equation}
\mu(V) = \int_{M}( \Dorf_{V}\Phi,\bar{\Phi} ),
\end{equation}
defines a moment map $\mu\colon\ZI{\SU{3,3}}\to\gdiff^*$. Here $\gdiff$ is the Lie algebra of generalised diffeomorphisms generated by the generalised Lie derivative. 

From~\eqref{eq:Lgen-spinor}, the integrability condition~\eqref{eq:dPhi} and~\eqref{eq:Clifford-adjoint} we have
\begin{equation}
\begin{split}
\mu(V) &= 
\int_{M} \left( \cliff{V} \dd \Phi + \cliff{V}\Phi,\dd\bar{\Phi} \right) 
= \int_{M} (\cliff{V} \cliff{A}\Phi,\Phi)+(\cliff{V}\Phi,\cliff{\bar{A}}\bar{\Phi}) \\ 
&= \int_{M}( \cliff{V}(\cliff{A}-\cliff{\bar{A}})\Phi, \bar{\Phi} )+((\cliff{A}-\cliff{\bar{A}})\cliff{V} \Phi, \bar{\Phi}) 
= 2\int_{M}\langle V, A-\bar{A}\rangle(\Phi, \bar{\Phi} ) ,
\end{split}
\end{equation}
where in going to the second line we have used $\cliff{A}\bar{\Phi}=\cliff{\bar{A}}\Phi=0$. Thus we see the moment map vanishes for all $V$ if and only if $A=\bar{A}=0$, that is, if the $\SU{3,3}$ structure is integrable.

Again, we consider two $\SU{3,3}$ structures that are related by a generalised diffeomorphism as equivalent and so the moduli space of $\SU{3,3}$ structures is a symplectic quotient.\footnote{As with the previous $\SL{3,\mathbb{C}}$ structures, this can be more nuanced. We refer the reader to section \ref{sec:GIT} for a discussion of this for $\SU7$ structures.}
Since the group action preserves the Kähler structure, we can view also view the moduli space as a quotient by the complexified group $\GDiff_{\mathbb{C}}$
\begin{equation}
\mathcal{M}_{\Phi} = \ZI{\SU{3,3}}\qquotient \GDiff 
\simeq \ZI{\SU{3,3}}\quotient \GDiff_{\mathbb{C}}.
\end{equation}
As before, if $\mathcal{I}$ is the complex structure on $\ZI{\SU{3,3}}$, we have
\begin{equation}
\label{eq:complex-gdiff-Phi}
    \imath_{\mathcal{I}\rho_V} \delta\Phi
       = - \imath_{\rho_V} (\mathcal{I}\delta\Phi)
       = \ii(\imath_{\rho_V} \delta\Phi)
       = \ii \,L_V \Phi 
       = - L_{\mathcal{J}V}\Phi + 2\ii \left<V,A\right> \Phi, 
\end{equation}
where in the last expression we have used~\eqref{eq:dPhi} and the fact that $\mathcal{J}V\circ\Phi=-\ii\,\imath_V\Phi$ and $\left<\mathcal{J}V,A\right>=\ii\left<V,A\right>$.  Thus, up to generalised diffeomorphisms, for each fixed complex structure, the action of $\GDiff_\bbC$ simply rescales $\Phi$ until $\dd\Phi=0$ and the moment map vanishes.

\section{Generalised \texorpdfstring{$\mathcal{N}=1$}{N = 1} structures}\label{sec:Gen N=1 Structures}

Our goal is to analyse generic Minkowski $\mathcal{N}=1$, $D=4$ flux compactifications of M-theory and type II supergravity. In this section, we will show that they define two closely related generalised $G$-structures, analogous to the $\GL{3,\mathbb{C}}$ and $\SL{3,\mathbb{C}}$ structures in conventional geometry we have just discussed. Remarkably, we will find that the supersymmetry conditions can be rephrased similarly as an involution condition and the vanishing of a moment map.  Conventional $\Gx2$ structures are of course a special case, corresponding to a compactification of M-theory with vanishing flux, as are the general type II solutions of GMTP~\cite{Grana06} and both will provide useful examples of generalised $\mathcal{N}=1$ structures in the following sections. 

Generic $\mathcal{N}=1$, $D=4$ Minkowski flux compactifications of M-theory have been analysed using conventional geometrical techniques several years ago~\cite{Kaste:2003zd,Behrndt:2003zg,DallAgata:2003txk,Lukas:2004ip}. The metric takes a warped form 
\begin{equation}
\label{eq:metric-ansatz}
    \dd s^2 = \ee^{2\Delta}\dd s^2(\bbR^{3,1}) + \dd s^2(M),
\end{equation}
where $M$ is the compactification manifold, the internal four-form flux is non-trivial and the eleven-dimensional Killing spinors take the form 
\begin{equation}
\label{eq:M-spinors}
\epsilon=\eta_{+}\otimes\ee^{\Delta/2}\zeta^{\text{c}}+\eta_{-}\otimes\ee^{\Delta/2}\zeta
,
\end{equation}
where $\eta_{\pm}$ are chiral spinors of $\Spin{3,1}$ and $\zeta$ is a complex $\Spin7$ spinor. Supersymmetry implies $\bar{\zeta}\zeta$ is constant and there is vanishing four-form flux on the non-compact Minkowski space. In the $\Gx2$ case $\zeta$ is real. The analogous type II backgrounds were analysed by GMPT~\cite{Grana06}. In this case the two type II Killing spinors take the form 
\begin{equation}
\label{eq:typeII-spinors}
\begin{aligned}
\epsilon_1 &= \eta_{+}\otimes\zeta_1^+ + \eta_{-}\otimes\zeta_1^- , \\
\epsilon_2 &= \eta_{+}\otimes\zeta_2^\mp + \eta_{-}\otimes\zeta_2^\pm , \\
\end{aligned}
\end{equation}
where $\zeta_i^\pm$ are chiral $\Spin6$ spinors, and the upper and lower choices of sign refer to type IIA and IIB respectively. One can again construct a constant-norm, eight-component spinor
\begin{equation}
    \zeta = \ee^{\Delta/2}\ee^{-\hat\varphi/6}
        \begin{pmatrix} \zeta_1^+ \\ \zeta_2^- \end{pmatrix} ,
\end{equation}
where $\hat\varphi$ is the dilaton. Note that in both the M-theory and type II compactifications, although $\zeta$ is nowhere vanishing, the individual $\Spin7$ components (the real and imaginary parts of $\zeta$) or $\Spin6$ components (the $\zeta_i^\pm$) may vanish, and hence do not define conventional (global) $G$-structures.

However, these backgrounds do make sense globally as generalised $G$-structures~\cite{PW08,CSW14b}. To specify the background one needs the bosonic fields on $M$ together with the Killing spinor $\zeta$. In exceptional generalised geometry the bosonic fields define a generalised metric $\GM$. For example in M-theory $\GM$ is equivalent to the set $\{\Delta,g,A,\tilde{A}\}$ where $g$ is the seven-dimensional metric, $A$ is the three-form potential on $M$ and $\tilde{A}$ is the six-form potential encoding the dual of the four-form field strength on the Minkowski space. Geometrically  $\GM$ defines an $\SU8/\bbZ_2\subset\ExR{7(7)}$ generalised structure. The spinor $\zeta$ then transforms as the $\rep 8$ representation of the double cover, $\SU8$. The stabiliser of such a nowhere-vanishing constant-norm element is $\SU7$.\footnote{This is analogous to a nowhere-vanishing $\Spin6\simeq\SU4$ spinor being stabilised by $\SU3$.} In this way, we see that a supersymmetric $\mathcal{N}=1$ background defines a generalised $\SU7$ structure. The differential conditions on the Killing spinor are then equivalent to the vanishing of the generalised intrinsic torsion of the $\SU7$ structure~\cite{CSW14b}.

\subsection{\texorpdfstring{$\SU7$}{SU(7)} and \texorpdfstring{$\mathbb{R}^{+}\times\Uni7$}{R+ x U(7)} structures}\label{sec: SU(7) & U(7) structures}

Rather than defining the $\SU7$ structure using the pair $(\GM,\zeta)$ one can also define it directly in terms of generalised tensors. In fact there will be two kinds of structure in $\ExR{7(7)}$ that will interest us~\cite{PW08}:
\begin{equation}
\begin{aligned}
    J & \ :\  \text{stabilised by $G=\mathbb{C}^{*}\times\SU7=\mathbb{R}^{+}\times\Uni7$,} \\
    \psi & \ :\ \text{stabilised by $G=\SU7$.} 
\end{aligned}
\end{equation}
We will refer to $J$ as an exceptional complex structure and $\psi$ as a generalised $\SU7$ structure. They are stabilised by the same $\SU7$, but $J$ is also invariant under an extra $\mathbb{C}^{*}$ action. This is directly analogous to the relation between an almost complex structure $I$ in six dimensions (a $\GL{3,\mathbb{C}}$ structure) and a complex three-form $\Omega$ (an $\SL{3,\mathbb{C}}$ structure), or an almost generalised complex structure $\mathcal{J}$ and an almost generalised Calabi--Yau structure $\Phi$. 

To see how these structures are defined, for definiteness consider the M-theory case. Recall that the generalised tangent space is given by
\begin{equation}
\begin{aligned}
    E &\simeq T \oplus \ext^2T^* \oplus \ext^5T^* \oplus (T^*\otimes\ext^7T^*) , \\
    V &= v + \omega + \sigma + \tau,
\end{aligned}
\end{equation}
where $V\in\Gamma(E)$ and $E$ transforms in the $\rep{56}_{\rep{1}}$ of $\ExR{7(7)}$. Here the bold subscript denotes the $\bbR^+$ weight, normalised so that the determinant bundle $\det T^*$ has weight $2$. We will occasionally denote the components of a generalised vector explicitly as $V^M$, where $M=1,\dots56$. One can then define~\cite{AW15} two $\Ex{7(7)}$-invariant maps 
\begin{equation}
    s\colon \ext^2 E \to \det T^* , \qquad
    q\colon S^4 E \to (\det T^*)^2 ,
\end{equation}
namely the symplectic invariant $s$ and symmetric quartic invariant $q$. We will also need the adjoint bundle
\begin{equation}
    \ad \tilde{F} \simeq \bbR \oplus (T\otimes T^*) \oplus \ext^3T \oplus \ext^3 T^*
       \oplus \ext^6T \oplus \ext^6T^* , 
\end{equation}
transforming in Lie algebra representation $\rep{133}_{\rep{0}}\oplus\rep{1}_{\rep{0}}$, as well as a bundle $K$, given for example in~\cite{CSW11}, which contains the torsion of a generalised connection and transforms in the $\rep{912}_{\rep{-1}}$ representation. We also recall that the generalised Lie derivative~\cite{PW08,CSW11}, or Dorfman bracket, generates infinitesimal diffeomorphisms and gauge transformations and takes the form
\begin{equation}
\label{eq:Dorf-def}
    \Dorf_V \alpha= \mathcal{L}_v \alpha- \left(\dd\omega + \dd\sigma\right)\cdot\alpha ,
\end{equation}
when acting on a arbitrary generalised tensor $\alpha$, where $\mathcal{L}$ is the conventional Lie derivative, $\dd\omega$ and $\dd\sigma$ are regarded as sections of $\ad\tilde{F}$ and $\cdot$ denotes the adjoint action. In the following it will also be useful to use the ``twisted'' generalised Lie derivative defined for $A\in\Gamma(\ext^3T^*M$ and $\tilde{A}\in\Gamma(\ext^6T^*M)$ via (see for example~\cite[Appendix D]{AW15})
\begin{equation}
\label{eq:twist-Dorf}
\begin{split}
    \Dorf^{F+\tilde{F}}_V \alpha &:= \ee^{-A-\tilde{A}}L_{\ee^{A+\tilde{A}}\cdot V}
       \left(\ee^{A+\tilde{A}}\cdot\alpha\right) \\
       &= \mathcal{L}_v\alpha -  \bigl(\dd\omega - \imath_v F + \dd\sigma - \imath_v\tilde{F} + \omega\wedge F\bigr)\cdot\alpha ,
\end{split}
\end{equation}
where $F=\dd A$ and $\tilde{F}=\dd\tilde{A}-\frac12 A\wedge F$. Given a generalised connection $D$ we can define the generalised torsion $T\colon\Gamma(E)\to\Gamma(\ad\tilde{F})$ via~\cite{CSW11b}
\begin{equation}
\label{eq:torsion}
L_{V}\alpha=L_{V}^{D}\alpha-T(V)\cdot \alpha,
\end{equation}
where
\begin{equation}
\label{eq:Dorf-D}
    L_{V}^{D}\alpha = D_V \alpha - (D\oadj V)\cdot \alpha,
\end{equation}
where $\oadj$ is a projection $\oadj\colon E^*\otimes E\to \ad\tilde{F}$ and $D_V=V^MD_M$ is the generalised derivative along $V$. One finds that this definition implies the torsion actually lies in $K\oplus E^*\subset E^*\otimes\ad\tilde{F}$. 

Let us turn first to defining the structure $J$. Recall that, at a point on the manifold, the generalised metric defines an $\SU8/\bbZ_2$ subgroup of $\ExR{7(7)}$, and the spinor $\zeta$ defines an $\SU7$ subgroup of $\SU8$. There is a $\Uni1\subset\SU8/\bbZ_2$ that commutes with this $\SU7$ subgroup. It is generated by an element of the $\su8$ Lie algebra conjugate to the diagonal matrix
\begin{equation}
\label{eq:u1-generator}
    \alpha =\diag(-1/2,-1/2,\ldots,7/2) \in \su8 \subset \ex{7(7)} \oplus \bbR . 
\end{equation}
The normalisation is chosen so that $\exp(\ii\theta J)$ with $0\leq\theta<2\pi$ generates a $\Uni1$ subgroup of $\SU8\quotient\mathbb{Z}_{2}$. Note that the commutant of this $\Uni1$ is an $\bbR^+\times\Uni7$ subgroup of $\ExR{7(7)}$. Globally the $\Uni1$ at each point will be generated by a section of the adjoint bundle $J\in\Gamma(\ad \tilde{F})$ that is conjugate to $\alpha$ at each point. This leads us to the definition:
\begin{defn*}
A \emph{generalised $\bbR^+\times U(7)$ structure} or \emph{almost exceptional complex structure} is a section $J\in\Gamma(\ad\tilde{F})$ that is conjugate at each point $p\in M$ to the element $\alpha\in\su8 \subset \ex{7(7)} \oplus \bbR$ given in~\eqref{eq:u1-generator}. 
\end{defn*}
\noindent
Since the maximal compact subgroup $\SU8/\bbZ_2\subset\Ex{7(7)}$ and the maximal torus of $\SU8$ are each unique up to conjugation, every reduction of the structure group of $E$ to $\bbR^+\times\Uni7$ should be included in the definition. Furthermore all such structures will be related by local $\ExR{7(7)}$ transformations. Hence, as discussed in \cite{PW08}, the choice of $J$ does not fill out all of the  $\rep{133}$ representation space but instead lies within a particular orbit. Concretely, decomposing $\Ex{7(7)}$ using explicit $\SU{8}$ indices (see~\cite{PW08} or~\cite{CSW14}) we have 
\begin{equation}
\rep{133} =\rep{63}\oplus\rep{70} \ni(\mu^{\alpha}{}_{\beta},\mu_{\alpha\beta\gamma\delta}), 
\end{equation}
and we can write $J$ using the spinor $\zeta$ as
\begin{equation}
    J^{\alpha}{}_{\beta} = 4\zeta^\alpha \bar{\zeta}_\beta 
        -\tfrac{1}{2} (\bar\zeta \zeta) \delta^\alpha{}_\beta,\qquad J_{\alpha\beta\gamma\delta}=0,
\end{equation}
where we have normalised $\bar\zeta\zeta=1$. For completeness, we note that further decomposing under $\SU7\times\Uni1$ we have 
\begin{equation}
\label{eq:ad-SU7}
    \rep{133} = \rep{1}_{0} \oplus \rep{48}_{0} 
        \oplus (\rep{7}_{-4} \oplus \repb{7}_{4})
        \oplus (\rep{35}_{2} \oplus \repb{35}_{-2})
\end{equation}
where now the subscripts denote the $\Uni1$ charge, and $J$ lies in the singlet $\rep{1}_{0}$ representation. 

Given $J$, in analogy with a conventional almost complex structure, we can use it to decompose the complexified generalised tangent space. Under the adjoint action of $J$ on sections of the generalised tangent bundle, decomposing under $\SU7\times\Uni1$, we find
\begin{equation}
\label{eq:gen_vec-SU7}
\begin{split}
E_{\mathbb{C}} & =L_{3}\oplus L_{-1}\oplus L_{1}\oplus L_{-3},\\
\rep{56}_{\mathbb{C}} & =\rep 7_{3}+\rep{21}_{-1}+\repb{21}_{1}+\repb{7}_{-3}. 
\end{split}
\end{equation}
Thus we get four rather than two subbundles, with $L_{-3}\simeq \bar{L}_3$ and $L_{-1}\simeq\bar{L}_1$. As we will see, $L_3$ will play the analogue of the role of $T^{1,0}$ in conventional complex geometry. As such, this leads to the alternative definition
\begin{defn*}
A almost exceptional complex structure is a subbundle $L_{3}\subset E_{\mathbb{C}}$ such that
	\begin{enumerate}
		\item[i)] $\dim_{\mathbb{C}}L_{3}=7$,
		\item[ii)] $L_{3}\times_{N}L_{3}=0$,
		\item[iii)] $L_{3}\cap\bar{L}_{3}=\{0\}$,
		\item[iv)] The map $h\colon L_{3}\times L_{3}\rightarrow (\det T^*)_\bbC$ defined by $h(V,W)=\ii \, s(V,\bar{W})$ is a definite hermitian inner product valued in $\det T^*$,
	\end{enumerate}
where $\times_{N}\colon E\times E \to N$, with $N$ the generalised tensor bundle transforming in the $\rep{133}_{\rep2}$ representation, is an $\ExR{7(7)}$ covariant map given in~\cite{CSW11}. In analogy with the generalised complex structure case we call a subbundle $L_3$ satisfying the first two conditions a \emph{(complex) exceptional polarisation}. 
\end{defn*}
\noindent
Note that the (complex) stabiliser groups in $\Ex{7,\bbC}$ of all exceptional polarisations are isomorphic. However the corresponding real stabiliser groups in $\ExR{7(7)}$ can differ. In particular, only almost exceptional complex structures are stabilised by a  subgroup $\Uni7\times\bbR^+\subset\ExR{(7(7)}$.

We now turn to the $\SU7$ structure $\psi$. Decomposing the $\rep{912}$ representation under $\SU7\times\Uni1\subset\SU8/\bbZ_2\subset\Ex{7(7)}$, we find
\begin{equation}
\label{eq:912intoSU7}
\begin{aligned}
    \rep{912} &= \rep{36} \oplus \rep{420} \oplus \text{c.c} , \\
    &= \rep{1}_{7} \oplus \rep{7}_{3} \oplus \rep{28}_{-1}
       \oplus \rep{21}_{-1} \oplus \rep{35}_{-5} \oplus \rep{140}_{3}
       \oplus \rep{224}_{-1} \oplus \text{c.c.}
\end{aligned}
\end{equation}
where the subscript denotes the $\Uni1$ charge. Consider the generalised tensor bundle transforming in the $\rep{912}_{\rep{3}}$ representation of $\ExR{7(7)}$ (where the bold subscript denotes the $\bbR^+$ weight; the reason for this particular choice will be discussed below)
\begin{equation}
\begin{split}
\tilde{K} = (\det T^{*})^{2}\otimes K
& \simeq\mathbb{R}\oplus\ext^{3}T^{*}\oplus(T^{*}\otimes\ext^{5}T^{*})\oplus(S^{2}T^{*}\otimes\ext^{7}T^{*})\\
& \eqspace\oplus(\ext^{3}T^{*}\otimes\ext^{6}T^{*})\oplus(\ext^3 T \otimes (\ext^7 T^*)^3)\oplus\ldots,
\end{split}
\end{equation}
where $K\subset E^{*}\otimes\ad\tilde{F}$ is the torsion bundle~\cite{CSW11}.  The $\SU7$ singlet in the decomposition~\eqref{eq:912intoSU7} implies that  each almost exceptional complex structure $J$ defines a unique line bundle $\mathcal{U}_J\subset\tilde{K}_\bbC$, satisfying 
\begin{equation}
\label{eq:psi-bundle}
V\bullet \psi = 0 \quad \forall \;V\in\Gamma(L_3), \qquad s(\psi,\bar{\psi})\neq 0, 
\end{equation}
where $\psi$ is a local section of $\mathcal{U}_J$, the product $V\bullet\psi$ is defined by the projection map $E\otimes \tilde{K}\to C$ where $C$ is the generalised tensor bundle transforming in the $\rep{8645}_{\rep{4}}$ representation\footnote{Note that this representation is just the next step in the tensor hierarchy~\cite{Bergshoeff:2007qi,deWit:2008ta} above $\rep{912}$.} of $\ExR{7(7)}$, and $s$ is the symplectic invariant on the $\rep{912}$ bundle $\tilde{K}\subset E\otimes E\otimes E$ induced from the symplectic invariant on the $\rep{56}$ bundle $E$. One can equivalently define a local section $\psi$ by the condition $J\psi=7\ii\psi$ under the adjoint action of $J$. In complete analogy with the almost complex and almost generalised complex cases we are then led to define 
\begin{defn*}
Given an almost exceptional complex structure $J$ with trivial line bundle $\mathcal{U}_J$, a \emph{generalised $SU(7)$ structure} is a global nowhere-vanishing section $\psi\in\Gamma(\mathcal{U}_J)$. 
\end{defn*}
\noindent
Again we expect all generalised geometries with $\SU7$ structure group will arise this way, and furthermore any two such structures will be related by a local $\ExR{7(7)}$ transformation. In particular, any two generalised $\SU7$ structures with the same almost exceptional complex structure $J$ will be related by a nowhere vanishing complex function $f$
\begin{equation}
    \psi' = f \psi . 
\end{equation}
Again~\cite{PW08}, $\psi$ parametrises a particular orbit in the $\rep{912}$ representation rather than filling out the whole representation space. One could always write down the (non-linear) conditions on $\psi$ (and $J$ for that matter) which define the relevant orbits, but we have not attempted to do so. This would give conditions that are the analogue of stability for a three-form $\Omega$ and non-degeneracy for a two-form $\omega$. Instead, we can always write $\psi$ concretely using the spinor $\zeta$ and generalised metric $\GM$. Under the decomposition in~\eqref{eq:912intoSU7} we can write a section of $\tilde{K}$ in explicit $\SU8$ indices~\cite{PW08,CSW14} as 
\begin{equation}
    \kappa = (\kappa^{\alpha\beta},\kappa^{\alpha\beta\gamma}{}_{\delta},\bar{\kappa}_{\alpha\beta},\bar{\kappa}_{\alpha\beta\gamma}{}^{\delta}) 
    \in \Gamma(\tilde{K}_\bbC) . 
\end{equation}
The $\SU7$ structure can then be written as
\begin{equation}
\label{eq:psi-zeta}
    \psi^{\alpha\beta} = \lambda (\vol_{\GM})^{3/2} \zeta^\alpha\zeta^\beta , \qquad
    \bar{\psi}_{\alpha\beta} = \psi^{\alpha\beta\gamma}{}_\delta 
       = \bar{\psi}_{\alpha\beta\gamma}{}^\delta = 0 , 
\end{equation}
where $\vol_{\GM}=\ee^{2\Delta}\sqrt{g}$ is the $\Ex{7(7)}$-invariant volume defined by the generalised metric~\cite{CSW11b,CSW14} and $\lambda$ is a non-zero complex number. 

Recall that, since $\SU7\subset\SU8$, the generalised structure $\psi$ also defines a generalised metric and so completely specifies the supergravity background. This is analogous to a $\Gx2$ structure in conventional geometry, where the invariant three-form $\varphi$ defines a metric. In this way, our construction gives what one might call a ``generalised $\Gx2$ structure''. However this obscures the fact that the stabiliser group is actually $\SU7$ and not $\Gx2$ or $\Gx2\times\Gx2$ as might be expected, so we do not follow this convention. Later we will see that for the example of a conventional $\Gx2$ structure, the invariant three-form $\varphi$ does indeed define both $\psi$ and $J$.

\subsection{Supersymmetry and integrability}\label{sec: SUSY + Intrinsic Torsion}

We now turn to the conditions imposed on the generalised structures $\psi$ and $J$ by supersymmetry. As shown in~\cite{CSW14b,CS15,CS16}, the vanishing of the generalised intrinsic torsion for the $\SU7$ structure is equivalent to $\mathcal{N}=1$ supersymmetry for the Minkowski space solution. 
In what follows it will be useful to consider the intrinsic torsion for both $\psi$ and $J$ as the conditions for a torsion-free $J$ are a subset of those for $\psi$. This will allow us to see that integrability for $\psi$ follows from integrability for $J$, phrased in terms of an involution condition plus the vanishing of a moment map for generalised diffeomorphisms, that is the group of diffeomorphisms and form-field gauge transformations. 

Following the analysis in~\cite{CSW14b,CS15,CS16}, it is easy to show that intrinsic torsion for each generalised structure lies in a sub-bundle of $\rep{912}$ torsion bundle $K$ transforming as
\begin{align}
W_{\SU7}^{\text{int}} :& 
    \quad \rep 1_{\rep{-7}} \oplus \bar{\rep{7}}_{\rep{-3}}
    \oplus \rep{21}_{\rep{-1}} \oplus \rep{35}_{\rep{-5}} \oplus\text{c.c.}
    \label{eq:su7_torsion}\\
W_{\mathbb{R}^{+}\times\Uni7}^{\text{int}} :& 
    \quad \rep 1_{\rep{-7}} \oplus \rep{35}_{\rep{-5}} 
    \oplus\text{c.c.}\label{eq:ru7_torsion}
\end{align}
where again, the subscript denotes that $\Uni1$ charge under the action of $J$. We saw earlier how integrability of a complex structure can be recast as involutivity of eigenspaces of the complex structure under the Lie bracket. It is thus natural to define:
\begin{defn*}
	A torsion-free $\mathbb{R}^{+}\times\Uni7$ structure $J$ or \emph{exceptional complex structure} is one satisfying involutivity of $L_{3}$ under the generalised Lie derivative 
	\begin{equation}
	\Dorf_{V}W\in\Gamma(L_{3}),\qquad V,W\in\Gamma(L_{3}). \label{eq:involution}
	\end{equation}
	Again in analogy with the generalised geometry case, we call the weaker case of an involutive exceptional polarisation, an \emph{exceptional Dirac structure}. 
\end{defn*}
\noindent
In general $\Dorf_VW\neq-\Dorf_WV$, however the definition of an exceptional polarisation implies
\begin{equation}
    \Dorf_VW = \llbracket V,W\rrbracket \qquad V,W\in\Gamma(L_{3}),
\end{equation}
where $\llbracket V,W\rrbracket=\frac12\left( \Dorf_VW- \Dorf_WV\right)$ is the antisymmetric Courant bracket, and in fact the involution condition could be equally well defined using the Courant bracket as the generalised Lie derivative.

To prove that involutivity is equivalent to vanishing intrinsic torsion of the $\mathbb{R}^{+}\times\Uni7$ structure, we first recall that we can always find a generalised connection $D$ that is compatible with the $\mathbb{R}^{+}\times\Uni7$ structure, so that $DJ=0$, but it will not necessarily be torsion free. Consider the definition of the torsion~\eqref{eq:torsion} with $V,W=\alpha\in\Gamma(L_{3})$. Compatibility of the connection with $J$ ensures $L_{V}^{D}W\in L_{3}$, so involutivity amounts to checking that $T(V)\cdot W$ is in $L_{3}$ only. Since the left-hand side of~\eqref{eq:torsion} does not depend on the choice of compatible connection, only the intrinsic torsion contributes to the components of $T(V)\cdot W$ not in $L_3$. Explicitly, the intrinsic torsion representations contribute to $T(V)\cdot W\in\Gamma(E)$ as
\begin{equation}
\label{eq:involution-torsion}
\begin{aligned}
\rep{1}_{-7}\otimes\rep{7}_{3}\otimes\rep 7_{3}
   &\supset \rep{21}_{-1} , \\
\rep{35}_{-5}\otimes\rep 7_{3}\otimes\rep 7_{3} 
   &\supset \repb{21}_{1} .
\end{aligned}
\end{equation}
In other words, a non-zero $\rep 1_{-7}$ component of the torsion would generate a $\rep{21}_{-1}=L_{-1}$ term in $L_{V}W$. Requiring $L_{V}W\cap L_{-1}=\{0\}\; \forall\,V,W\in \Gamma(L_{3})$ thus sets the $\rep{1}_{-7}$ component of the torsion to zero. In a similar way, one sees that the $\rep{35}_{-5}$ component is set to zero by $L_{V}W\cap L_{1}=\{0\}$. One has $L_{V}W\cap L_{-3}=\{0\}$ identically just by counting the $\Uni1$ charges. 

We now need to consider the remaining conditions that imply we have a torsion-free $\SU7$ structure and hence an  $\mathcal{N}=1$, $D=4$ background. 
Comparing the representations that appear in the intrinsic torsion for the $\mathbb{R}^{+}\times\Uni7$ and $\SU7$ structures~\eqref{eq:su7_torsion} and~\eqref{eq:ru7_torsion}, we see there must then be an additional condition that sets the $\rep 7$ and $\rep{21}$ components of the $\SU7$ intrinsic torsion to zero. As we will now show these appear as the vanishing of a moment map $\mu$ for the action of generalised diffeomorphisms on the space of $\psi$ structures.

One first notes that the space of $\SU7$ structures on $M$ admits a natural pseudo-Kähler metric. This is a consequence of viewing the theory as a rewriting of the ten- or eleven-dimensional theory so that only four supercharges are manifest (the analogous situation for $\mathcal{N}=2$ theories was described in~\cite{GLW06,GLW07,GLSW09,AW15}). In analogy to~\cite{DN86}, the local $\SO{9,1}$ or $\SO{10,1}$ Lorentz symmetry is broken and the four-dimensional scalar degrees of freedom, that is the space of generalised $\SU7$ structures, can be packaged into $\mathcal{N}=1$, $D=4$ chiral multiplets~\cite{PW08}. As such they must admit a Kähler metric, albeit infinite-dimensional. The explicit construction is as follows. At a point $p\in M$, a choice of $\psi$ is equivalent to picking a point in the coset
\begin{equation}
\label{eq:QSU7}
\psi|_{p}\in Q_{\SU7} = \frac{\ExR{7(7)}}{\SU7},
\end{equation}
so that an $\SU7$ structure on $M$ corresponds to a section of the fibre bundle
\begin{equation}
Q_{\SU7}\to \mathcal{Q}_{\SU7}\to M.
\end{equation}
We can then identify
\begin{equation}
    \text{space of $\SU7$ structures, $\ZS{\SU7}$}
    \simeq \Gamma(\mathcal{Q}_{\SU7}) . 
\end{equation}
The key point is that the space $Q_{\SU7}$ admits a homogeneous pseudo-Kähler metric of signature $(70,16)$, picked out by supersymmetry. One first notes that the related space $\Ex{7(7)}/\Uni7$ admits a homogeneous pseudo-Kähler metric by a classic result of Borel~\cite[Proposition~2]{Borel54}. The metric is unique up to an overall scale~\cite{dorfmeister1991fine}. The space $Q_{\SU7}$ can be viewed as a complex line bundle $\lb$ over $\Ex{7(7)}/\Uni7$, with the zero section removed, since we only have an $\bbR^+$ action.  There is then a natural one-parameter family of conical, homogeneous Kähler metrics on $Q_{\SU7}$, distinguished by the relative size of the $\Uni1$ circle relative to the  $\Ex{7(7)}/\Uni7$ base. The infinite-dimensional space of structures $\ZS{\SU7}$ then inherits a pseudo-Kähler structure from the pseudo-Kähler structure on $Q_{\SU7}$. Our choice of $\bbR^+$ weight for $\psi$ picks out a particular Kähler metric within the one-parameter family with the explicit Kähler potential given by
\begin{equation}\label{eq:kahler_potential}
\mathcal{K}=\int_{M}\left(\ii\,s(\psi,\bar{\psi})\right)^{1/3},
\end{equation}
where $\psi$ can be viewed as a complex coordinate on the space of structures, or more precisely as a holomorphic embedding $\psi\colon\ZS{\SU7}\hookrightarrow \Gamma(\tilde{K}_\bbC)$. Given the $\bbR^+$ weight of the $\tilde{K}$ bundle, we need to take the $1/3$-power so that the integrand in~\eqref{eq:kahler_potential} is a section of $\det T^*$ and hence can be integrated over $M$. A different choice of weight would have led to a different power in $\mathcal{K}$ and hence a different Kähler metric. 

In analogy to the $\mathcal{N}=2$ case described in~\cite{GLW06,GLW07,GLSW09,AW15}, the existence of the Kähler structure follows from supersymmetry. As we mentioned above, one can consider rewriting the ten- or eleven-dimensional theory so that only four supercharges are manifest. Similar to~\cite{DN86}, the local $\SO{9,1}$ or $\SO{10,1}$ Lorentz symmetry is broken and the internal degrees of freedom can be packaged into $\mathcal{N}=1$, $D=4$ chiral multiplets~\cite{PW08} coupled to four-dimensional supergravity. Note that there are an infinite number of fields as no Kaluza--Klein truncation is performed: one keeps all modes on the internal space. The scalar degrees of freedom should hence parametrise an infinite-dimensional Kähler space, but from our discussion, this is just the space of generalised $\SU7$ structures.\footnote{As we discuss below, the chiral multiplet space is strictly a $\bbC^*$ quotient of the space of structures.} In this context, the particular weight of $\psi$, and hence Kähler metric, is fixed by the four-dimensional supersymmetry. In particular, as we will see in section~\ref{sec:super-pot}, the power of $1/3$ is required for the $D=4$, $\mathcal{N}=1$ superpotential on the space of chiral fields parametrising $\ZS{\SU7}$ to be a holomorphic function of $\psi$. 

We can write the symplectic structure corresponding to~\eqref{eq:kahler_potential} very explicitly as follows. Using $\varpi =\ii\,\partial'\bar{\partial}'\mathcal{K}$, we have, contracting two vectors $\alpha, \beta\in\Gamma(T\ZS{\SU7})$ into the symplectic form,
\begin{equation}\label{eq:symplectic_form}
\begin{split}
\imath_{\beta}\imath_{\alpha}\varpi
& =\ii\int_{M}\frac{1}{3}\frac{1}{\bigl(\ii\,s(\psi,\bar{\psi})\bigr)^{2/3}}\biggl(\ii\,s(\imath_{\alpha}\delta\psi,\imath_{\beta}\delta\bar{\psi})-\ii\,s(\imath_{\beta}\delta\psi,\imath_{\alpha}\delta\bar{\psi})\\
&\phantom{=\ii\int_{M}\frac{1}{3}}-\frac{2}{3}\frac{\ii\,s(\imath_{\alpha}\delta\psi,\bar{\psi})\,\ii\,s(\psi,\imath_{\beta}\delta\bar{\psi})}{\ii\,s(\psi,\bar{\psi})}+\frac{2}{3}\frac{\ii\,s(\imath_{\beta}\delta\psi,\bar{\psi})\,\ii\,s(\psi,\imath_{\alpha}\delta\bar{\psi})}{\ii\,s(\psi,\bar{\psi})}\biggr).
\end{split}
\end{equation}
Note that if we define a new \emph{non-holomorphic} parametrisation
\begin{equation}
    \phi = \left(\ii s(\psi,\bar{\psi})\right)^{-1/3} \psi , 
\end{equation}
which transforms in the $\rep{912}_{\rep{1}}$ representation, the symplectic structure takes the simple form 
\begin{equation}
\imath_{\beta}\imath_{\alpha}\varpi=-\frac{1}{3}\int_{M}\bigl(s(\imath_{\alpha}\delta\phi,\imath_{\beta}\delta\bar{\phi})-s(\imath_{\beta}\delta\phi,\imath_{\alpha}\delta\bar{\phi})\bigr),
\end{equation}
that is, it is just the pull-back $\varpi=\frac13\phi^*s$ of the symplectic form on the space of $\phi$. 

One can also restrict to the subspace of structures that define an (integrable) exceptional complex structure, so that $L_3$ is involutive,
\begin{equation}
    \ZI{\SU7} 
    = \{ \psi\in \ZS{\SU7} \; | \; \text{$J$ is integrable} \} .
\end{equation}
As we will show in section~\ref{sec:super-pot}, the integrability condition is holomorphic as a function of $\psi$ and so $\ZI{\SU7}$ inherits a Kähler metric from the one on $\ZS{\SU7}$, with the same Kähler potential. 

Finally we can turn to the remaining integrability conditions for the $\SU7$ structure. As in our previous examples, the Kähler structure on $\ZI{\SU7}$ is invariant under generalised diffeomorphisms. Infinitesimally these are generated by the generalised Lie derivative, and parametrised by generalised vectors $V\in\Gamma(E)$. As deformations in the space of structures, this defines a vector field $\rho_V\in\Gamma(T\ZI{\SU7})$
\begin{equation}
    \imath_{\rho_V} \delta\phi = \Dorf_V \phi ,
\end{equation}
where for convenience we are using the non-holomorphic structure $\phi$. We then have \begin{equation}\label{eq:vary_omega}
\begin{split}
\imath_{\rho_{V}}\imath_{\alpha}\varpi & =\frac{\ii}{3}\int_{M}\bigl(\ii\,s(\imath_{\alpha}\delta\phi,\Dorf_{V}\bar{\phi})-\ii\,s(\Dorf_{V}\phi,\imath_{\alpha}\delta\bar{\phi})\bigr)\\
& =-\frac{\ii}{3}\int_{M}\bigl(\ii\,s(\Dorf_{V}\imath_{\alpha}\delta\phi,\bar{\phi})+\ii\,s(\Dorf_{V}\phi,\imath_{\alpha}\delta\bar{\phi})\bigr)\\
& =\imath_{\alpha}\delta\left(\tfrac{1}{3}\int_{M}s(\Dorf_{V}\phi,\bar{\phi})\right)\\
& =\imath_{\alpha}\delta\mu(V),
\end{split}
\end{equation}
where we have used compactness to integrate by parts and have defined 
\begin{equation}\label{eq:moment_map_su7}
\begin{split}
    \mu(V) &= \tfrac{1}{3}\int_{M}s(\Dorf_{V}\phi,\bar{\phi}) \\
       &= \tfrac{1}{3}\int_{M}s(\Dorf_{V}\psi,\bar{\psi})
           (\ii\,s(\psi,\bar{\psi}))^{-2/3} ,
\end{split}
\end{equation}
where in going to the second line we use $\int_M\Dorf_V(\cdots)=0$. This gives a moment map $\mu\colon\ZI{\SU7}\to\gdiff^*$, where, as before, $\gdiff^*$ is the dual of the Lie algebra of generalised diffeomorphisms. 

We now want to prove that integrability of $\psi$ is equivalent to the vanishing of the moment map~\eqref{eq:moment_map_su7}. Let $D$ be a (torsionful) generalised connection compatible with the $\SU7$ structure, that is $D\psi=0$ (and hence $D\phi=0$). Using the definition of torsion~\eqref{eq:torsion}, we have 
\begin{equation}
\begin{split}
    \mu(V) &= \tfrac{1}{3}\int_{M}s((\Dorf^D_{V}\phi,\bar{\phi}) 
           - s(T(V)\phi,\bar{\phi}) \\
       &= \tfrac{1}{3}\int_{M}s(D_{V}\phi,\bar{\phi})
           -s((D\oadj V)\phi,\bar{\phi}) - s(T(V)\phi,\bar{\phi}) \\
       &= - \tfrac{1}{3}\int_{M}s(T(V)\phi,\bar{\phi}) ,
\end{split}
\end{equation}
where in moving to the last line we integrate the middle term by parts and use $D\phi=D\bar{\phi}=0$. Since the definition of $\mu$ is independent of any choice of connection, only the $\SU7$ intrinsic torsion can contribute in the last expression. Given that the generalised vector $V\in\Gamma(E)$ transforms in the $\rep 7+\rep{21}+\text{c.c.}$ representation, and $\phi$ is an $\SU7$ singlet, only the $\rep 7+\rep{21}+\text{c.c.}$ representations of the $\SU7$ intrinsic torsion can appear\footnote{Note that there could in principle be a further kernel in the map from the intrinsic torsion to $\mu$ so that only one of the $\rep{7}$ and $\rep{21}$ representations appeared. However it is easy to show that both representations are in fact present.} in $\mu$. However, from~\eqref{eq:ru7_torsion} and~\eqref{eq:su7_torsion}, we see these are precisely the additional components that must be set to zero for an exceptional complex structure to be an integrable $\SU7$ structure. Thus we have shown that the following definition is consistent:
\begin{defn*}
A torsion-free generalised $\SU7$ structure is one where $L_3$ is involutive and the moment map~\eqref{eq:moment_map_su7} vanishes. 
\end{defn*}

Since two $\SU7$ structures that are related by diffeomorphisms and gauge transformations give physically equivalent backgrounds, the moduli space of $\SU7$ structures is naturally a symplectic quotient by generalised diffeomorphisms $\GDiff$,
or equivalently a quotient by the complexified group $\GDiff_\mathbb{C}$:
\begin{equation}\label{eq:com_quotient}
\mathcal{M}_{\psi}= \ZI{\SU7}\qquotient \GDiff \simeq \ZI{\SU7}\quotient \GDiff_\mathbb{C}.
\end{equation}

Recall that the moduli space of $\Gx2$ holonomy manifolds in M-theory is associated with $\mathcal{N}=1$, $D=4$ chiral superfields~\cite{Papadopoulos:1995da,Harvey:1999as,Gutowski:2001fm,Beasley:2002db}. For a generic $\mathcal{N}=1$, $D=4$ background, supersymmetry again implies that the moduli space of integrable (torsion-free) generalised $\SU7$ structures should again define fields in chiral multiplets. However, note that not all deformations of $\psi$ deform the physical fields on the internal space. In particular, only those within the coset $\ExR{(7(7)}/(\SU8/\bbZ_2)$ are physical (deformations that change the generalised metric). First note that, from the warped product form~\eqref{eq:metric-ansatz}, shifts of the warp factor $\Delta\to\Delta+c$ for some constant $c$ can be absorbed in the four-dimensional metric. Second note that any modulus that lies in $\SU8/\SU7$ would correspond to a change of Killing spinor $\zeta$ for the same physical background. However this just implies that the background admits a second Killing spinor and so really preserves $\mathcal{N}=2$ supersymmetry. The exception to this is the change of $\zeta$ by a constant phase, that is by the $\Uni1$ generated by $J$, since this too can always be reabsorbed into the four-dimensional spinors in the ansatzë~\eqref{eq:M-spinors} and~\eqref{eq:typeII-spinors}. Thus for honest $\mathcal{N}=1$ backgrounds we only need to consider the action of this $\Uni1$ and the shift in $\Delta$. As we note from the form of $\psi$ in~\eqref{eq:psi-zeta}, shifting $\Delta$ simple rescales $\psi$, in fact via the $\bbR^+$ action. Put together we see that the physical moduli space is given by 
\begin{equation}
\text{Moduli space of $\mathcal{N}=1$ background} 
   = \mathcal{M}_\psi\qquotient \Uni1 \simeq  \mathcal{M}_\psi\quotient\bbC^*,\nonumber
\end{equation}
where the $\bbC^*$ action acts as 
\begin{equation}
    \psi \to \lambda^3 \psi ,
\end{equation}
where we have normalised the $\bbC^*$ action to match the $\bbR^+$ action on $\psi$ which implies $\mathcal{K}\to|\lambda|^2\mathcal{K}$. Under the symplectic quotient, the physical moduli space has a K\"ahler potential $\tilde{\mathcal{K}}$ given by
\begin{equation}\label{eq:physical_kahler}
\tilde{\mathcal{K}} = - 3 \log \mathcal{K}.
\end{equation}
This is the K\"ahler potential that determines the metric on the moduli space of the supergravity background. For example, in the $\Gx2$ case that we will discuss in section \ref{sec: G2}, this reduces to the known result that the K\"ahler potential $\tilde{\mathcal{K}}=-3\log\int_M\vol$ describes the coupling of moduli in the four-dimensional effective theory, where $\vol$ is determined by the $\Gx2$ structure~\cite{Harvey:1999as,Gutowski:2001fm,Beasley:2002db}. 

Note that, strictly, one should check that the kinetic terms and potentials in the $D=4$ effective theory are given by $\tilde{\mathcal{K}}$ (specifically checking that the coefficient of $-3$ is correct). One check is to compare with the $\Gx2$ holonomy case, as we do in the next section. Alternatively, we can note that the quotient is simply the standard relation between the Kähler geometry in superconformal supergravity~\cite{Ferrara:1977ij, Kaku:1978nz,Kaku:1978ea}, using a compensator field, and the standard supergravity formalism where a gauge for the compensator is chosen. This fixes the $\bbC^*$ scaling of $\mathcal{K}$ and the factor of $-3$ comes from the standard normalisation of the gravitational coupling constant (as reviewed for example in~\cite{Kallosh:2000ve}).

\section{Examples of integrable generalised \texorpdfstring{$\SU7$}{SU(7)} structures}\label{sec:examples}

We now present two classic examples of $\mathcal{N}=1$, $D=4$ backgrounds and describe how they can be understood as integrable generalised $\SU7$ structures. We discuss $\Gx2$ backgrounds in M-theory and $\mathcal{N}=1$ GMPT backgrounds in type II theories. In both of these cases, we will see that involutivity of the $L_3$ subbundle reproduces a subset of the known differential conditions these backgrounds must satisfy. The final differential conditions come from the vanishing of the moment map. In particular, this gives a completely new way of viewing $\Gx2$ manifolds that, intriguingly, closely mirrors the discussion of complex structures. 

\subsection{\texorpdfstring{$\Gx2$}{G2} structures in M-theory}\label{sec: G2}

Recall that a $\Gx2$ structure is defined by a nowhere-vanishing three-form $\varphi\in\Gamma(\ext^{3}T^{*}M)$, which can be written in a local frame as
\begin{equation}
\varphi=e^{246}-e^{235}-e^{145}-e^{136}+e^{127}+e^{347}+e^{567}.
\end{equation}
This defines a metric $g=e_{a}\otimes e_{a}$ and an orientation $\vol=e^{1\ldots7}=\star1$. The Hodge dual of $\varphi$ is
\begin{equation}
\star\varphi=e^{1234}+e^{1256}+e^{3456}+e^{1357}-e^{1467}-e^{2367}-e^{2457},
\end{equation}
so that $\varphi\wedge\star\varphi = 7 \vol$. The structure is integrable, that is we have a $\Gx2$ holonomy manifold, if and only if
\begin{equation}
    \dd \varphi=\dd \star\varphi = 0 . 
\end{equation}
Compactifying M-theory on a $\Gx2$ holonomy manifold with $\Delta=0$ gives a $\mathcal{N}=1$, $D=4$ background. One can also include non-trivial three-form potential $A$ such that $\dd A=0$.

We would like to first identify how a $\Gx2$ structure defines a generalised $\SU7$ structure. Before doing so it is useful to define the notion of ``type'' for almost exceptional complex structures in M-theory in an analogous way to the type of generalised complex structures given in \cite{Gualtieri04b}:
\begin{defn*}
	The \emph{type} of an almost exceptional complex structure $L_{3}\subseteq E_{\mathbb{C}}$ is the (complex) codimension of its projection onto the tangent bundle $T_\bbC$. That is, if $\pi\colon E\rightarrow T$ is the anchor map then
	\begin{equation}
	\operatorname{type} L_{3} \coloneqq  \operatorname{codim}_\bbC \pi(L_{3}) = 7-\dim_\bbC \pi(L_{3}).
	\end{equation}
\end{defn*}

A generic\footnote{Generic in the sense that the set of all seven-dimensional subspaces not of this type is measure zero in the Grassmannian.} seven-dimensional subspace of a fibre of $E$ will have a surjective projection onto the tangent space $T$, and hence a generic exceptional complex structure is type-0. We can always write such a space as
\begin{equation}
L_{3} = \ee^{\alpha+\beta}T_\bbC \qquad \text{for }\alpha\in \Gamma(\ext^{3}T^{*}_\mathbb{C}),\, \beta\in\Gamma( \ext^{6}T^{*}_\mathbb{C}).
\end{equation}
This identically satisfies the first two conditions for an almost exceptional complex structure, and one gets simple constraints on the polyform $\alpha+\beta$ from $L_3\cap\bar{L}_3=\{0\}$ and the requirement that $\ii \,s(V,\bar{W})$ for $V,W\in\Gamma(L_3)$ defines a definite hermitian inner product. Note that $T_\bbC$ is the simplest example of an exceptional Dirac structure (following the definition given in section~\ref{sec: SUSY + Intrinsic Torsion}), but is not an exceptional complex structure since, for example, $T_\bbC\cap\overline{T_\bbC}\neq0$. In terms of the Killing spinor $\zeta$, viewed as a complexified $\Spin7$ spinor, the requirement that the structure is type-0 is that the scalar $\zeta^\trsp\zeta$ is nowhere vanishing.\footnote{Note that this condition involves $\zeta^\text{T}\zeta$ and not $\bar\zeta \zeta$, which is what defines the $\SU7$ structure (see below \eqref{eq:M-spinors}). Given that $\psi$ is of the form \eqref{eq:psi-zeta}, this condition amounts to requiring that the $1\in\mathbb{R}$ component of $\psi$ is non-vanishing.} This is precisely the case discussed in~\cite{Lukas:2004ip} where the real and imaginary parts of $\zeta$ have different normalisations (and/or are non-orthogonal). The analyses in~\cite{KMT03} and~\cite{DallAgata:2003txk}, on the other hand, fix equal norms and orthogonal real spinors and so define a structure that is strictly not type-0. 

A $\Gx2$ structure, embedded in generalised geometry, defines the simplest example of a type-0 almost exceptional complex structure. Taking $\alpha=\ii\varphi$ and $\beta=0$, we have 
\begin{equation}
L_{3}=\ee^{\ii\varphi}T_\bbC ,
\end{equation}
so that a section of $L_3$ takes the form (using the ``$j$-notation'' of~\cite{CSW11})
\begin{equation}
\begin{split}
   \Gamma(L_3) &\ni
   v+\ii\,\imath_{v}\varphi- \tfrac12\varphi\wedge\imath_v\varphi
      - \tfrac16\ii\,j\varphi\wedge\varphi\wedge\imath_v\varphi \\
   &= v+\ii\,\imath_{v}\varphi-\star\imath_{v}\varphi-\ii\,v^{\flat}\otimes\vol , 
\end{split}    
\end{equation}
for some $v\in\Gamma(T_\bbC)$. The condition on $\ii\, s(V,\bar{W})$ for $V,W\in\Gamma(L_3)$ is equivalent to the weighted metric
\begin{equation}
\label{eq:tilde-g}
    \tilde g(v,w) = \imath_v\varphi\wedge\imath_{w}\varphi\wedge\varphi
\end{equation}
being positive definite for $v,w\in\Gamma(T)$. However, this is just the condition that $\varphi$ is a (positive) stable form in the sense of Hitchin~\cite{Hitchin00}. (It also implies $L_3\cap\bar{L}_3=\{0\}$.) The $\mathbb{R}^{+}\times\Uni 7$ structure $J$ is given by
\begin{equation}
J=\varphi^{\sharp}-\varphi,
\end{equation}
where $\varphi^{\sharp}$ is obtained from $\varphi$ by raising its indices using the inverse metric $g^{-1}$ defined by $\varphi$. One can check that this satisfies $JL_{3}=3\ii L_{3}$ using the action of the $\rep{133}$ on the $\rep{56}$ given in~\cite[Appendix C]{CSW11}. 

We now turn to the integrability condition on $J$. Involutivity of $L_{3}$ is simple to show using the properties of the generalised Lie derivative. Writing generic sections of $L_3$ as $V=\ee^{\ii\varphi}v$ and $W=\ee^{\ii\varphi}w$, given two vectors $v,w\in\Gamma(T_\bbC)$, we then have
\begin{equation}\label{eq:twist_Dorf_g2}
    \Dorf_{V}W=\Dorf_{\ee^{\ii\varphi}v}(\ee^{\ii\varphi}w)
       =\ee^{\ii\varphi}\Dorf_{v}^{\ii\,\dd\varphi+\frac12\varphi\wedge\dd\varphi}w
       = \ee^{\ii\varphi}\bigl([v,w]
       +\imath_{w}\imath_{v}(\ii\,\dd\varphi+\tfrac12\varphi\wedge\dd\varphi)\bigr) ,
\end{equation}
where we have used the twisted generalised Lie derivative~\eqref{eq:twist-Dorf}. 
The second term must vanish for the right-hand side to be a section of $L_{3}$ only. As this is defined for arbitrary $v$ and $w$, the bundle $L_3$ is involutive on $L_{3}$ if and only if we have a \emph{closed} $\Gx 2$ structure
\begin{equation}
\text{involutivity of $L_3$ :} \qquad \dd\varphi=0.
\end{equation}
This condition is weaker than a torsion-free $\Gx 2$ structure (which requires $\dd\star\varphi=0$ as well). A theorem due to Bryant states that, like symplectic structures, all closed $\Gx 2$ structures are equivalent, taking a standard form in a local patch~\cite{Bryant03}.

Now we examine the moment map for this example. To do so, we first need to define the $\SU7$ structure $\psi$. Recall that $\psi$ is a section of 
\begin{equation}
\begin{split}
\tilde{K} 
& \simeq\mathbb{R}\oplus\ext^{3}T^{*}\oplus(T^{*}\otimes\ext^{5}T^{*})\oplus(S^{2}T^{*}\otimes\ext^{7}T^{*})\\
& \qquad \oplus(\ext^{3}T^{*}\otimes\ext^{6}T^{*})\oplus(\ext^3 T \otimes (\ext^7 T^*)^3)\oplus\ldots ,
\end{split}
\end{equation}
and also that $V\bullet \psi=0$ for all $V\in\Gamma(L_3)$. Since $\bbR$ is the lowest degree term in $\tilde{K}$, we note that, taking $1\in\Gamma(\tilde{K})$, we must have $v\bullet 1=0$ for any vector $v\in\Gamma(T_\bbC)$ viewed as a section of $\Gamma(E_\bbC)$. Since $L_3=\ee^{\ii\varphi}T_\bbC$ we see this means we can construct $\psi$ as
\begin{equation}
\psi=\ee^{\ii\varphi}\cdot1,
\end{equation}
where the exponential acts on $1\in\mathbb{R}$ via the adjoint action. The components of $\psi$ have the form
\begin{equation}
\psi\sim(1,\varphi,j\varphi\wedge\varphi,\tilde{g},\ldots) .
\end{equation}
Recall that $s(\psi,\bar{\psi})\in\Gamma((\det T^{*})^{3})$, so it has $3\times7=21$ indices. Given that $\varphi\in\Gamma(\ext^3T^*)$, it 
hence must be degree $7$ in $\varphi$, meaning the Kähler potential~\eqref{eq:kahler_potential}
is degree $7/3$. This is precisely the same scaling as the $\Gx 2$ Hitchin functional~\cite{Hitchin00a,Hitchin01} so that, up to an overall constant, we must have
\begin{equation}\label{eq:g2_kahler}
\mathcal{K}=\int_{M}\bigl(\ii\,s(\psi,\bar{\psi})\bigr)^{1/3}
    \propto \int_{M}\varphi\wedge\star\varphi.
\end{equation}
One can check this is indeed the case by an explicit calculation.
Using the the twisted generalised Lie derivative and invariance of the symplectic form under a complexified $\Ex{7(7)}$ transformation, we can then calculate the moment map~\eqref{eq:moment_map_su7}
\begin{equation}
\begin{split}
\mu(V) & =\tfrac{1}{3}\int_{M}s\left(\Dorf_{V}(\ee^{\ii\varphi}\cdot1),\ee^{-\ii\varphi}\cdot1\right)(\ii\,s(\psi,\bar{\psi}))^{-2/3}\\
& =\tfrac{1}{3}\int_{M}s(\ee^{\ii\varphi}\Dorf_{\ee^{-\ii\varphi}V}^{\ii\,\dd\varphi+\frac12\varphi\wedge\dd\varphi}1,\ee^{-\ii\,\varphi}\cdot1)(\ii\,s(\psi,\bar{\psi}))^{-2/3}\\
& =\tfrac{1}{3}\int_{M}s(\Dorf_{\ee^{-\ii\varphi}V}^{\ii\,\dd\varphi+\frac12\varphi\wedge\dd\varphi}1,\ee^{-2\ii\varphi}\cdot1)(\ii\,s(\psi,\bar{\psi}))^{-2/3}.
\end{split}
\end{equation}
As $\ee^{-\ii\varphi}$ has no kernel, we can relabel $\ee^{-\ii\varphi}V\to V$ to give
\begin{equation}
\mu(\ee^{\ii\varphi}V)  =\tfrac{1}{3}\int_{M}s(\Dorf_{V}^{\ii\,\dd\varphi+\frac12\varphi\wedge\dd\varphi}1,\ee^{-2\ii\varphi}\cdot1)(\ii\,s(\psi,\bar{\psi}))^{-2/3}.
\end{equation}
Given $V=v+\omega+\sigma+\tau$ and $\dd\varphi=0$
\begin{equation}
\begin{split}
\Dorf_{V}^{\ii\,\dd\varphi+\frac12\varphi\wedge\dd\varphi}1 = \Dorf_V1 
     = - (\dd\omega+\dd\sigma)\cdot 1 
     = - \dd\omega - j\dd\sigma .  
\end{split}
\end{equation}
For general $\gamma_{mnp}\in\Gamma(\ext^3T^*)$ and $\pi_{m,n_1\dots n_5}\in\Gamma(T^*\otimes\ext^5T^*)$ we have 
\begin{equation}
\label{eq:gamma-pi}
\begin{split}
   s(\gamma+\pi,&\ee^{-2\ii\varphi}\cdot1)(\ii\,s(\psi,\bar{\psi}))^{-2/3}_{m_1\dots m_7} \\
      &= \text{const}\times \gamma_{[m_1m_2m_3}(\star\varphi)_{m_4m_5m_6m_7]}
         + \text{const}\times g^{np}\pi_{n,p[m_1m_2m_3m_4}\varphi_{m_5m_6m_7]} , 
\end{split}
\end{equation}
where, rather than evaluate the expression directly, we have used the facts that it must be linear in $\gamma$ and $\pi$ and a top form, and that the only $\Gx2$-invariant tensors are $\varphi$, $\star\varphi$, the metric $g$ and its inverse. However for $\pi=j\dd\sigma$ the second term vanishes.
We  thus have\footnote{Note that an analogous argument gives the same expression for the variation of the Kähler potential for $\delta\varphi=\dd\omega$. (This gives a reason for why the coefficient of the first term in~\eqref{eq:gamma-pi} cannot vanish; one knows the generic variation of the Hitchin function is non-zero.) As we will discuss in section~\ref{sec:kahler_potential}, this reflects the fact that the vanishing of the moment map is the same as the extremisation of the Kähler potential.}
\begin{equation}
\begin{split}
\mu(\ee^{\ii\varphi}V) 
& \propto\int_{M}\dd\omega\wedge\star\varphi\\
& \propto\int_{M}\omega\wedge\dd\star\varphi,
\end{split}
\end{equation}
where we have assumed that $M$ is compact to integrate by parts. 
\begin{equation}
\text{vanishing of moment map:} \quad \dd\star\varphi=0,
\end{equation}
so that the $\Gx 2$ structure must be torsion free. 

We can extend this example to include fluxes by including them in the complexified transformation as
\begin{equation}
L_{3}=\ee^{\tilde{A}+A}\ee^{\ii\varphi}T=\ee^{\tilde A -\tfrac{1}{2}\ii A\wedge\varphi+A+\ii\varphi} T,
\end{equation}
where $A$ and $\tilde{A}$ are three- and six-form potentials. The real $\Ex{7(7)}$ transformation by $A+\tilde{A}$ amounts to turning on four-form and seven-form fluxes, given by
\begin{equation}
F=\dd A,\qquad \tilde F = \dd \tilde A - \tfrac12 A\wedge F.
\end{equation}
The involutivity condition is now
\begin{equation}
[v,w]+\imath_{w}\imath_{v}(F+\ii\,\dd\varphi+\tilde{F}+\tfrac{1}{2}\varphi\wedge\dd\varphi-\ii F\wedge\varphi)\in\Gamma(TM),
\end{equation}
which holds if and only if
\begin{equation}
    \dd\varphi=F=\tilde{F}=0.
\end{equation}
In other words, involutivity of $L_3$ forces the $\Gx2$ structure to be closed and the fluxes to vanish. Note that one could include a warp factor by including $\ee^ \Delta$ in the definition of $L_3$ -- one would then find that involutivity also forces the warp factor to be constant. Since all the fluxes vanish, the twisted generalised Lie derivative is equal to the ordinary Lie derivative and the analysis of the $\mu=0$ condition is exactly as before, that is, it simply implies $\dd\star\varphi=0$, and the $\Gx2$ structure is integrable. We have thus reproduced the standard conditions for a supersymmetric compactification of M-theory on a $\Gx2$ manifold. For the $\SU7$ structure there is strictly one extra degree of freedom, since we can always rescale $\psi$ by a complex constant. As we discussed at the end of section~\ref{sec: SUSY + Intrinsic Torsion}, this rescaling is not physical.

Recall that $\SU7$ structures are equivalent if they differ by generalised diffeomorphisms. The gauge symmetries will simply shift
\begin{equation}
    A \to A + \dd \omega , \qquad \tilde{A} \to \tilde{A} + \dd\sigma , 
\end{equation}
thus the physical gauge degrees of freedom parametrise the de Rham cohomology classes $H_\dd^3(M,\bbR)$ and $H_\dd^6(M,\bbR)$. The conventional diffeomorphisms on the other hand simply relate diffeomorphic $\Gx2$ structures. Locally the moduli space of integrable $\Gx2$ structures is diffeomorphic to a open set of $H_\dd^3(M,\bbR)$ (see for example~\cite{Joyce00}). Furthermore $H_\dd^6(M,\bbR)=0$. As we will show in section~\ref{sec:moduli}, all deformations of the $\SU7$ structure either deform the $\Gx2$ structure $\varphi$, deform $A$ in $H_\dd^3(M)$ or correspond to rescaling $\psi$. Thus, dropping the non-physical rescaling, the physical moduli space is 
\begin{equation}
    \mathcal{M}_\psi\quotient\bbC^* \simeq H_\dd^3(M,\bbC) \qquad \text{(locally)} ,
\end{equation}
with the Kähler metric given by~\eqref{eq:physical_kahler}, as in~\cite{Harvey:1999as,Gutowski:2001fm,Beasley:2002db}. We will comment more on a formal way to treat this moduli problem in section \ref{sec:moduli}.

In summary, we have shown that by embedding the problem in $\ExR{7(7)}$ generalised geometry, the $\Gx2$ manifold has an intriguing reinterpretation, as a sort of generalised complex structure. There is an involutive complex subbundle whenever $\dd\varphi=0$, and the final condition $\dd\star\varphi=0$ comes from a moment map. 

\subsection{GMPT structures in type II}\label{sec:GMPT}

The GMPT solutions give $\mathcal{N}=1$ compactifications of type II supergravities and were first analysed in~\cite{GMPT05} and further studied in~\cite{Tomasiello07}. While the solutions are not completely general,\footnote{The construction requires that the two internal spinors $\{\eta^{1},\eta^{2}\}$ in~\eqref{eq:typeII-spinors} are nowhere vanishing. An example that falls outside of this classification is an NS5-brane wrapping a Calabi--Yau. As shown in \cite{AW15}, this class of solution can be embedded within exceptional generalised geometry.} they do cover a large class of compactifications in which the internal manifold has an $\SU{3}$ structure, an $\SU{2}$ structure, or an intermediate case where the two $\SU{3}$ structures can degenerate. The key observation of~\cite{GMPT05} was that these three cases are examples of $\SU{3}\times \SU{3}$ structures on the generalised tangent bundle $\EO=T\oplus T^{*}$ and can all be described as generalised Calabi--Yau manifolds  admitting two pure spinors~\cite{Hitchin02,Gualtieri04b}. We begin with a brief review of the key aspects of the GMPT solutions before embedding them into the $\SU{7}$ structures we have described above. We will use this formulation of the solutions to find their moduli in section \ref{GMPT Moduli}.

The GMPT solutions admit two non-vanishing, compatible pure spinors $\{\Phi_{+},\Phi_{-}\}$ with associated generalised complex structures $\{\mathcal{J}_{+},\mathcal{J}_{-}\}$ satisfying
\begin{equation}
(\Phi_{+},\cliff{V}\Phi_{-})=( \Phi_{+},\cliff{V}\bar{\Phi}_{-} ) = 0 \quad \forall\;V\in \EO \quad \Leftrightarrow \quad [\mathcal{J}_{+},\mathcal{J}_{-}] = 0,
\end{equation}
where $[\,,\,]$ is the usual commutator and the slash denotes the Clifford action, as it will for the remainder of this section. This is a special case of the generalised K\"{a}hler structures defined in \cite{Gualtieri10b} and gives an $\SU{3}\times \SU{3}$ structure.
The two pure spinors are constructed as bilinears of the Killing spinors $\{\zeta^\pm_1,\zeta^\pm_2\}$ given in~\eqref{eq:typeII-spinors}.  
The Killing spinor equations in terms of $\Phi^\pm$ were first given in~\cite{GMPT05}. Here it will be useful to use an equivalent form derived by Tomasiello~\cite{Tomasiello07}
\begin{gather}
    \dd\Phi_{\pm} = 0, \qquad \qquad 
    F = -8\,\dd^{\mathcal{J}_{\pm}}(\ee^{-3\Delta}\im\Phi_{\mp}), \label{eq:Tom a} \\
    \dd(\ee^{-\Delta}\re \Phi_{\mp}) = 0, \label{eq:Tom b}
\end{gather}
where $\dd^{\mathcal{J}}=[\dd,\mathcal{J}]$, the upper/lower sign is for type IIA/B, $\Delta$ is the warp factor in the string frame and $F$ is the Ramond--Ramond flux. The spinors are normalised so that 
\begin{equation}\label{eq:pure_norm}
    (\Phi_{+},\bar{\Phi}_+) = (\Phi_-,\bar{\Phi}_-) = \tfrac{1}{8}\ee^{6\Delta-2\dil}\vol,  
\end{equation}
where $(\cdot,\cdot)$ is the Mukai pairing~\eqref{eq:mukai}, $\dil$ is the dilaton and $\vol$ is the volume form defined by the string-frame metric.  Note that in~\cite{Tomasiello07} the twisted differential $\dd_H=\dd-H\wedge$ is used. Here we will use the convention that the $B$-field is included in the definition of the spinors and RR flux (that is they are twisted by $\ee^{-B}$ relative to those in~\cite{Tomasiello07}) and hence the usual differential $\dd$ appears. 

We now show how to embed these solutions into the framework of generalised $\SU7$ structures. We start by defining $L_{3}$ as\footnote{Such a procedure for going from generalised to exceptional complex structures was originally formulated in an $\Ex{6(6)}$ context by two of the current authors (AA and DW) with Michela Petrini and Edward Tasker~\cite{APTW}.} 
\begin{equation}\label{eq:gmpt_l3}
L_3 = \ee^{C+8\,\ii \,\ee^{-3\Delta}\im\Phi_{\mp}}(L^{\mathcal{J}_\pm}_{1}\oplus \mathcal{U}_{\mathcal{J}_{\pm}}).
\end{equation}
Here the upper/lower signs correspond to type IIA/B respectively, $L^{\mathcal{J}_\pm}_{1}\subset \EO_\bbC \simeq(T\oplus T^{*})_\bbC$ is the $+\ii$-eigenspace of $\mathcal{J}_{\pm}$, $\mathcal{U}_{\mathcal{J}_{\pm}}$ is the pure spinor line bundle defined by $\mathcal{J}_\pm$ 
and $C$ is the polyform potential for the RR flux $F$.
Note that in writing \eqref{eq:gmpt_l3}, we are implicitly using an embedding of the $\Orth{6,6}$ structures into the $\Ex{7(7)}$ generalised tangent bundle and adjoint bundle: this is given in appendix~\ref{Appendix:Conventions} for type IIB.\footnote{The powers of $\Delta$ in the normalisation~\eqref{eq:pure_norm} imply that $\Phi_\pm$ are sections of a weight-three bundle under the $\bbR^+$ action. The adjoint bundle is weight-zero, hence the $\ee^{-3\Delta}$ factor in~\eqref{eq:gmpt_l3}.} We will focus on type IIB for definiteness but analogous results hold in IIA with the appropriate embedding. It is relatively straightforward to check that $L_3$ satisfies the necessary and sufficient conditions to define an almost exceptional complex structure. 

Now we turn to the involutivity condition. We will show first that the untwisted bundle $L^{\mathcal{J}_-}_1\oplus \mathcal{U}_{\mathcal{J}_{-}}$ is involutive if and only if $\mathcal{J}_-$ is integrable. One can check that $\ii \, s(V,\bar{W})$ is not positive definite, thus it defines only an exceptional Dirac structure, but not an exceptional complex structure. We find that the modified bundle $L_3$ is involutive provided an extra condition on the twisting factor $C+8\,\ii \,\ee^{-3\Delta}\im\Phi_{-}$ is satisfied. Let 
\begin{equation}
    V = W + \alpha\,\Phi_{-} \in \Gamma(L^{\mathcal{J}_-}_1\oplus \mathcal{U}_{\mathcal{J}_{-}}) ,
\end{equation}
where $W\in\Gamma(L^{\mathcal{J}_-}_1)$ and $\alpha\in C^\infty(M,\bbC)$, and similarly for $V'$. Requiring 
\begin{equation}
    \Dorf_VV' = \Dorf_{W+\alpha\Phi_-}(W'+\alpha'\Phi)\in\Gamma(L^{\mathcal{J}_-}_1\oplus \mathcal{U}_{\mathcal{J}_{-}}),
\end{equation}
implies first that  
\begin{equation}
    \Dorf_WW' \in L^{\mathcal{J}_-}_1, \qquad \forall\; W,W'\in L^{\mathcal{J}_-}_1,
\end{equation}
that is, the generalised complex structure $\mathcal{J}_{-}$ associated to $\Phi_{-}$ must be integrable. From~\eqref{eq:dPhi} and $\cliff{W}\Phi_-=\cliff{W'}\Phi_-=0$ we then immediately have 
\begin{equation}
\begin{aligned}
    \Dorf_W(\alpha'\Phi_-) 
       &= (\mathcal{L}_v\alpha') \Phi_- + \alpha'\Dorf_W\Phi_-
       = \left< W,\dd\alpha'+2A\right> \Phi_- \in \mathcal{U}_{\mathcal{J}_{-}} , \\
    \Dorf_{\alpha\Phi_-}W'
       &= - \dd(\alpha\Phi_-)\cdot W' 
       = - \left< W',\dd\alpha+2A\right> \Phi_- \in \mathcal{U}_{\mathcal{J}_{-}} ,
\end{aligned}
\end{equation}
as required (in the second line $\dd(\alpha\Phi_-)$ acts via the $\Ex{7(7)}$ adjoint action). For the final term we have
\begin{equation}
    \Dorf_{\alpha\Phi_-} (\alpha'\Phi_-) 
       = - \dd(\alpha\Phi_-)\cdot(\alpha'\Phi_-)
       = - \alpha'[(\cliff{\dd\alpha}+\cliff{A})\Phi_-]\cdot\Phi_- = 0 
\end{equation}
identically, as can be seen simply by counting the $\mathcal{J}_-$ charge. Hence we see
\begin{equation}
    \text{involutive $L^{\mathcal{J}_-}_1\oplus \mathcal{U}_{\mathcal{J}_{-}}$}
    \qquad \Leftrightarrow \qquad \text{integrable $\mathcal{J}_-$} .
\end{equation}

We now define 
\begin{equation}
    \Sigma=C+8\,\ii\,\ee^{-3\Delta}\im\Phi_{+},
\end{equation}
so that $\ee^\Sigma V\in\Gamma(L_3)$ if $V\in\Gamma(L^{\mathcal{J}_-}_1\oplus \mathcal{U}_{\mathcal{J}_{-}})$. We then have\footnote{Note that the generalised Lie derivative is antisymmetric when $L_3$ is involutive, so checking involutivity with the generalised Lie derivative is equivalent to checking it with the Courant bracket. The condition that $L_{3}\times_{N}L_{3}=0$ ensures this.}
\begin{equation}
    \Dorf_{\ee^\Sigma V}(\ee^\Sigma V') = \ee^{\Sigma}\Dorf^{\dd\Sigma}_VV' 
       = \ee^\Sigma \left[
           \Dorf_VV' - (\cliff{W}\dd\Sigma)\cdot V' \right] , 
\end{equation}
where $\Dorf^{\dd\Sigma}$ is the twisted generalised Lie derivative (for type IIB, see~\cite{AW15}) and $(\cliff{W}\dd\Sigma)$ acts on $V'$ via the $\ExR{7(7)}$ adjoint action. To be involutive we need the term in brackets to be an element of $L^{\mathcal{J}_-}_1\oplus \mathcal{U}_{\mathcal{J}_{-}}$. Since the first term is differential and the second algebraic in $V$ and $V'$ this can only be true if each term is separately a section of $L^{\mathcal{J}_-}_1\oplus \mathcal{U}_{\mathcal{J}_{-}}$. We have already analysed the first term. For the second term it means $\cliff{W}\dd\Sigma\in\Gamma(\ad\tilde{F})$ must stabilise $L^{\mathcal{J}_-}_1\oplus \mathcal{U}_{\mathcal{J}_{-}}$. For the $W$ component we have
\begin{equation}
    - (\cliff{W}\dd\Sigma)\cdot W' = \cliff{W'}\cliff{W} \dd\Sigma . 
\end{equation}
If we will split the spinor bundle $S^-$ into its $\mathcal{J}_{-}$ $n\ii$-eigenspaces, $S_{n}$, where $n=-3,-1,1,3$, and denote by a subscript $n$ the projection of a polyform to $S_{n}$, this implies 
\begin{equation}
\cliff{W'} \cliff{W} \dd\Sigma \in S_{3} \quad \Leftrightarrow \quad (\dd\Sigma)_{-1}=(\dd\Sigma)_{-3} = 0.
\end{equation}
Combining these conditions with their complex conjugates we find
\begin{equation}
F=-8\,\dd^{\mathcal{J}_{-}}(\ee^{-3\Delta}\im\Phi_{+}). \label{eq:GMPT Integrability 2}
\end{equation}
Moreover, just counting the $\mathcal{J}_-$ charges, we see that this condition is enough to imply $(\cliff{W}\dd\Sigma)\cdot\Phi_-=0$. Taken together, we see that the first two equations in~\eqref{eq:Tom a} are necessary and sufficient conditions for involutivity of $L_{3}$:
\begin{equation}
    \text{involutivity of $L_3$ :}  \qquad  
    \dd\Phi_-=\cliff{A}\Phi_, \qquad
    F=-8\,\dd^{\mathcal{J}_{-}}(\ee^{-3\Delta}\im\Phi_{+}).
\label{eq:GMPT Full Integrability a}
\end{equation}

As expected, involutivity does not provide a full solution to the supersymmetry equations. Instead we find that it implies essentially the first two~\eqref{eq:Tom a} of the three conditions found in~\cite{Tomasiello07}. From section \ref{sec: SUSY + Intrinsic Torsion} we know that these equations only imply the vanishing of part of the intrinsic torsion, and that the vanishing of the rest of the intrinsic torsion, here given by the final equation $\dd(\ee^{-\Delta}\re \Phi_{+})=0$, is implied by the vanishing of the moment map~\eqref{eq:moment_map_su7}. In other words we have 
\begin{equation}
    \text{vanishing of moment map :}  \qquad  
    A=0_, \qquad \dd(\ee^{-\Delta}\re \Phi_{+}) = 0 ,
\label{eq:GMPT Full Integrability b}
\end{equation}
that is, the generalised complex structure $\mathcal{J}_-$ is promoted to a generalised Calabi--Yau structure, and in addition the third condition of~\cite{Tomasiello07} is satisfied. Since the full set of equation~\eqref{eq:Tom a} and~\eqref{eq:Tom a} are equivalent to supersymmetry, the proof in~\cite{CSW14b,CS16} that supersymmetry is equivalent to vanishing intrinsic torsion is sufficient for these last conditions to  to indeed be equivalent to the vanishing of the the moment map. Thus, rather than give all the details, let us simply sketch below how the relevant conditions arise.

Since it defines an exceptional polarisation, the $L^{\mathcal{J}_-}_1\oplus \mathcal{U}_{\mathcal{J}_{-}}$ subbundle will have an associated singlet in the $\tilde{K}_\bbC$ bundle, just as for an almost exceptional complex structure. Given the decomposition under $\Orth{6,6}\times \SL{2,\bbR}\subset\Ex{7(7)}$
\begin{equation}
    \rep{912} = (\rep{352}',\rep{1})+(\rep{220},\rep{2})+(\rep{12},\rep{2})+(\rep{32},\rep{3}) ,
\end{equation}
the only $\SU{3,3}\subset\Orth{6,6}$ singlet appears in the $\rep{32}$ representation, given by, up to $\det T^*$ factors, $\Phi_-$ itself. In fact, the $\bbR^+$ weight of $\tilde{K}$ is such that singlet is simply $\Phi_-\in\Gamma(\tilde{K}_\bbC)$. It has the property that $V\bullet\Phi_-=0$ for all $V\in\Gamma(L^{\mathcal{J}_-}_1\oplus \mathcal{U}_{\mathcal{J}_{-}})$. Given the twisting of $L_3$ it is then easy to see that the corresponding $\SU7$ structure is simply 
\begin{equation}
\label{eq:gmpt-psi}
    \psi = \ee^\Sigma\cdot\Phi_- 
    = \ee^{C+8\,\ii \,\ee^{-3\Delta}\im\Phi_{+}}\cdot\Phi_{-} ,
\end{equation}
where, since $\Phi_-$ is already naturally the section of a weight-three bundle under the  $\bbR^+$ action, we do not expect any additional powers of $\ee^\Delta$. 
Turning to the moment map, we can repeat the same steps in the analysis for $\Gx2$ structures in section~\ref{sec: G2} to derive
\begin{equation}
\begin{split}
    \mu(\ee^\Sigma V)      
        &= \tfrac{1}{3}\int_{M}s(\Dorf_{V}^{\dd\Sigma}\Phi_-,\ee^{-\Sigma}\ee^{\bar{\Sigma}}\cdot\bar{\Phi}_-)(\ii\,s(\psi,\bar{\psi}))^{-2/3} \\
        &= \tfrac{1}{3}\int_{M}s(\Dorf_{V}^{\dd\Sigma}\Phi_-,\ee^{-2\ii\im\Sigma}\cdot\bar{\Phi}_-)(\ii\,s(\psi,\bar{\psi}))^{-2/3} ,
\end{split}
\end{equation}
where in the second line we have use the property that we can always choose a gauge for $C$ such that $\Sigma$ and $\bar{\Sigma}$ commute. We also have 
\begin{equation}
\label{eq:L-flux-Phi}
    \Dorf_{V}^{\dd\Sigma}\Phi_-
       = \Dorf_V\Phi_- - (\cliff{V}\dd\Sigma)\cdot\Phi_- = \Dorf_V\Phi_- 
       = \Dorf_Z\Phi_- - (\dd\Lambda_-+\dd\tilde{\Lambda})\cdot\Phi_-,
\end{equation}
where we have used the fact, derived from the involutivity condition, that $\dd\Sigma$ stabilises $L^{\mathcal{J}_-}_1\oplus \mathcal{U}_{\mathcal{J}_{-}}$ and hence the singlet $\Phi_-\in\Gamma(\tilde{K})$, and in the last expression have split $V=Z+\Lambda_++(\tilde{\Lambda}+\tau)$ using the decomposition~\eqref{eq:Odd-decomp}. One can argue that the terms that survive in the moment map are of the form
\begin{equation}
   \mu(V) \sim \text{const} \int (\Dorf_Z\Phi_-,\bar{\Phi}_- ) 
      + \text{const} \int (\dd\Lambda_-,\ee^{-\Delta}\re\Phi_{+} ) .
\end{equation}
This form follows from keeping track of the $\Uni1\subset\SL{2,\bbR}$ charge in the $\Orth{6,6}\times\SL{2,\bbR}\subset\Ex{7(7)}$ decomposition, noting the $\bbR^+$ weight to get the correct $\ee^\Delta$ factor, and recalling the algebraic relations between $\mathcal{J}_\pm$ and $\Phi_\pm$. In particular, the $\Uni1$ charge implies that the second term arises from the third-order term in $\im \Phi_{+}$ exponential. As was first noticed in~\cite{Hitchin02}, one can determine the real part of a pure spinor from the imaginary part\footnote{In fact, in \cite{Hitchin02} they show that $\im \Phi$ can be obtained from $\re \Phi$. However the converse statement is also true.} as a third-order expression in $\im \Phi_{+}$, hence the appearance of $\re\Phi_+$. The first term vanishes if and only if $A=0$, while, integrating by parts 
on the second term gives $\dd(\ee^{-\Delta}\re\Phi_{+})=0$, the final condition in~\eqref{eq:Tom b}.


\subsubsection{Calabi--Yau as $\mathcal{N}=1$}
It is straightforward to describe the usual Calabi--Yau compactifications in our formalism. While these actually give $\mathcal{N}=2$ compactifications, we can still write them in our $\mathcal{N}=1$ language. In this case, the internal spinors are equal, $\zeta_1 = \zeta_2$, and can be used to construct a complex three-form $\Omega$ and a real two-form $\omega$. Given vanishing flux one finds that the dilaton and warp factor must be constant. The Killing spinor equations then imply
\begin{equation}
\dd\Omega =0, \qquad \dd\omega = 0.
\end{equation}
These objects can be embedded as generalised complex structures as
\begin{alignat}{8}
\Phi_{-} &=\tfrac{1}{8}\ee^{3\Delta-\dil} \Omega,\qquad\quad & 
L_1^{\mathcal{J}_-} &= T^{0,1}\oplus T^{*1,0},\\
\Phi_{+} &= \tfrac{1}{8}\ee^{3\Delta-\dil}\ee^{\ii\,\omega}, & 
L_1^{\mathcal{J}_+} &=T\lrcorner (1-\ii\,\omega).
\end{alignat}
where these are chosen such that they have the correct normalisation according to \eqref{eq:Tom a} and \eqref{eq:Tom b}.

Focussing on type IIB, we take
\begin{equation}
L_3 = \ee^{\ii\,\ee^{-\dil}(\omega-\frac{1}{6}\omega\wedge \omega\wedge\omega)}(T^{0,1}\oplus T^{*1,0}\oplus \mathbb{C}\,\ee^{3\Delta-\varphi}\Omega).
\end{equation}
Integrability of $L_{3}$ then implies
\begin{equation}
\dd\Omega = A\wedge\Omega, \qquad \dd^{I}\omega = \dd^{I}\dil\wedge \omega,
\end{equation}
where $I$ is the (integrable) complex structure associated to $\Omega$ and $\dd^{I}=[\dd,I]$. Clearly these are not the full set of integrability conditions for Calabi--Yau. Imposing the vanishing of the moment map, we find that $A=\dd \Delta=\dd\dil=0$ and hence the above become
\begin{equation}
\dd\Omega = 0, \qquad \dd^{I}\omega=0 \quad \Leftrightarrow \quad \dd\omega=0.
\end{equation}
Finally note that we could have instead taken the pure spinors to be
\begin{equation}
\Phi_{-} = \tfrac{1}{8}\ee^{\ii\alpha}\ee^{3\Delta-\dil}\Omega, \qquad \Phi_{+} = \tfrac{1}{8}\ee^{\ii\beta}\ee^{3\Delta-\dil}\ee^{\ii\omega},
\end{equation}
where $\alpha,\beta$ are two real constants. This would not change the normalisation condition or the generalised metric, but would affect what we mean by the real and imaginary parts of $\Phi_{\pm}$ and hence would rearrange which terms appear in the involutivity and moment map conditions. This amounts to choosing which $\mathcal{N}=1\subset\mathcal{N}=2$ we want to make manifest.

\section{The superpotential, the Kähler potential and extremisation}
\label{sec:WK-GIT}

As we discussed in section~\ref{sec: SUSY + Intrinsic Torsion}, the existence of the Kähler metric on the space $\ZS{\SU7}$  of generalised $\SU7$ structures is really just a reflection of fact that one can rewrite the full ten- or eleven-dimensional supergravity in $D=4$, $\mathcal{N}=1$ language, in line with the $\mathcal{N}=2$ discussion of~\cite{GLW06,GLW07,GLSW09,AW15}. The internal degrees of freedom parametrised by $\psi$ lie in chiral multiplets and hence parametrise a Kähler manifold. By including the unphysical constant overall scaling and phase of $\psi$ we are in the superconformal formulation of the supergravity. The $D$-term (or more strictly the Killing prepotential $\mathcal{P}$) is just the moment map $\mu$ for the action of the $\GDiff$ gauge symmetry, with $V\in\Gamma(E)$ giving a parametrisation of $\gdiff$:
\begin{equation}
\begin{aligned}
    &\text{Kähler potential :} & &&
    \mathcal{K} &= \int_{M}\left(\ii\,s(\psi,\bar{\psi})\right)^{1/3}, \\
    & \text{$D$-term :} & &&
    \mathcal{P} &= \tfrac{1}{3}\int_{M}s(\Dorf_{V}\psi,\bar{\psi})
           (\ii\,s(\psi,\bar{\psi}))^{-2/3} .
\end{aligned}
\end{equation}

To complete the description of the chiral multiplet sector we need the generic superpotential $\mathcal{W}$ in terms of $\psi$. This was first discussed in~\cite{PW08}. The $D=4$, $\mathcal{N}=1$ supersymmetry conditions are the vanishing of the $D$-term, namely $\mathcal{P}=0$, and the superpotential conditions $\delta\mathcal{W}/\delta\psi=\mathcal{W}=0$. In terms of our previous discussion this means that the superpotential conditions should imply the involutivity of $L_3$. A missing ingredient thus far in our discussion is to show that involutivity is a holomorphic condition in terms $\psi$. In this section, we will extend the analysis of~\cite{PW08} to give the expression for $\mathcal{W}$ for a generic $\mathcal{N}=1$ background. We will see that it is indeed a holomorphic function of $\psi$ and furthermore show that, in the special cases of a $\Gx2$ structure and GMPT, it matches the standard expressions in the literature. 

Recall also that the moment map picture implies that formally the moduli space of integrable $\SU7$ structures can be viewed as as a quotient by the complexification $\GDiff_\bbC$ of the generalised diffeomorphism group. As for the complex and generalised complex structure cases, the complexification does not really exist as a group, and instead what is really meant is modding out by the complexification of the orbits generated by the action of $\GDiff$. The other focus of this section is to investigate this action and show that it gives a (generalised) reinterpretation of Hitchin's picture of integrable $\Gx2$ structures as extremising a particular functional. We will also comment very briefly on how this might suggest notions of stability for $\Gx2$ manifolds and their generalisations.

\subsection{The superpotential}
\label{sec:super-pot}

In this section we will derive a general form for the superpotential $\mathcal{W}$, building on work on superpotentials in the presence of flux first proposed in~\cite{GVW00,Gukov:1999gr}, and the generalised geometry expressions given in~\cite{PW05}. A natural conjecture is that $\mathcal{W}$ is given by the singlet part of the intrinsic torsion for the $\SU 7$ structure integrated over the internal manifold. As we will see, one can pick out this singlet by a projection that is holomorphic in terms of $\psi$, meaning that the superpotential is a holomorphic function of $\psi$, justifying $\psi$ as the holomorphic coordinate on $\ZS{\SU7}$. 

As mentioned above we expect the supersymmetry conditions $\delta\mathcal{W}/\delta\psi=\mathcal{W}=0$ to imply the involution condition on $L_3$. We note that the variations of the $\SU 7$ structure $\psi$ transform as $\rep 1_{7}$, $\rep 7_{3}$ and $\repb{35}_{5}$, and so $\delta\mathcal{W}/\delta\psi=0$ will constrain the dual $\rep 1_{-7}$, $\repb 7_{-3}$ and $\rep{35}_{-5}$ components of the intrinsic torsion. This means $\delta\mathcal{W}/\delta\psi=0$ implies $\mathcal{W}=0$ (as $\mathcal{W}$ itself is the singlet) and furthermore is a slightly stronger condition than $L_3$ being involutive, which only constrained the $\rep{1}_{-7}$ and $\rep{35}_{-5}$ components. 

Before turning to the superpotential itself, it is useful to show that one can rephrase involutivity as a holomorphic condition on $\psi$. Suppose $V\in\Gamma(L_3)$ and $D$ is a compatible generalised connection, that is $D\psi=0$. From the definition~\eqref{eq:torsion} we find
\begin{equation}
\label{eq:DL3psi}
    \Dorf_V \psi = - T(V)\cdot \psi \qquad \text{for $V\in\Gamma(L_3)$} . 
\end{equation}
Note that this expression is linear in $V$. For any other $\bbR^+$ weight we would have gotten an additional factor of the form $(D\cdot V)\psi$ where $D\cdot V=D_M V^M$, and hence a non-linear expression. Since $\Dorf_V\psi$ is independent of $D$, only the intrinsic torsion contributes to $T(V)\cdot\psi$. From the $\Uni1\times\SU7$ representations it is easy to check that the $\rep{1}_{-7}$, $\repb{7}_{-3}$, and $\rep{35}_{-5}$ components of the intrinsic torsion~\eqref{eq:su7_torsion} appear, precisely the components in $\delta\mathcal{W}/\delta\psi$. This gives us an alternative formulation of the involutivity condition\footnote{Note that in the conventional and generalised complex structure cases we could equally well have formulated the conditions~\eqref{eq:dOmega} and~\eqref{eq:dPhi} as $\mathcal{L}_V\Omega=(\imath_VA)\Omega$ and $\Dorf_V\Phi=2\left<V,A\right>\Phi$ for all $V\in\Gamma(L_1)$.} (i.e.~the vanishing of the $\rep{1}_{-7}$ and $\rep{35}_{-5}$ components)\footnote{Note that relations of this form were first noted in the context of integrable structures in $\Ex{6(6)}$ generalised geometry by Edward Tasker (private communication).}:
\begin{equation}
\label{eq:alt-involutive}
    \text{involutive $L_3$} \qquad \Leftrightarrow \qquad
    \Dorf_V\psi = A(V)\, \psi  \quad \forall\; V\in L_3 , 
\end{equation}
where $A\in\Gamma(L_3^*)$ is the $\repb{7}_{-3}$ component of the $\SU7$ intrinsic torsion, and $A(V)=A_M V^M$ is just the natural pairing between sections of $E^*$ and $E$. We also see that we expect
\begin{equation}
\label{eq:W-conds}
    \frac{\delta\mathcal{W}}{\delta\psi}=0 \qquad \Leftrightarrow \qquad
    \Dorf_V\psi = 0 \quad \forall \;V\in L_3 . 
\end{equation}
In analogy with the complex structure and generalised complex structure cases, we expect that we can always take a $\psi$ satisfying the involutive condition and rescale by a complex function $\psi'=f\psi$ so that the stronger superpotential condition is satisfied. 

Crucially both of these conditions are linear in $V$ and so can be viewed as a holomorphic expressions in $\psi$. (Note from~\eqref{eq:psi-bundle} that $L_3$ is fixed by $V\bullet\psi=0$ and so also only depends holomorphically on $\psi$.) If we had chosen a structure $\psi'$ with a different $\bbR^+$-weight we would have had an additional $(D\cdot V)\psi'$ term. For the involutivity condition we could still have phrased the condition in the holomorphic form $\Dorf_V\psi'\propto\psi'$, however the $\delta\mathcal{W}/\delta\psi'=0$ condition would not be holomorphic because it would have to be written as $\Dorf_V[(\ii s(\psi',\bar{\psi'})^p\psi']=0$ for some suitable power $p$. Thus we anticipate that the superpotential $\mathcal{W}$ is a holomorphic function only if we take $\psi$ transforming in $\rep{912}_{\rep{3}}$. 

Returning to the definition of the superpotential, why is it natural to conjecture that it is the singlet torsion $\rep{1}_{-7}$? Consider the AdS case for a moment. We know from~\cite{CSW14b} that the cosmological constant appears as a singlet of the intrinsic torsion when decomposed under $\SU8$ and this descends to the singlet for the $\SU7$ structure (since there is only one singlet). The supersymmetry conditions for an AdS background include the vanishing of derivatives of the superpotential (the F-terms) but the superpotential itself does not vanish. Instead, requiring the superpotential to vanish is the final condition for a Minkowski solution. Thus it is reasonable that the superpotential itself is simply the singlet of the torsion.

To see this more concretely, we conjecture 
\begin{equation}
    \mathcal{W} \coloneqq \int_M W \sim \int_M \ii\, s(\psi,T) ,
\end{equation}
where $T$ is the intrinsic torsion of the structure. The symplectic product with $\psi$ projects onto the singlet component (specifically the $\rep{1}_{-7}$ component). We also note that $\psi$ is weight-3 and $T$ is weight-$(-1)$ with respect to the $\bbR^+$ action. This means $\ii\, s(\psi,T)$ is weight-2 and hence is a volume form which can be integrated over the manifold. From~\eqref{eq:DL3psi} we know that the $\rep{1}_{-7}$ component of the torsion is a holomorphic function of $\psi$, and hence the superpotential is holomorphic. 

We can make the $\psi$ dependence more manifest as follows. It was shown in~\cite{PW08}, using the Killing spinor equations, that $W$ can be written as\footnote{Technically, in~\cite{PW08} a specific choice of the connection $D$ was taken. We show that the operator appearing here is independent of this choice at the end of this section.}
\begin{equation}
(D \oadj \psi) \cdot \psi \sim W \psi,
\end{equation}
where $D$ is now a torsion-free $\SU8$ connection (not $\SU7$), $\oadj$ is a projection to the adjoint representation $\rep{133}$, so that $D\oadj\psi$ transforms in the $\rep{133}_{\rep{2}}$ representation, and $W$ is the desired singlet component of the intrinsic torsion of the structure defined by $\psi$. Clearly we can project onto $W$ by calculating
\begin{equation}
\label{eq:E7-W}
\frac{s( \bar\psi, (D \times_{\text{ad}} \psi) \cdot \psi)}{s(\bar\psi,\psi)}\sim W.
\end{equation}
%
%
At first sight, this appears to depend on $\bar\psi$ and so will not be holomorphic on $\ZS{\SU7}$. However, this apparent dependence factors out. Consider an infinitesimal variation of the structure
\begin{equation}
\begin{aligned}
\delta \psi &\sim c\, \psi + a \cdot \psi + \tilde a \cdot \psi ,\\
\delta \bar\psi &\sim \bar{c} \,\bar\psi + \bar{a} \cdot \bar\psi + \bar{\tilde a} \cdot \bar\psi, 
\end{aligned}
\end{equation}
where we are acting with the Lie algebra $\ex{7(7)}\oplus\bbR$. Decomposing under $\SU7$ as in~\eqref{eq:ad-SU7}, $c$ is a complex singlet coming from  $\rep{1}_{0}$ and the $\bbR$ action, while $a$ and $\tilde{a}$ transform in $\rep{7}_{-4}$ and $\repb{35}_{-2}$ representations respectively. 
Of the antiholomorphic parameters $(\bar{c}, \bar{a}, \bar{\tilde a})$, only $\bar c$ can appear in the relevant projection
\begin{equation}
s( \delta\bar\psi, (D \times_{\text{ad}} \psi) \cdot \psi)
\sim W \,s( \delta\bar\psi, \psi)
= \bar{c}\,W\, s(\bar\psi, \psi)
\end{equation}
as the parts involving $\bar{a}$ and  $\bar{\tilde a}$ are non-singlet and thus projected out. Thus we are left with only a $\bbC^*$ scaling of $\bar\psi$ by the antiholomorphic factor $\ee^{\bar{c}}$. However this scaling clearly factors out of~\eqref{eq:E7-W} and hence $W$ is indeed a holomorphic function of $\psi$. 



In conclusion, the general expression for the superpotential of a generic $D=4$, $\mathcal{N}=1$ background up to an overall constant is
\begin{equation}
\label{eq:E7-superpotential}
\mathcal{W} = \int_M W \sim \int_M \frac{s( \bar\psi, (D \times_{\text{ad}} \psi) \cdot \psi)}{s(\bar\psi,\psi)}
\sim \int_M \tr\left(J,(D\times_{\text{ad}}\psi)\right).
\end{equation}
We have included an alternative expression in a slightly simpler form that has the benefit of being easier to calculate explicitly. However, it is less obvious to see that it does not depend on antiholomorphic variations of the structure. 

For completeness we should check that our expressions for $W$ are well defined, in the sense that they do not depend on the parts of the torsion-free $\SU{8}$ connection $D$ which are not determined by the generalised metric $G$. These undetermined components form the $\rep{1280}+\repb{1280}$ parts of the connection, and they \emph{do} appear in the unprojected operator $D\oadj\psi$, which thus depends on the choice of the connection $D$. To see that they do \emph{not} appear in our expressions for the superpotential above, note that $J$, $\psi$ and the operators $\tr(J (D\times_{\text{ad}} \psi))$ and $s( \bar\psi, (D \times_{\text{ad}} \psi) \cdot \psi)$ are all $\SU{7}$ singlets. This means that only $\SU{7}$ singlet parts of the connection can appear in them. A routine decomposition reveals that there are no singlets in the $\SU{7}$ decomposition of the $\rep{1280}+\repb{1280}$ representation of $\SU{8}$, and thus these parts of the connection cannot appear in our expressions. As such, these operators represent a complex $\SU{7}$ singlet part of the intrinsic torsion, as claimed.

\subsubsection{$\Gx2$ in M-theory}

In the $\Gx2$ case, it is straightforward to calculate the superpotential directly and compare with the existing literature. 
As discussed in section~\ref{sec: G2}, the $\SU7$ structure corresponding to a $\Gx2$ structure with flux has the form 
\begin{equation}
\begin{split}
\psi&=\ee^{\tilde{A}+A}\ee^{\ii\varphi}\cdot 1=\ee^{\tilde A -\tfrac{1}{2}\ii A\wedge\varphi+A+\ii\varphi}\cdot 1=\ee^\gamma\cdot 1,\\
L_{3}&=\ee^{\tilde A -\tfrac{1}{2}\ii A\wedge\varphi+A+\ii\varphi}\cdot T 
= \ee^\gamma \cdot T_\bbC,
\end{split}
\end{equation}
where we have defined $\gamma=\tilde A -\tfrac{1}{2}\ii A\wedge\varphi+A+\ii\varphi$ as a sum of six- and three-forms. The Dorfman derivative of $\psi$ along $V=\ee^{\gamma}v\in \Gamma(L_{3})$ satisfies
\begin{equation}
\label{eq:Lpsi-G2}
\begin{split}
\Dorf_{V}\psi & = \Dorf_{\ee^{\gamma}\cdot v}(\ee^\gamma\cdot 1)
=\ee^{\gamma}\cdot\Dorf_v^{\Gamma}1
=\ee^{\gamma}\cdot(\mathcal{L}_{v}1- \imath_v\Gamma\cdot1)
=-\ee^{\gamma}\cdot \imath_v\Gamma\cdot1 ,
\end{split}
\end{equation}
where the complex flux
\begin{equation}
    \Gamma = F+\ii\,\dd\varphi+\tilde{F}+\tfrac{1}{2}\varphi\wedge\dd\varphi-\ii F\wedge\varphi
    \in\Gamma(\ext^4 T^* \oplus \ext^7 T^*)
\end{equation}
can be viewed as a section of the torsion bundle $K$. Using the various actions of $\gamma$ as an adjoint element, we also have
\begin{equation}
\label{eq:Tpsi-G2}
\begin{split}
T(V)\cdot\psi = T(\ee^{\gamma}v)\cdot\ee^\gamma\cdot 1
=\ee^\gamma\cdot (\ee^{-\gamma}\cdot T)(v)\cdot 1
=\ee^\gamma \cdot \imath_v (\ee^{-\gamma}\cdot T)\cdot 1 . 
\end{split}
\end{equation}
Finally we note that 
\begin{equation}
    s(\psi,T) = s(\ee^\gamma\cdot 1,T)
       = s(1,\ee^{-\gamma}\cdot T) \sim (\ee^{-\gamma}\cdot T)_{(7)},
\end{equation}
where $(\ee^{-\gamma}\cdot T)_{(7)}$ is the seven-form component of $(\ee^{-\gamma}\cdot T)$. However, using~\eqref{eq:DL3psi} and comparing~\eqref{eq:Lpsi-G2} and~\eqref{eq:Tpsi-G2}, we see that $(\ee^{-\gamma}\cdot T)_{(7)}=\Gamma_{(7)}$ and hence
\begin{equation}
\begin{split}
\mathcal{W} \propto \int_M \ii\, s(\psi,T)
&\propto\int_M \left(\tilde{F}+\tfrac{1}{2}\varphi\wedge\dd\varphi-\ii F\wedge\varphi\right).
\end{split}
\end{equation}
The superpotential is simply the integral of the seven-form component of the complex flux. 

We can compare this expression to those that have already appeared the literature. Beasley and Witten considered the M-theory superpotential on manifolds of $\Gx2$ holonomy~\cite{Beasley:2002db} -- this means we should assume $\dd\varphi=0$ to match their results. In addition, they take $\int_M \tilde F = -\tfrac{1}{2}\int_M A\wedge F$.\footnote{As discussed by Beasley and Witten, this comes about as the Page charge (the integral of $\tfrac{1}{(2\pi)^2}\dd\tilde A$) is quantised. Since $\tfrac{1}{(2\pi)^2}\tfrac{1}{2}\int_M A\wedge F$ is only defined modulo an integer~\cite{Witten:1996hc}, one can take $\int_M (\tilde F+\tfrac{1}{2}A\wedge F)=0$ without introducing extra ambiguities.}  Using these assumptions, the above superpotential can be rewritten as
\begin{equation}
\mathcal{W} \propto \int_M \left(\tfrac{1}{2} A + \ii\varphi\right)\wedge F,
\end{equation}
which matches that given in~\cite{Beasley:2002db}. More generally, the M-theory superpotential on manifolds with $\Gx2$ structure with flux has been discussed in a number of places~\cite{Acharya:2000ps,House:2004pm,Lambert:2005sh,DallAgata:2005zlf}. Following~\cite{Lambert:2005sh}, we define
\begin{equation}
P_0=\int_M (\tilde F+\tfrac{1}{2}A\wedge F)\quad\in(2\pi)^2\,\mathbb{Z},
\end{equation}
which allows us to rewrite our superpotential as
\begin{equation}
\begin{split}
\mathcal{W}&\propto P_0 + \int_M \left(- \left(\tfrac{1}{2} A + \ii\varphi\right)\wedge F+\tfrac{1}{2}\varphi\wedge\dd\varphi\right)\\
&\propto P_0 -\tfrac{1}{2} \int_M (A+\ii\varphi)\wedge\dd(A+\ii\varphi).
\end{split}
\end{equation}
This matches the expression found in \cite{Lambert:2005sh} up to an overall multiplicative constant.

Let us make one further comment. Recall that involutivity for a $\Gx2$ structure implied $\dd\varphi=\dd A=\dd\tilde{A}=0$ and so $\dd\gamma=0$. From~\eqref{eq:Lpsi-G2} this means $\Dorf_V\psi=0$ for all $V\in\Gamma(L_3)$ -- in other words $\delta\mathcal{W}/\dd\psi=0$. This is a result of our choice of normalisation of $\psi$. If we had scaled by a complex function $f$ so that $\psi'=\ee^\gamma\cdot f$, we would have had an additional one-form contribution to the intrinsic torsion $T$ and $\Dorf_V\psi$ would not vanish, consistent with the comments below~\eqref{eq:W-conds}. 

\subsubsection{GMPT}

We can repeat the same analysis to give the superpotential in the GMPT case. The $\SU7$ structure has the form given in~\eqref{eq:gmpt-psi} and~\eqref{eq:gmpt_l3}
\begin{equation}
\psi = \ee^\Sigma\cdot \Phi_-, \qquad 
L_{3} = \ee^\Sigma (L_1^{\mathcal{J}_-}\oplus\mathcal{U}_{\mathcal{J}_-}),
\end{equation}
where $\Sigma=C+8\,\ii \,\ee^{-3\Delta}\im\Phi_+$. Using~\eqref{eq:L-flux-Phi}, we then have 
\begin{equation}
\label{eq:Lpsi-gmpt}
\begin{split}
\Dorf_{V}\psi &= \ee^{\Sigma}\cdot\left[
   \Dorf_{Z+\alpha\Phi-} \Phi_- - (\cliff{Z}\dd\Sigma)\cdot \Phi_- \right) \\
   &= \ee^{\Sigma}\cdot\bigl[ \cliff{Z}\dd\Phi_- - \alpha(\dd\Phi_-)\cdot\Phi_-
       -(\cliff{Z}\dd\Sigma)\cdot \Phi_-\bigr] , 
\end{split}
\end{equation}
where we take $Z\in \Gamma(L_1^{\mathcal{J}_-})$ so that $V=\ee^\Sigma(Z+\alpha\Phi_-)\in\Gamma(L_3)$ and have used the algebraic property $(Z+\alpha\Phi_-)\bullet\Phi_-=0$. As in~\eqref{eq:Tpsi-G2} we have for the torsion
\begin{equation}
\label{eq:Tpsi-gmpt}
T(V)\cdot\psi =\ee^\Sigma\cdot \bigl(\ee^{-\Sigma}\cdot T\bigr)(Z+\alpha\Phi_-)\cdot \Phi_- . 
\end{equation}
Finally we have
\begin{equation}
    s(\psi,T) = s(\ee^\Sigma\cdot\Psi,T)
       = s(\Psi_-,\ee^{-\Sigma}\cdot T) 
       \sim \left(\Phi_-,(\ee^{-\Sigma}\cdot T)_-\right),
\end{equation}
where in the last expression we have the Mukai pairing of $\Phi_-$ and the odd-polyform component $(\ee^{-\Sigma}\cdot T)_-$ of the torsion. However, using~\eqref{eq:DL3psi} and comparing~\eqref{eq:Lpsi-gmpt} and~\eqref{eq:Tpsi-gmpt}, we see that $(\ee^{-\Sigma}\cdot T)_-=\dd\Sigma$ and hence
\begin{equation}
\begin{split}
\mathcal{W} \propto \int_M \ii\, s(\psi,T)
&\propto\int_M \left(\Phi_-,F+8\,\ii\,\dd(\ee^{-3\Delta}\im\Phi_+)\right).
\end{split}
\end{equation}
Taking into account the normalisations~\eqref{eq:pure_norm}, we see that this is in precise agreement with the $\Orth{6,6}$ generalised geometry expressions given in~\cite{GLW06,Benmachiche:2006df,Koerber:2007xk,Koerber:2008sx}.

\subsection{The Kähler potential, the moment map and extremisation}
\label{sec:kahler_potential}

Almost twenty years ago Hitchin~\cite{Hitchin00a} gave an intriguing reformulation of integrable $\Gx2$ structures as corresponding to stationary points of a suitable functional on the space of closed structures, that is those satisfying $\dd\varphi=0$, taking the variation within the cohomology class of $\varphi$. In this section we will show that the Kähler potential $\mathcal{K}$ gives a natural generalised geometry extension of Hitchin's functional for $\SU7$ structures. In particular, we show that the moment map conditions $\mu=0$ can be rephrased as stationary points of $\mathcal{K}$ when varying over the space of complexified generalised diffeomorphisms $\GDiff_\bbC$. In the case of $\Gx2$ structures we show that this is identical to Hitchin's variational problem. 

We start by recalling that an infinitesimal generalised diffeomorphism defines a vector field $\rho_V\in\Gamma(T\ZS{\SU7})$ on the space $\ZS{\SU7}$ of generalised $\SU7$ structures given by\footnote{Note that here $\mathcal{L}_{\rho_V}$ is the Lie derivative along $\rho_V$ in the space of structures $\ZS{\SU7}$, whereas $\Dorf_V$ is the generalised Lie derivative on the manifold $M$.}
\begin{equation}
\label{eq:rhoV}
    \mathcal{L}_{\rho_V}\psi = \imath_{\rho_V}\delta\psi=\Dorf_V\psi .
\end{equation}
The symplectic form $\varpi$ on $\ZS{\SU7}$ given in~\eqref{eq:symplectic_form} is invariant under the action of $\GDiff$, that is $\mathcal{L}_{\rho_V}\varpi=0$, and $\mu$ in~\eqref{eq:moment_map_su7} is the corresponding moment map defined by  $\imath_{\rho_V}\varpi=-\delta\mu(V)$. Note that it is straightforward to check that $\imath_{\rho_W}\delta\mu(V)=\mu(\llbracket V,W\rrbracket)$, where $\llbracket V,W\rrbracket$ is the Courant bracket, and hence the moment map is equivariant. We also immediately note $\mathcal{L}_{\rho_V}\psi=\Dorf_V\psi$ is holomorphic in $\psi$ hence the $\GDiff$ action also preserves the complex structure on $\ZS{\SU7}$. 

It is a standard result from the supergravity literature that the moment map (or $D$-term) can be solved in terms of the Kähler potential~\cite{Wess:1992cp}. Explicitly, if $\rho_V$ generates the symmetry, one has, by definition,
\begin{equation}
    \delta\mu(V) = - \imath_{\rho_V} \varpi 
       = - \imath_{\rho_V} \left(\tfrac12\delta \delta^\mathcal{I} \mathcal{K}\right)
       = - \tfrac12\mathcal{L}_{\rho_V} \left(\mathcal{I}\delta\mathcal{K}\right)
           + \tfrac12\delta\left(\imath_{\rho_V}\mathcal{I}\delta\mathcal{K} \right) ,
\end{equation}
where $\delta^\mathcal{I}=[\mathcal{I},\delta]$ and $\mathcal{I}$ is the complex structure on $\ZS{\SU7}$. But we have $\mathcal{L}_{\rho_V}\mathcal{I}=0$, so, assuming we choose the Kähler potential such that it is also invariant, that is $\mathcal{L}_{\rho_V}\mathcal{K}=0$, the first term vanishes. Using $\imath_{\mathcal{I}\rho_V}\delta \mathcal K = -\imath_{\rho_V}\mathcal{I}\delta \mathcal K$, one then has (up to closed terms which are fixed to vanish by the requirement of equivariance) 
\begin{equation}
\label{eq:mu-K-relation}
    \mu(V) = - \tfrac12\imath_{\mathcal{I}\rho_V}\delta\mathcal{K} 
    = - \tfrac12\mathcal{L}_{\mathcal{I}\rho_V}\mathcal{K} . 
\end{equation}
To check this relation explicitly in our case, we first calculate $\mathcal{I}\rho_V$. Since $\psi$ is holomorphic, splitting the exterior (functional) derivative on $\ZS{\SU7}$ into holomorphic and antiholomorphic parts $\delta=\del'+\bar{\del}'$, we have 
\begin{equation}
    \mathcal{L}_{\mathcal{I}\rho_V}\psi
       = \imath_{\mathcal{I}\rho_V}\del'\psi
       = \ii \imath_{\rho_V}\del'\psi
       = \ii \mathcal{L}_{\rho_V}\psi = \ii \Dorf_V \psi.  
\end{equation}
We then have 
\begin{equation}
\begin{split}
\mathcal{L}_{\mathcal{I} \rho_V} \mathcal{K}
& = \int_M \tfrac{1}{3}\left(\ii\, s(\psi,\bar\psi)\right)^{-2/3}\left( \ii\,s(\imath_{\mathcal{I}\rho_V}\delta\psi,\bar\psi)+ \ii\,s(\psi,\imath_{\mathcal{I}\rho_V}\delta\bar\psi)\right)\\
& = - \int_M \tfrac{1}{3}\left(\ii\, s(\psi,\bar\psi)\right)^{-2/3}\left( s(\Dorf_V \psi,\bar\psi)-s(\psi,\Dorf_V \bar\psi)\right)\\
& = -\int_M \tfrac{2}{3}\left(s(\psi,\bar\psi)\right)^{-2/3}\, \,s(\Dorf_V \psi,\bar\psi)\\
& = -2\, \mu(V),
\end{split}
\end{equation}
where we used an integration by parts and compactness to reach the final line. This is in complete agreement with~\eqref{eq:mu-K-relation}. For completeness, using the non-holomorphic structure $\phi$, we can also check the invariance of $\mathcal{K}$:
\begin{equation}
\begin{split}
    \mathcal{L}_{\rho_{V}}\mathcal{K} 
    & =\ii\,\int_{M}s(\imath_{\rho_{V}}\delta\phi,\bar{\phi})
        +s(\phi,\imath_{\rho_{V}}\delta\bar{\phi})
    =\ii\,\int_{M}s(\Dorf_{V}\phi,\bar{\phi})+s(\phi,\Dorf_{V}\bar{\phi}) \\
    & =\ii\,\int_{M}\Dorf_{V}s(\phi,\bar{\phi})
=0,
\end{split}
\end{equation}
where the action of $\Dorf_{V}$ on a top-form reduces to the Lie derivative, which then vanishes due to compactness of $M$. 

The relation~\eqref{eq:mu-K-relation} is striking because it shows that the zeros of the moment map can be equally well thought of as critical points of $\mathcal{K}$
\begin{equation}
    \mu = 0 \qquad \Leftrightarrow \qquad \text{critical point of $\mathcal{K}$ under $\GDiff_\bbC$ action} .
\end{equation}
The group $\GDiff$ does not really complexify, so what is really meant here is motion on the orbits generated by $\rho_V$ and $\mathcal{I}\rho_V$. Since $\mathcal{K}$ is invariant under the former, the extremisation is really over $\ii\GDiff$ generate by $\mathcal{I}\rho_V$. For the set of critical points to form a nice moduli space after quotienting by $\GDiff$, as in the symplectic quotient, strictly one needs to show that a critical point of the Kähler potential is non-degenerate transverse to the orbit of $\GDiff$~\cite{Hitchin00a}. It is a general result that the Hessian for the imaginary transformations is given by 
\begin{equation}
\label{eq:Hessian}
   \mathcal{L}_{\mathcal{I} \rho_V} \mathcal{L}_{\mathcal{I} \rho_W} \mathcal{K} 
      = -2\,\mathcal{L}_{\mathcal{I} \rho_V}\mu(W)
      = -2\,\imath_{\mathcal{I}\rho_V}\delta\mu(W)
      = 2\,\imath_{\mathcal{I}\rho_V}\imath_{\rho_W}\varpi
      = 2\,\tilde{g}(\rho_V,\rho_W) ,
\end{equation}
where $\tilde{g}$ is the pseudo-Kähler metric on $\ZS{\SU7}$. Because the metric is pseudo-Kähler, it is possible that $\tilde{g}(\rho_V,\rho_W)$ could vanish for all $\rho_W$ and this not imply that $\rho_V=0$. Since we want to mod out by real generalised diffeomorphisms, the non-degeneracy condition we require is that, at the extremum, 
\begin{equation}
    \tilde{g}(\rho_V,\rho_W) = 0 \quad \forall\; W\in\Gamma(E) \qquad \implies \qquad 
    \exists\; U\in\Gamma(E) : \quad \ii\Dorf_V\psi = \Dorf_U \psi .
\end{equation}
In other words, any degeneracy in the direction of an imaginary $\GDiff$ transformation is always equivalent to a real $\GDiff$ transformation. One can rephrase this condition in terms of the operators discussed in section~\ref{sec: Generic moduli problem}. 
However, at this point, we do not understand them well enough to check if the non-degeneracy is generically true. That said, from a physical perspective, since the equations of motion of supergravity are elliptic and supersymmetry implies the equations of motion, we would expect there to be a sensible finite-dimensional moduli space.

The extremisation of $\mathcal{K}$ is a generalised geometry extension of Hitchin's extremisation of a $\Gx2$ functional~\cite{Hitchin00a} as we will now see. We saw in section~\ref{sec: G2} that for $\Gx2$ structures, the Kähler potential is proportional to the $\Gx2$ Hitchin functional $V(\varphi)$ 
\begin{equation}
\label{eq:Hitchin}
    \mathcal{K}(\psi) \propto V(\varphi) = \int_M \varphi\wedge\star\varphi \qquad 
    \text{for $\psi=\ee^{\tilde{A}+A}\ee^{\ii\varphi}\cdot1$} .
\end{equation}
Furthermore, under an imaginary $\GDiff$ transformation it is straightforward to calculate
\begin{equation}
\begin{split}
    \imath_{\rho_V}\delta\psi = \ii\Dorf_V\psi 
       = \ii\mathcal{L}_v\psi - \ii (\dd\omega+\dd\sigma)\cdot\psi , 
       = - \dd(\imath_v\varphi)\cdot \psi 
          - \ii\left( \dd\omega'+\dd\sigma'\right)\cdot \psi . 
\end{split}
\end{equation}
where $\omega'=\omega-\imath_vA$ and $\sigma'=\sigma-\imath_v\tilde{A}-\tfrac{1}{2}A\wedge\imath_vA + \tfrac{1}{2}\varphi\wedge\imath_v\varphi$ and we have used the involutivity conditions $\dd\varphi=\dd A=\dd\tilde{A}=0$. We see that, up to real generalised diffeomorphisms, an imaginary $\GDiff$ is equivalent to an imaginary gauge transformation. Exponentiating, again up to real gauge transformations, we get
\begin{equation}
    \psi \mapsto \psi' = \ee^{\tilde{A}+A}\ee^{\ii\dd\sigma'}\ee^{\ii(\varphi+\dd\omega')}\cdot 1 
    = \ee^{\tilde{A}+A}\ee^{\ii(\varphi+\dd\omega')}\cdot(1
       + \text{const}\times j\dd\sigma' + \dots ) , 
\end{equation}
where $j\dd\sigma'$ denotes $\dd\sigma_{m,n_1\dots n_5}\in\Gamma(T^*\otimes\ext^5T^*)$ and the dots denote higher-order terms in $\dd\sigma'$. In particular, we see the $\Gx2$ three-form is shifted within its cohomology class. We now want to extremise $\mathcal{K}$ with respect to the $\sigma'$ and $\omega'$ variations. First note that it is independent of $A$ and $\tilde{A}$ since it is a $\ExR{7(7)}$-invariant. Next, we first show that $\dd\sigma'=0$ is an extremum with respect to the $\sigma'$ variations. Writing the modified $\Gx2$ structure as $\varphi'=\varphi+\dd\omega'$, linearising in $\pi=j\dd\sigma'$ we then have, using the same arguments that led to~\eqref{eq:gamma-pi},  
\begin{equation}
    \delta\mathcal{K} = \int_M \kappa \qquad \text{where} \qquad
    \kappa_{m_1\dots m_7} 
        = \text{const}\times g'^{np}\pi_{n,p[m_1\dots m_4}\varphi'_{m_5m_6m_7]} .
\end{equation}
However, the antisymmetry of $\dd\sigma'$ implies $\kappa$ vanishes and hence $\delta\mathcal{K}=0$. This means we are back to extremising $\mathcal{K}(\psi')$ in~\eqref{eq:Hitchin} with $\varphi$ replaced with $\varphi'=\varphi+\dd\omega'$. But this is exactly the extremisation introduced by Hitchin~\cite{Hitchin00a}. For a variation $\delta\varphi'=\dd\omega'$ it gives
\begin{equation}
\delta V(\varphi') \propto \int_M \delta\varphi' \wedge\star\varphi' = \int_M \dd\omega'\wedge\star\varphi'.
\end{equation}
Integrating by parts shows that $V$ has a critical point for $\dd\star\varphi'=0$, recovering the condition from the vanishing of the moment map as we expected.

\subsection{Moduli spaces, GIT and stability}
\label{sec:GIT}

The fact that the moduli space can be viewed either as a symplectic quotient or a quotient by the complexified group is a general result for group actions that preserve a Kähler structure (see for example the discussion in~\cite{HKLR87}). For the case in hand, we have
\begin{equation}
\label{eq:GIT}
    \mathcal{M}_\psi = \ZI{\SU7} \qquotient \GDiff \simeq \ZI{\SU7}^{\text{ps}} \quotient \GDiff_\bbC . 
\end{equation}
There is a subtlety we have glossed over previously which is that for the complex quotient one needs to consider not the full space of structures but a subset $\ZI{\SU7}^\text{ps}\subset\ZI{\SU7}$ of ``polystable'' points. The equivalence of quotients in~\eqref{eq:GIT} is the Kempf--Ness theorem. This is part of ``Geometric Invariant Theory'' or GIT, as reviewed for example in~\cite{Thomas06}. The point is that not all complex orbits will intersect the space of zeros of the moment map $\mu^{-1}(0)$. If $\psi$ lies on an orbit that fails to meet $\mu^{-1}(0)$ it is called unstable and is excluded from $\ZI{\SU7}^\text{ps}$. Our setup is typical of a number of classic geometric problems: one has an infinite-dimensional Kähler manifold with a group action such that the vanishing of a moment map corresponds to the solution of a differential equation. For example, it appears in Atiyah and Bott's work on flat connections on Riemann surfaces~\cite{AB83}, in the ``hermitian Yang--Mills'' equations of Donaldson--Uhlenbeck--Yau~\cite{Donaldson85,UY86,UY89}, and the equations of Kähler--Einstein geometry~\cite{Fujiki90,Donaldson97,KEproof}. Famously, in each case, developing the correct GIT notion of stability allows one to translate the question of existence of solutions to the differential equation into algebraic conditions arising from the analysis of the complex orbits.

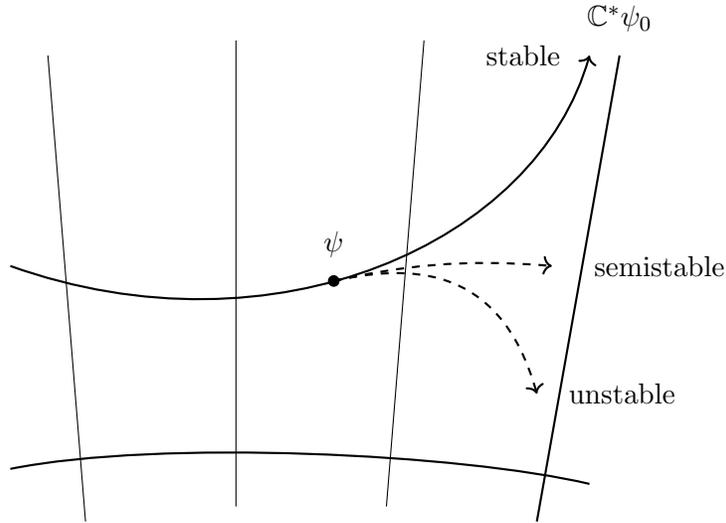
\begin{figure}
    \centering
    \begin{tikzpicture}
        \draw[thick,->] (-4,1.7) .. controls (-0.5,0.45) and (3,2) .. (3.7,4.5) node[left=7pt]{stable};
        \draw[thick,dashed,->] (0.3,1.5) node[above=5pt]{$\psi$} .. controls (1.2,1.7) and (2,1.8) .. (3.2,1.7) node[right=12pt]{semistable};
        \draw[thick,dashed,->] (0.3,1.5) .. controls (1.2,1.7) and (2.5,1.8) .. (3,0) node[right=8pt]{unstable};
        \draw[thick] (-4,-1) .. controls (-2,-0.6) and (2,-0.8) .. (3.7,-1.2);
        \draw[thick] (3,-1.7) --  (4.1,4.5) node[above=5pt]{$\bbC^*\psi_0$};
        \draw (-3,-1.7) -- (-3.5,4.5);
        \draw (-1,-1.5) -- (-1,4.7);
        \draw (1,-1.5) -- (1.5,4.7);
        \filldraw (0.3,1.5) circle (2pt);
    \end{tikzpicture}    
    \caption{Stability for a 1-PS orbit of $\psi$.}
    \label{fig:stability}
\end{figure}

In this section, we will sketch how our description of integrable $\SU7$ structures might translate into the GIT picture, and discuss the form of the moduli space. In general, stability can be understood in the following way. Consider a $\Uni1$ subgroup of the group action. For us this is some $\Uni1\subset\GDiff$ generated by some vector field $\rho_V\in\Gamma(T\ZI{\SU7})$. Under complexification this gives a $\bbC^*$ action on the space of involutive structures $\ZI{\SU7}$. Starting at some point $\psi\in\ZI{\SU7}$ the $\bbC^*$ action generates an orbit of structures $\psi(\nu)$ parameterised by $\nu\in\bbC^*$. If the space of structures were compact up to overall scalings of the $\SU7$ structure of the form $\psi\to \lambda^3\psi$ with $\lambda\in\bbC^*$, then in the limit $\nu\to 0$, the two $\bbC^*$ actions must coincide, giving a fixed line of structures (see figure~\ref{fig:stability})
\begin{equation}
\label{eq:stability1}
    \lim_{\nu\to0} \psi(\nu) = \nu^{3w(\psi)} \psi_0 \quad 
    \Rightarrow \quad
    \lim_{\nu\to0} \mathcal{K} = |\nu|^{2w(\psi)} \mathcal{K}_0 , \qquad
    \text{where $w(\psi) \in \bbZ$} ,
\end{equation}
where the weight $w(\psi)$ depends on the orbit (and hence the original structure $\psi$) and is necessarily quantised since we have a $\Uni1\subset\bbC^*$ action.\footnote{We have normalised the $\Uni1$ charges relative to the $\bbR^+$ action, hence the factor of three in~\eqref{eq:stability1}.}
Considering all such $\Uni1$ subgroups, or ``one-parameter subgroups'' (1-PS), one then defines\footnote{More generally one can define stability for the action of the whole of the complexified group (in our case $\GDiff_\bbC$) but the Hilbert--Mumford criterion implies that stability for all the 1-PS is an equivalent condition.} 
\begin{equation}
    \begin{aligned}
        & \text{if $w(\psi)<0$ for all 1-PS then $\psi$ is stable,} \\
        & \text{if $w(\psi)\leq0$ for all 1-PS then $\psi$ is semistable,} \\
        & \text{if $w(\psi)>0$ for some 1-PS then $\psi$ is unstable.} 
    \end{aligned}
\end{equation}
The beautiful observation is then that if the function $\mathcal{K}$ is convex with respect to varying $|\nu|$, and is stable in both directions (that is for 1-PS generated by $\rho_V$ and the inverse 1-PS generated by $-\rho_{V}$), then it must have a (unique) minimum. But we have already seen from~\eqref{eq:mu-K-relation} that a minimum of $\mathcal{K}$ is equivalent to the vanishing of the moment map $\mu(V)=0$ for this particular $V$. Since stability is for all 1-PS it implies there is a unique minimum where $\mu(V)=0$ for all $V$. Hence if $\psi$ is stable\footnote{The actual condition is the slightly more subtle notion of ``polystability'' which includes equivalence classes of semistable orbits, at the boundary between stable and unstable orbits.} then there is a unique solution of the moment map in the orbit of $\psi$ generated by $\GDiff_\bbC$. In the language of GIT we are identifying 
\begin{equation}
    \text{norm functional $=$ Kähler potential $\mathcal{K}$} 
\end{equation}
which as we saw above is the $\ExR{7(7)}$ extension of Hitchin's $\Gx2$-functional. 

In the Kähler--Einstein context, Yau~\cite{Yau78} originally introduced the notion of a functional that is the integral of the square of the scalar curvature, and in the moment map picture is the integral of the square of the moment map. Critical points of the Yau functional are called ``extremal metrics''. In our context, the $\mathcal{N}=1$, $D=4$ supergravity picture gives a simple interpretation of the analogous object. Recall that the potential of the supergravity is given by
\begin{equation}
    \mathcal{V} = \ee^\mathcal{K} \left( 
        \hat{g}^{i\bar{j}} D_i\mathcal{W} D_{\bar{j}} \bar{\mathcal{W}} 
          - 3\mathcal{W}\bar{\mathcal{W}} \right)
          + \tfrac12 (\re\tau)^{ab} \mathcal{P}_a\mathcal{P}_b , 
\end{equation}
where $\hat{g}_{i\bar{j}}$ is the Kähler metric on the space of chiral fields $\Phi^i$, $D_i\mathcal{W}=\del_i\mathcal{W}-(\del_i\mathcal{K})\mathcal{W}$, and $\re\tau_{ab}$ is an invariant metric on the Lie algebra of the moment map symmetry. If we consider $\SU7$ structures that are involutive (or strictly the slightly stronger condition that the superpotential is extremised~\eqref{eq:W-conds}) the term in parentheses vanishes. The metric on the Lie algebra is fixed by the generalised metric $G_{MN}$ (see for example~\cite{Hohm:2013uia}) and we are left with 
\begin{equation}
    \mathcal{V} 
    \sim \int_M \vol_G^{-1} G^{MN} \mathcal{P}_N\mathcal{P}_M \sim \int_M \vol_G \mathcal{R} ,
\end{equation}
where $\vol_G$ is the $\Ex{7(7)}$-invariant volume form defined by the generalised metric. (Note that the factor of $\vol_G^{-1}$ in the first term comes from the fact that $\mathcal{P}\in\Gamma(\det T^*\otimes E^*)$.) We see that the potential is the square of the moment map. Furthermore, from the reformulation of supergravity in terms of $\ExR{7(7)}$ generalised geometry~\cite{CSW11,CSW14}, the potential is the supergravity action on $M$ which is just the integral of the generalised Ricci scalar $\mathcal{R}$ as we write in the second term. Thus we have 
\begin{equation}
    \text{Yau functional\ } \sim \int_M \vol_G \mathcal{R} .
\end{equation}
We see that extremising the Yau functional corresponds to generalised Ricci-flat solutions, that is generic solutions of the supergravity equations. 

Central to the equivalence of stability and the vanishing of the moment map is the condition that the norm functional is convex. This is usually a consequence of the general result~\eqref{eq:Hessian} that the second derivative is given by the Kähler metric 
\begin{equation}
    \mathcal{L}_{\mathcal{I} \rho_V} \mathcal{L}_{\mathcal{I} \rho_V} \mathcal{K} 
      = 2\,\tilde{g}(\rho_V,\rho_V) .
\end{equation}
A positive-definite metric then implies convexity. As we have already mentioned, a key difference for $\SU7$ structures is that we have a pseudo-Kähler metric and so we can no longer guarantee that the norm functional is convex under the action of $\ii\GDiff$. Thus a stable orbit may have more than one solution of the moment map, and unstable orbits may still include solutions, implying stability is only a sufficient condition for the existence of solutions. This problem is closely related to the degeneracy question, mentioned above, as to whether critical points of $\mathcal{K}$ form a nice moduli space. 

The pseudo-Kähler structure raises other potential subtleties with the description of the moduli space of integrable $\SU7$ structures as we have presented it. First, the holomorphic involutivity condition might define a null subspace within the space of structures $\ZS{\SU7}$, meaning there is no guarantee that the subspace $\ZI{\SU7}$ inherits a Kähler metric (since the pullback of the metric can be degenerate). Secondly, if the group action defining the moment map is null, there is similarly no guarantee that there is a Kähler metric on the symplectic quotient. Although we have not checked directly, physically we might expect that neither problem arises, the point being that supersymmetry implies that there must be a Kähler metric on the final moduli space, since it is a space of chiral superfields. Furthermore, unless the background secretly admits more supersymemtries, this metric must be positive definite (since it gives the kinetic terms of the four-dimensional fields). If there are extra supersymmetries these appear as deformations which change the $\SU7$ structure but not the generalised metric, and hence are unphysical. 

This makes one wonder if there could be a more standard GIT picture underlying the conditions. Recall that at a point $p\in M$ the tangent space $TQ_{\SU7}$ to the $\ExR{7(7)}/\SU7$ coset space~\eqref{eq:QSU7} decomposes under $\SU7\times\Uni1$ as 
\begin{equation}
\label{eq:TQ}
    TQ_{\SU7}\ : \quad (\rep{1}_{0} \oplus \rep{1}_{0})
        \oplus (\rep{7}_{-4} \oplus \bar{\rep{7}}_{4} )
        \oplus (\rep{35}_{2} \oplus \bar{\rep{35}}_{-2} ) ,
\end{equation}
where the first two terms are generated by the action of $J$ and the $\bbR^+$ scaling. The complex structure on $TQ_{\SU7}$ pairs the representations in parentheses, with a positive definite metric on $\rep{35}_{2} \oplus \bar{\rep{35}}_{2}$ and a negative definite metric on the remaining directions giving a signature $(70,16)$, which is then inherited by the full space of structures $\ZS{\SU7}$. Focusing on the $\Gx2$ case, or perhaps more generally the type-0 case, we will now discuss how the negative deformations can potentially be removed. First one considers the space of exceptional complex structures $J$ (rather than $\ZS{\SU7}$), which removes the two singlet components in~\eqref{eq:TQ}. Then one takes the symplectic quotient by the normal subgroup of gauge transformations generated by five-forms which removes the remaining $\rep{7}\oplus\bar{\rep{7}}$ components. 

An exceptional complex structure $J$ determines $\psi$ up to rescaling by a function $\psi\to f\psi$. Thus we can define the space of exceptional complex structures as a symplectic quotient
\begin{equation}
    \text{$\XI$, space of exceptional complex structures} 
        = \ZI{\SU7} \qquotient H .
\end{equation}
The Lie algebra of $H$ is given by $\mathfrak{h}\simeq C^\infty(M)$ and $\alpha\in\mathfrak{h}$ acts via $\rho_\alpha(\psi) = \ii \alpha \psi$, giving the moment map
\begin{equation}
    \mu_H(\alpha) = \int_M \alpha \left(\ii\, s(\psi,\bar\psi) \right)^{1/3},
\end{equation}
and in the quotient we set $\mu_H=\vol_0$ for some fixed reference volume form. Since the action preserves the Kähler structure on $\ZI{\SU7}$ there is then also a Kähler metric on $\XI$ though now based on the coset space $\Ex{7(7)}/\Uni7$ with signature $(70,14)$. The corresponding Kähler potential is given by choosing an arbitrary section $\psi\in\Gamma(\mathcal{U}_J)$ and calculating 
\begin{equation}
    \tilde{K} = \int_M \log\left( \ii s(\psi,\bar{\psi})/\vol_0^3\right) \vol_0 . 
\end{equation}
The action of $\GDiff$ descends to $\XI$ (strictly we need to restrict to the subgroup $\GDiff_0\subset\GDiff$ that preserves $\vol_0$, that is in the Lie algebra $\Dorf_V\vol_0=0$, but we will ignore this subtlety). Hence one can define a corresponding moment map $\tilde{\mu}$ on $\XI$ given by
\begin{equation}
    \tilde{\mu}(V) = \int_M \frac{s(\Dorf_V\psi,\bar{\psi})}{\ii\, s(\psi,\bar{\psi})} \vol_0 , 
\end{equation}
and define the quotient moduli space
\begin{equation}
    \Mphys = \XI \qquotient \GDiff .
\end{equation}
We claim that this is isomorphic the physical moduli space $\mathcal{M}_\psi/\bbC^*$, where the $\bbC^*$ action is the constant rescaling $\psi\to\lambda^3\psi$. The point is that the vanishing of the moment map $\tilde{\mu}(V)=0$ on $\XI$ implies the vanishing of the moment map $\mu(V)=0$ on $\ZI{\SU7}$ except for those transformations that preserve $J$, that is $\Dorf_VJ=0$. However, such transformations simply rescale $\psi$. The effect is that for each $J$ satisfying $\tilde{\mu}=0$ the addition conditions from $\mu=0$ simply fix the particular section $\psi\in\Gamma(\mathcal{U}_J)$. Up to an overall $\bbC^*$ rescaling $\psi\to\lambda^3\psi$, we expect one such solution for each $J$, and hence $\mathcal{M}_J$ is isomorphic to the physical moduli space $\mathcal{M}_\psi/\bbC^*$. (This is completely analogous to the $\SL{3,\bbC}$ structure case.) 

If we focus on $\Gx2$ structures, fixing an integrable $J$, the compatible $\psi$ can be written as 
\begin{equation}
    \psi = \ee^{\tilde{A}+A}\ee^{\ii\varphi}\cdot f,
\end{equation}
for some function $f$ and with $\dd\varphi=\dd A=\dd\tilde{A}=0$. We note that the group $G_\sigma\subset\GDiff$ of five-form gauge transformations forms a normal subgroup. Thus we can do the symplectic reduction by stages, first reducing by $G_\sigma$ and then by the quotient group $\GDiff'=\GDiff/G_\sigma$. As we saw in section~\ref{sec:kahler_potential}, the form of $\psi$ we have written already satisfies $\mu(\sigma)=0$ for all five-forms $\sigma$. Hence the symplectic quotient just identifies $\tilde{A}\sim\tilde{A}+\dd\sigma$. Taking $H^6_\dd(M,\bbR)=0$, we have $\tilde{A}\sim 0$. By moving to the quotient space 
\begin{equation}
    \XI_\sigma = \XI\qquotient G_\sigma , 
\end{equation}
we have effectively removed 14 of the allowed deformations. Direct calculation in the $\Gx2$ case implies that this removes precisely the negative directions in the metric, so that the Kähler metric on $\XI_\sigma$ is positive definite. Thus we have a conventional picture of stability with
\begin{equation}
    \Mphys \simeq \XI_\sigma \qquotient \GDiff'
       \simeq \XI_\sigma^{\text{ps}} \quotient \GDiff'_\bbC .
\end{equation}
This suggests that, at least formally, the space of integrable $\Gx2$ structures, complexified by including the closed three-form potential $A$, can be viewed as a GIT quotient of the space of closed $\Gx2$ structures. The  $\Ex{7(7)}$ extension of Hitchin's $\Gx2$ functional $\mathcal{K}$ plays the role of the norm functional. 

A choice of 1-PS in this case should be a diffeomorphism corresponding to circle actions on $M$ since the gauge transformations in $\GDiff$ are always non-compact. If the diffeomorphism is generated by $\xi\in\Gamma(T)$, fixed points of the 1-PS amount to solutions to
\begin{equation}
\label{eq:fixedJ}
    \Dorf_\xi J = 0  \quad \text{where} \quad 
    J = \ee^A \bigr( \varphi^\sharp - \varphi\bigl),
\end{equation}
where we have allowed for a non-trivial three-form potential. The value of the moment map at the fixed point, suitably normalised, should give an integer invariant. This will be the analogue of the Futaki invariant in K\"ahler--Einstein geometry~\cite{Futaki1983}. Furthermore, these should be obstructions to the existence of solutions to the moment map. The simplest solution to~\eqref{eq:fixedJ}, is to take $\mathcal{L}_\xi \varphi=\mathcal{L}_\xi A=0$. In this case, the $\SU7$ structure $\psi\in\Gamma(\mathcal{U}_J)$ can only depend on the circle action through the function $f$. One would expect that the integer invariants would thus encode the topology of the line bundle $\mathcal{U}_J$, since the moment map is independent of the choice of section. The obstruction is thus that the bundle must be trivial, as we expect for the existence of a globally defined $\psi$. More interestingly however, the 1-PS motion may lead to other types of solution to~\eqref{eq:fixedJ}, most notably exceptional complex structures with type-changes, perhaps associated to circle actions with fixed points. These are structures $J$ which are no longer type-0 in the whole of $M$. This is possible since although the $\bbC^*$ action generated by $\xi$ preserves the cohomology class of $\varphi$ and $A$, the forms themselves may vanish or become singular at points in $M$. The moment map evaluated on such solutions should again give some integral invariant of the closed $\Gx2$ structure. Naively, understanding such configurations would be key to formulating any notion of stability.

\section{Moduli of \texorpdfstring{$\mathcal{N}=1$}{N=1} backgrounds}\label{sec:moduli}

The generalised $\SU7$ structure we have described characterises generic $\mathcal{N}=1$ flux backgrounds with a four-dimensional Minkowski factor. A natural question to ask is what is the moduli space of these backgrounds? If the background is to be used for phenomenology, this will tell us about the massless chiral superfields in the four-dimensional effective theory (ignoring extra M-theory or stringy massless excitations localised at singularities, since we are in the supergravity limit). Although the answer is well-known for $\Gx2$ compactifications, very little is known about generic supersymmetric flux compactifications. In this section we will use the generalised geometrical description to show how the moduli are related to particular cohomologies. For $\Gx2$ this reproduces the well-known result that the number of chiral fields is counted by the third de Rham cohomology $H^3_\dd(M,\bbC)$. The analysis trivially extends to generic type-0 $\SU7$ structures giving the local moduli space as $H^3_\dd(M,\bbC)\oplus H^6_\dd(M,\bbC)$.
Remarkably it also gives a complete description of the moduli for the GMTP solutions, completing an analysis first considered in~\cite{Tomasiello07}. 

As we have seen, the moduli space of a $D=4$, $\mathcal{N}=1$ background is given by $\Mphys=\mathcal{M}_\psi\quotient \mathbb{C}^{*}$, where $\mathcal{M}_\psi $ is the space of torsion-free $\SU7$ structures modulo generalised diffeomorphisms. In section~\ref{sec:GIT} we argued that if the infinite-dimensional GIT picture is valid this is equivalent to to $\XI/\GDiff_\bbC$, where $\XI$ is the space of exceptional complex structures. If we have a solution $J$, the local moduli space thus corresponds to a finding the integrable deformations of $J$ modulo complexified generalised diffeomorphisms. As we noted, strictly the GIT picture is not necessarily equivalent because the the metric on $\XI$ is not positive definite.  However, assuming the critical points of $\mathcal{K}$ are non-degenerate transverse to the orbit of $\GDiff$,  infinitesimally this will produce the correct moduli space. A generic deformation then defines an element of the intrinsic torsion that must vanish for the deformation to be integrable. The complexified generalised diffeomorphisms will be generated by the Dorfman derivative and are necessarily integrable. This sets up a problem in cohomology and it is this that we aim to understand better. We will start with a quick review of the moduli of conventional complex structures as this will illustrate many of the key ideas that we will use in analysing the deformations of $\SU7$ structures.

\subsection{Review of the moduli space of complex structures}
\label{sec:deformations}

Let us recall how the moduli space of integrable $\SL{3,\bbC}$ structures arises. One starts by considering deformations of an integrable $\GL{3,\mathbb{C}}$ structure. Define $Q_{\GL{3,\mathbb{C}}}=\GL{6,\mathbb{R}}\quotient\GL{3,\mathbb{C}}$ as the space of (almost) complex structures at a point $p\in M$. This can be viewed as
\begin{equation}
Q_{\GL{3,\mathbb{C}}}=\GL{6,\mathbb{R}}\quotient\GL{3,\mathbb{C}}=\GL{6,\bbR}\cdot I_{0}=\GL{6,\mathbb{C}}\quotient P,
\end{equation}
where $\GL{6,\bbR}\cdot I_{0}$ is the orbit of a fixed complex structure $I_{0}$ under $g\in\GL{6,\mathbb{R}}$, and $P$ is the parabolic subgroup of $\GL{6,\mathbb{C}}$ that stabilises $L_{1}$
\begin{equation}
P=\Stab L_{1}=(\GL{3,\mathbb{C}}\times\GL{3,\mathbb{C}})\ltimes\mathbb{C}^{9}.
\end{equation}
The orbit picture means that deformations of the complex structure are parametrised by a choice of element of $\gl{6,\bbC}\quotient\mathfrak{p}$ at each point in the manifold. In other words, one takes a section of the vector bundle
\begin{equation}
\mathfrak{gl}_{6,\mathbb{C}}\quotient \mathfrak{p} \to \mathfrak{Q}_{\GL{3,\mathbb{C}}}\to M.
\end{equation}
In practice one can view $\mathfrak{Q}\subset \ad\tilde{F}_{\mathbb{C}}$ by choosing an embedding $\gl{6,\mathbb{C}}\quotient \mathfrak{p} \hookrightarrow \gl{6,\mathbb{C}}$. In particular, using the real structure one can decompose 
\begin{equation}
\begin{aligned}
    \gl{6,\mathbb{C}}
       &= \gl{3,\mathbb{C}}\oplus\gl{3,\mathbb{C}}
          \oplus\mathfrak{q}\oplus\bar{\mathfrak{q}} , \\
    \mathfrak{p} &= \gl{3,\mathbb{C}}\oplus\gl{3,\mathbb{C}}
          \oplus\mathfrak{q} ,
\end{aligned}
\end{equation}
where we identify the (nilpotent) subalgebra $\mathfrak{q}\simeq\Gamma(T_{p}^{1,0}\otimes T_{p}^{*0,1})$. The pair of $\gl{3,\mathbb{C}}$ algebras and $\mathfrak{q}$ preserve $L_{1}=T^{1,0}\subset T_{\mathbb{C}}$. This means a deformation of $L_{1}$ at a point $p\in M$ can formally be parametrised by $\bar{\alpha}_p\in\bar{\mathfrak{q}}$ alone, that so one can identify $\mathfrak{Q}_{\GL{3,\mathbb{C}}}\simeq T^{0,1}\otimes T^{*1,0}$. The deformed subbundle is then
\begin{equation}
    L'_{1} = \ee^{\bar{\alpha}}L_{1} = (1+\bar{\alpha})L_{1}.
\end{equation}
$L_{1}'$ can then be used to define $L'_{-1}\subset T_{\mathbb{C}}$ via $L'_{-1}=\bar{L}'_{1}$ provided $L_{1}'\cap L_{-1}'=0$. Note that nilpotency of $\mathfrak{q}$ implies $\exp\bar{\alpha}=1+\bar{\alpha}$.

As before, this new subbundle is integrable if and only if
\begin{equation}
[L'_{1},L'_{1}]\subset L'_{1}.
\end{equation}
One can check that for an arbitrary deformation parametrised by $\bar\alpha$ we have 
\begin{equation}
\left(\ee^{-\bar{\alpha}}[\ee^{\bar{\alpha}}V,\ee^{\bar{\alpha}}W]\right)^{m} = \underbrace{(1+\bar{\alpha})^{p}{}_{q}(V^{q}\dd_{p}W^{m}-W^{q}\dd_{p}V^{m})}_{\in L_{1}} + \underbrace{V^{q}W^{r}(\partial\bar{\alpha}+[\bar{\alpha},\bar{\alpha}])^{m}{}_{qr}}_{\in L_{-1}}.
\end{equation}
This gives the well-known result that a complex structure deformation is integrable if and only if $\bar{\alpha}$ satisfies the Maurer--Cartan equation.
\begin{equation}
\partial\bar{\alpha} + [\bar{\alpha},\bar{\alpha}] = 0.
\end{equation}

If one is just interested in the infinitesimal moduli at this point, taking $\bar\alpha=\epsilon\bar{\beta}$, in the limit $\epsilon\to 0$ the condition is simply $\partial\bar{\beta} = 0$. In general there may be an obstruction to extending this solution for finite $\epsilon$, although the Kodaira--Nirenburg--Spencer theorem states there is no obstruction if the cohomology class $H_{\del}^{2,0}(M,T^{0,1})$ vanishes. For the moduli space one should mod out by deformations generated by diffeomorphisms. Infinitesimally, that is of the form 
\begin{equation}
    L'_{1} = (1+\epsilon\mathcal{L}_{v})L_{1} \qquad v\in \Gamma(T) .
\end{equation}
Writing $v=x+\bar{x}$ for a unique $x\in \Gamma(T^{1,0})$, one finds
\begin{equation}
    L'_{1} = (1+\epsilon\partial \bar{x})L_{1},
\end{equation}
where one views $\partial \bar{x}\in \Gamma(\mathfrak{Q}_{\GL{3,\mathbb{C}}})$. A deformation is then trivial if $\bar{\beta} = \partial \bar{x}$ for some $\bar{x}\in \Gamma(T^{0,1})$. Hence we get the result that the infinitesimal moduli of $\GL{3,\mathbb{C}}$ structures is given by
\begin{equation}
    H^{1,0}_{\partial}(M,T^{0,1}) .
\end{equation}
Finally we note that it is simple to connect this picture to the moduli space of $\SL{3,\bbC}$ structures. An integrable complex structure $I$ defines a line of $\SL{3,\bbC}$ structures $\mathcal{U}_I$. Up to a constant $\bbC^*$ rescaling $\Omega\to\lambda\Omega$ there is a unique integrable structure $\Omega\in\Gamma(\mathcal{U}_I)$ (that is one satisfying $\dd\Omega=0$) for each complex structure $I$. Hence, we get the standard result that the moduli space of integrable $\SL{3,\bbC}$ structures is just $H^{1,0}_{\partial}(M,T^{0,1})\oplus\bbC\simeq H^{1,2}_\del(M)\oplus H^{0,3}_\del(M)$. 

\subsection{Moduli space of \texorpdfstring{$\SU{7}$}{SU(7)} structures}

Now let us turn to the moduli space $\mathcal{M}_\psi$ of $\SU7$ structures $\psi$. As we have discussed, locally the physical moduli space $\mathcal{M}_\psi/\bbC^*$ can be identified with the space of deformations of $J$ that remain integrable, modulo complex diffeomorphisms.

First, let us introduce some notation as we did in the previous section. We consider the space $Q_{\Uni{7}\times \mathbb{R}^{+}}$ of almost exceptional complex structures at a point $p\in M$. This can be viewed as
\begin{equation}
Q_{\Uni{7}\times \mathbb{R}^{+}} = \Ex{7(7)}\quotient \Uni{7} = \Ex{7(7)}\cdot J_{0} = \Ex{7,\mathbb{C}}\quotient P,
\end{equation}
where $\Ex{7(7)}\cdot J_{0}$ is the orbit of a fixed almost exceptional complex structure $J_{0}$ under $\Ex{7(7)}$ at some fixed point on the manifold, and $P$ is the parabolic subgroup that stabilises $L_{3}$
\begin{equation}
P= \Stab L_{3} = \GL{7,\mathbb{C}}\ltimes\mathbb{C}^{42}.
\end{equation}
By considering the orbit of $J_{0}$ at all points on the manifold, we see that infinitesimal deformations of the structure can be viewed as a sections of the vector bundle
\begin{equation}
\ex{7,\mathbb{C}}\quotient \mathfrak{p} \to \mathfrak{Q}_{\Uni7\times\mathbb{R}^{+}} \to M.
\end{equation}
Again, in practice we will embed $\mathfrak{Q}_{\Uni7\times{\mathbb{R}^{+}}}{\hookrightarrow} \mathrm{ad}\,\tilde{F}_{\mathbb{C}}$ by choosing an embedding $\ex{7,\mathbb{C}}\quotient \mathfrak{p}\hookrightarrow \ex{7,\mathbb{C}}$. Explicitly, we write a generic infinitesimal deformation of $L_{3}$ as
\begin{equation}
L_{3}\rightarrow L_{3}'=(1+\epsilon A)\cdot L_{3},
\end{equation}
where we view $A\in\Gamma(\mathfrak{Q}_{\Uni7\times{\mathbb{R}^{+}}})$ as a map
\begin{equation}
A\colon L_{3} \rightarrow E_{\mathbb{C}}\quotient L_{3},
\end{equation}
and then make a choice of embedding $E_{\mathbb{C}}\quotient L_{3}\hookrightarrow E_{\mathbb{C}}$. As the original subbundle $L_3$ is involutive, the intrinsic torsion vanishes. For a generic deformation $A$, $L'_{3}$ will have some non-zero intrinsic torsion that appears as an obstruction to the involutivity of $L_3'$ with respect to the generalised Lie derivative (or equivalently the Courant bracket). Expanding to first order in $\epsilon$ we get a differential map $\dd_2$
\begin{equation}
\dd_{2}\colon \Gamma(\mathfrak{Q}_{\Uni7\times\mathbb{R}^{+}}) \rightarrow \Gamma(W^{\text{int}}_{\UniR7}),
\end{equation}
where sections of $W^{\text{int}}_{\UniR7}$ are the intrinsic torsion for the deformed almost exceptional complex structure.\footnote{From the discussion around~\eqref{eq:involution-torsion}, note that here $W^{\text{int}}_{\UniR7}$ is strictly a complex bundle transforming in the $\rep{1}_{-7}\oplus\rep{35}_{-5}$ representation of $\UniR7$.} The $L_3'$ subbundle will be involutive if the intrinsic torsion vanishes, and so the deformed structure will be integrable if and only if $A\in \ker \dd_{2}$. 

We also have the notion of a trivial deformation. As we have discussed, this corresponds to the action of the complexified generalised diffeomorphism group $\GDiff_\bbC$. To linear order, such deformations are given by the action of the Dorfman derivative. That is, we consider $L'_{3}$ to be equivalent to $L_{3}$ if
\begin{equation}
L'_{3} = (1+\epsilon\Dorf_{V})L_{3} \quad 
\text{for some $V\in\Gamma(E_{\mathbb{C}})$}.
\end{equation}
This defines a second differential map $\dd_1$
\begin{equation}
\dd_{1}\colon \Gamma(E_{\mathbb{C}})\rightarrow\Gamma( \mathfrak{Q}_{\UniR7}) .
\end{equation}
A trivial deformation should automatically be torsion free -- by the  Leibniz property of the generalised Lie derivative we have 
\begin{align}
\begin{split}
    \Dorf_{W+\epsilon\Dorf_VW}(W'+\epsilon\Dorf_VW')
      &= \Dorf_W W' + \epsilon \left(
          \Dorf_{\Dorf_VW} W' + \Dorf_W(\Dorf_VW') \right) + O(\epsilon^2) \\
      &= \left(1+\epsilon\Dorf_V\right) \Dorf_W W' + O(\epsilon^2) 
\end{split}
\end{align}
and hence any trivial deformation is indeed integrable. This is precisely the statement that $\dd_{2}\circ \dd_{1}=0$ and so we have a three-term complex
\begin{equation}
\label{eq:complex 1}
\begin{tikzcd}
\Gamma(E_{\mathbb{C}}) \arrow[r,"\dd_{1}"] & \Gamma(\mathfrak{Q}_{\UniR7})
    \arrow[r,"\dd_{2}"] & \Gamma( W^{\text{int}}_{\UniR7}). 
\end{tikzcd}
\end{equation}
Assuming there are no obstructions, the local moduli space of the $\SU{7}$ structure is modelled on the cohomology of this complex.

In the rest of this section we will calculate this cohomology for both the $\Gx{2}$, generic type-zero and GMPT structure examples. In the $\Gx{2}$ case we recover the known result that the moduli are counted by the third de Rham cohomology of the underlying manifold. In the GMPT case, we find new results -- the full set of moduli were previously unknown. We will see that in both cases the ability to calculate the cohomology of~\eqref{eq:complex 1} relies on finding a nice parametrisation of the embeddings $E_{\mathbb{C}}\quotient L_{3}\hookrightarrow E_{\mathbb{C}}$ and $\mathfrak{Q}_{\UniR7}\hookrightarrow \ad\tilde{F}_{\mathbb{C}}$. This then leads to a description of the moduli in terms of cohomologies defined by differentials that are naturally associated to the problem. For the $\Gx{2}$ (and general type-zero) case this is the de Rham differential, while for the GMPT solutions it is the generalised Dolbeault operator associated to the integrable generalised complex structure. One may hope that the general case could be solved in terms of some natural differential associated to the $L_{3}$ bundle -- we make some comments on this at the end of this section in~\ref{sec: Generic moduli problem}, noting some of the complications that arise. 

\subsection{Example 1: \texorpdfstring{$\Gx{2}$}{G2} and type-0 geometries}

Recall that we can embed $\Gx{2}$ structures into the language of exceptional complex structures via the definition
\begin{equation}
L_{3} = \ee^{\ii\varphi}\cdot T_\bbC.
\end{equation}
The involutivity of this bundle then gives $\dd\varphi=0$. A useful parametrisation of the quotient spaces as subspace of $E_\bbC$ and $\ad\tilde{F}_\bbC$ is given by
\begin{equation}
\label{eq:EQ-G2}
\begin{aligned}
    E_{\mathbb{C}}\quotient L_{3} 
    &\simeq \ext^{2}T_\bbC^{*}\oplus \ext^{5}T_\bbC^{*}\oplus (T_\bbC^{*}\otimes \ext^{7}T_\bbC^{*}) ,\\
    \mathfrak{Q}_{\UniR7} 
    &\simeq \ext^{3}T_\bbC^{*}\oplus \ext^{6}T_\bbC^{*} .
\end{aligned}
\end{equation}
It is worth noting that these are not eigenspaces of the exceptional complex structure $J$ and hence this is a different parametrisation to that given in \eqref{eigenspace param 2} below. They instead come from the natural deformations of the underlying exceptional Dirac structure defined by $T$. They are invariant under the map $\ee^{\ii\varphi}$, meaning they can equally well be viewed as defining deformations of $L_3$. In the same way, we can also identify the space of the intrinsic torsion as
\begin{equation}
\label{eq:Wint-G2}
    \Wint_{\UniR7} \simeq \ext^4 T^*_\bbC \oplus \ext^7T^*_\bbC ,
\end{equation}
that is the space of intrinsic torsion of the Dirac structure. 

If we take $\alpha\in \Gamma(\ext^{3}T^{*}_\mathbb{C})$ and $\beta\in \Gamma(\ext^{6}T^{*}_\mathbb{C})$, the infinitesimal deformation is given by 
\begin{equation}
\label{eq:defL-g2}
    L'_{3} = \bigl(1+\epsilon(\alpha+\beta)\bigr)\cdot\ee^{\ii\varphi}\cdot T_\bbC 
       = \ee^{\ii\varphi+\epsilon(\alpha+\tilde{\beta})}\cdot T_\bbC + O(\epsilon^2) ,
\end{equation}
where $\tilde{\beta}=\beta-\frac{1}{2}\varphi\wedge\alpha$. Repeating the calculation in~\eqref{eq:twist_Dorf_g2}, we can use the twisted Dorfman derivative and $\dd\varphi=0$ to find
\begin{equation}
    \text{involutive $L'_3$}
    \quad \Leftrightarrow \quad \dd\alpha=\dd\beta = 0.
\end{equation}
Hence integrable deformations are given by \emph{closed} three-forms and six-forms. For the trivial deformations, writing  $V=v+\omega+\sigma+\tau\in \Gamma(E_{\mathbb{C}})$ we have, since $\mathcal{L}_v T=0$,
\begin{equation}
\begin{split}
    L'_{3} &= (1+\epsilon\Dorf_{V})\,\ee^{\ii\varphi}\cdot T_\bbC \\
        &= \left( 1 - \epsilon(\dd\omega - \dd\sigma 
             - \ee^{\ii\varphi}\mathcal{L}_v \ee^{-\ii\varphi})\right)
             \cdot \ee^{\ii\varphi}\cdot T_\bbC \\
        &=  \left( 1 + \epsilon(\dd\tilde\omega + \dd\tilde\sigma)\right) L_3 ,
\end{split}
\end{equation}
where $\tilde{\omega}=-\omega+\ii\,\imath_v\varphi$ and $\tilde{\sigma}=-\sigma-\frac{1}{2}\varphi\wedge\imath_v\varphi$. Hence the complex~\eqref{eq:complex 1} becomes
\begin{equation}
\label{eq:complex-g2}
\begin{tikzcd}
\Gamma(\ext^{2}T_\bbC^{*}\oplus \ext^{5}T_\bbC^{*}) 
\arrow[r,"\dd"] 
& \Gamma(\ext^{3}T_\bbC^{*}\oplus \ext^{6}T_\bbC^{*})
\arrow[r,"\dd"] 
& \Gamma(\ext^4 T^*_\bbC \oplus \ext^7T^*_\bbC). 
\end{tikzcd}
\end{equation}
where $\dd$  is the exterior derivative, and the inequivalent deformations are counted by 
\begin{equation}
\frac{\{\alpha\in \Gamma(\ext^{3}T^{*}_{\mathbb{C}}),\beta\in\Gamma(\ext^{6}T^{*}_{\mathbb{C}})\, | \, \dd\alpha=\dd\beta=0  \}}{ \{\alpha=\dd\tilde\omega, \beta=\dd\tilde\sigma \}} = H_{\dd}^{3}(M,\mathbb{C})\oplus H^{6}_{\dd}(M,\mathbb{C}).
\end{equation}
That is, the inequivalent deformations are counted by the third and sixth de Rham cohomologies. For a $\Gx{2}$ manifold, the sixth de Rham cohomology is trivial and hence the cohomology of \eqref{eq:complex 1} is counted by $H^{3}_{\dd}(M,\mathbb{C})$ alone. The imaginary elements are deformations of the $\Gx{2}$ structure while the real elements shift the gauge potential such that the flux remains zero. This is in complete agreement with standard analysis of the moduli space of $\Gx2$ compactifications of M-theory~\cite{Harvey:1999as,Beasley:2002db,Gutowski:2001fm}.

It is also clear from the way we have written these deformations that they are unobstructed. The action of complex gauge potentials $\alpha+\beta$ can be exponentiated for finite $\epsilon$ as in the final term of~\eqref{eq:defL-g2}, such that the linearised closure condition is enough to imply the deformation is integrable. Thus the moduli space looks like $H^3_\dd(M,\bbC)$ in a finite patch. Formally this is the statement that there is an open subset of the moduli space $\mathcal{V}\subseteq\Mphys$ containing this exceptional complex structure, an open subset $\mathcal{U}\subseteq H^{3}_{\dd}(M,\mathbb{C})$ containing $\boldsymbol{0}$, and a diffeomorphism $\mathcal{V}\rightarrow \mathcal{U}$.

Finally we note that the $\Gx2$-structure calculation extends straightforwardly to a generic type-0 structure. Recall these take the form 
\begin{equation}
    L_3 = \ee^{\alpha+\beta}\cdot T_\bbC,
\end{equation}
where $\alpha\in\Gamma(\ext^3T^*_\bbC)$, $\beta\in\Gamma(\ext^6T^*_\bbC)$ and involutivity implies $\dd\alpha=\dd\beta=0$. 
By following the same analysis as above, one sees that the deformations of this structure will again be given by
\begin{equation}
H^{3}_{\dd}(M,\mathbb{C})\oplus H^{6}_{\dd}(M,\mathbb{C}).
\end{equation}
This gives the moduli space of the class of supersymmetric backgrounds discussed in~\cite{Lukas:2004ip}, complementary to those analysed in~\cite{Kaste:2003zd,DallAgata:2003txk}. It would be interesting to analyse further the conventional geometry of these solutions. 

\subsection{Example 2: GMPT geometries}\label{GMPT Moduli}

As we saw in section~\ref{sec:GMPT}, we can write the GMPT solutions as
\begin{equation}
\label{eq:L-gmpt2}
L_{3} = \ee^{\Sigma}[L^{\mathcal{J}_{\pm}}_{1}\oplus\mathcal{U}_{\mathcal{J}_{\pm}}], \qquad \Sigma= C+8\,\ii\,\ee^{-3A}\im\Phi_{\mp},
\end{equation}
where the upper/lower signs correspond to type IIA/B respectively and the $\Orth{6,6}$ bundles are appropriately embedded into $\Ex{7(7)}\times\mathbb{R}^+$. As before, we will work in type IIB for concreteness but similar results hold for type IIA. We will use the notation set out in section~\ref{sec:GMPT}. In particular, recall that the generalised complex structure $\mathcal{J}_-$ defines a decomposition of the generalised spinor bundles into $\ii n$-eigenspaces $S^+=S_2\oplus S_0\oplus S_{-2}$ and $S^-= S_3\oplus S_1\oplus S_{-1} \oplus S_{-3}$ where $S_3\simeq \mathcal{U}_{\mathcal{J}_-}$. We can always choose $C$ such that the twisting $\Sigma$ lies in $S_0\oplus S_2$ since any component in $S_{-2}$ acts trivially on $L_3$. 

We take the parametrisation
\begin{equation}
\begin{aligned}
E_\mathbb{C} / L_3 &= L^{\mathcal{J}_{-}}_{-1} \oplus (S_{1} \oplus S_{-1}\oplus S_{-3}) \oplus \ext^{5}T^{*}_\bbC , \\
\mathfrak{Q}_{\UniR7} &= 
\ext^{2}(L^{\mathcal{J}_{-}}_{-1})^{*}\oplus (S_{0}\oplus S_{-2}) \oplus \ext^{6}T^{*}_\bbC .
\end{aligned}
\end{equation}
As before, these are not eigenspaces of $J$. Instead the are the spaces of natural deformations of the underlying exceptional Dirac structure defined by $L^{\mathcal{J}_{\pm}}_{1}\oplus\mathcal{U}_{\mathcal{J}_{-}}\subset E_\bbC$. Since $\Sigma\in\Gamma(S_0\oplus S_2)$, these spaces are invariant under the action of $\ee^\Sigma$ and hence can be used to describe deformations of the twisted bundle~\eqref{eq:L-gmpt2}. One can similarly identify the intrinsic torsion
\begin{equation}
    \Wint_{\UniR7} \simeq \ext^{3}(L^{\mathcal{J}_{-}}_{-1})^{*} \oplus (S_{-1}\oplus S_{-3}) 
\end{equation}
as a subbundle of $K$. 

We leave the details of the calculation to appendix \ref{Appendix: GMPT moduli} but to summarise, we note that we deform the $L_{3}$ bundle by $\varepsilon\in \Gamma(\ext^{2}(L^{\mathcal{J}_{-}}_{-1})^{*}$), $\chi=\chi_{0}+\chi_{-2}\in \Gamma(S_{0}\oplus S_{-2})$ and $\Theta\in \Gamma(\ext^{6}T^{*})$, then assuming the $\dd\dd^{\mathcal{J}}$-lemma~\eqref{ddJ-lemma}, one can show that the integrable moduli are counted by
\begin{equation}\label{eq:GMPT_moduli}
[\varepsilon] \in H^{2}_{\dd_{L}}(M), \qquad [\chi] \in H^{0}_{\bar{\partial}}(M)\oplus H^{-2}_{\bar{\partial}}(M), \qquad [\Theta]\in H^{6}_{\dd}(M,\mathbb{C}).
\end{equation}
The differentials $\dd_{L}$ and $\bar{\partial}$ are operators associated to the generalised complex structure given by $\Phi_{-}$ in the IIB case, and are defined in~\cite{Gualtieri04b}. The operator $\dd_{L}$ is the differential associated to the Lie algebroid structure $L^{\mathcal{J}_{-}}_{-1}$. The operators $\bar{\partial}$ are not the Dolbeault operators but are the generalised Dolbeault operators defined on the spinor bundles by the decomposition of $\dd=\del+\delb$. Hence we have 
\begin{equation}
\dd_L\colon \ext^{p}(L^{\mathcal{J}_{-}}_{-1})^{*} \rightarrow \ext^{p+1}(L^{\mathcal{J}_{-}}_{-1})^{*}, 
\qquad \bar{\partial}\colon S_{n}\rightarrow S_{n-1} . 
\end{equation}
We see that the operators in the complex~\eqref{eq:complex 1} are both given by $\dd_L+\delb+\dd$ acting on the appropriate bundles. The second comhomology group of $\dd_L$ counts the deformations of the $\mathcal{J}_-$ generalised complex structure~\cite{Gualtieri04b}. The $\delb$ cohomology groups count the deformations of $F$ and $\im \Phi_{+}$. Since $M$ is a generalised Calabi--Yau manifold, the cohomologies of $\dd_{L}$ and $\bar{\partial}$ are actually isomorphic. We see that apart from the top form (which just measures the Wilson line for the dual NSNS six-form potential $\tilde{B}$), all of the moduli are counted by natural differentials associated to the integrable $\SU{3,3}$ structure of the GMPT solutions.


This includes and extends the results of~\cite{Tomasiello07}, where the moduli of $\Phi_{+}$ keeping $\Phi_{-}$ fixed (and vice versa) were examined. It was also suggested that one might be able to find the full moduli space by varying $\Phi_{-}$ and $\re\Phi_{+}$ independently while satisfying their closure conditions. It was hoped that one could then find a solution to the $\im\Phi_{+}$ equation by examining critical points of a modified Hitchin functional by varying over a fixed cohomology class. This allows an estimate of an upper bound for the number of moduli in this case. In contrast, we are able to find the exact number of moduli by finding variations of $\Phi_{-}$ and $\im\Phi_{+}$ such that
\begin{equation}
\dd\Phi_{-} = 0, \qquad F=-8\,\dd^{\mathcal{J}_{-}}(\ee^{-3A}\im\Phi_{+}).
\end{equation}
The final condition $\dd(\ee^{-A}\re\Phi_{+})=0$ is imposed by the vanishing of the moment map. However, as we have mentioned, imposing this is equivalent to quotienting by $\GDiff_\bbC$ and hence we get it without imposing a further differential condition. As we have noted several times, this construction works only away from sources and hence these deformations do not account for deformations of branes or orientifolds.

We can see how each of these deformations affects the form of $L_{3}$:
\begin{align}
\Phi_{-}'&= (1+\cliff{\varepsilon})\Phi_{-}, \\ 
F'&= F+\tfrac{1}{2}\dd\bigl(\re(\cliff{\varepsilon}\mu+\chi)\bigr), \\
\im\Phi_{+}'&= \im\Phi_{+} +\tfrac{1}{8}\ee^{3A} \im(\cliff{\varepsilon}\mu +\chi).
\end{align}
Here $\mu$ is a polyform in $\Gamma(S_{2})$, related to $\Sigma$ and defined in appendix~\ref{mu definition}. As noted by Hitchin~\cite{Hitchin02}, $\re\Phi_{+}$ is determined by $\im\Phi_{+}$, and hence these deformations determine the full solution $\{\Phi_+,\Phi_-,F\}$. Note that a small deformation of a GMPT solution remains within the GMPT class. GMPT describes all $\mathcal{N}=1$ solutions for which the two internal spinors are nowhere vanishing -- this is an open condition and hence will not be changed by small deformations~\cite{GMPT05,Tomasiello07}.

Finally we consider the existence of obstructions to the linear deformations described above. We begin with the observation that a polyform deformation can be lifted to a finite deformation simply by promoting it to an exponential. Indeed this is precisely what we have done in the derivation above. The real question then is whether there are any obstructions to the generalised complex structure deformation $\varepsilon\in \Gamma(\ext^{2} (L^{\mathcal{J}_{-}}_{-1})^{*})$. A result due to Hitchin~\cite{Hitchin02} states that all deformations of generalised Calabi--Yau structures are unobstructed. Since we have a global $\Phi_{-}$ that satisfies $\dd\Phi_{-}=0$, we have a generalised Calabi--Yau structure defined by $\mathcal{J}_{-}$. Taken together, this would seem to imply that the moduli are unobstructed, much like in the previous $\Gx2$ case. 

\subsubsection{Calabi--Yau as $\mathcal{N}=1$}

As we saw in section~\ref{sec:GMPT}, we can embed a Calabi--Yau compactification in type IIB via
\begin{equation}
L_{3} = \ee^{\ii\,\ee^{-\varphi}(\omega - \tfrac{1}{6}\omega\wedge\omega\wedge\omega)}[T^{0,1}\oplus T^{*1,0}\oplus \mathbb{C}\,\ee^{3A-\varphi}\Omega].
\end{equation}
As is shown in \cite{Gualtieri04b}, for $\Phi_-\propto\Omega$ the generalised Dolbeault operator $\bar{\partial}$ reduces to the usual Dolbeault operator associated to the complex structure defined by $\Omega$. It is also shown that
\begin{align}
H^{2}_{\dd_{L}}(M) &= H^{2}_{\bar{\partial}}(M,\mathbb{C})\oplus H^{1}_{\bar{\partial}}(M,T^{1,0}_{\mathbb{C}}) \oplus H^{0}_{\bar{\partial}}(M,\ext^{2}T^{1,0}_{\mathbb{C}}), \\
H^{0}_{\bar{\partial}}(M) &= \bigoplus_{i=0}^{3} H^{i,i}_{\bar{\partial}}(M,\mathbb{C}), \\
H^{-2}_{\bar{\partial}}(M) &= H^{0,2}_{\bar{\partial}}(M,\mathbb{C})\oplus H^{1,3}_{\bar{\partial}}(M,\mathbb{C}),
\end{align}
where the cohomologies on the left-hand side are with respect to the generalised Dolbeault operators and those on the right-hand side are with respect to the usual Dolbeault operators. Using the isomorphism provided by the three-form $\Omega$, we see that the moduli of such a solution are counted by the Hodge numbers
\begin{equation}
h^{2,1}+(h^{0,0}+h^{1,1}+h^{2,2}+h^{3,3})+h^{3,3}.
\end{equation}
Note that these are the complex dimensions. Here $h^{2,1}$ corresponds to the deformations of the complex structure associated to $\Omega$. The real part of the Dolbeault groups in the parentheses corresponds to shifts in the RR polyform potential $C$. The imaginary part corresponds to shifts in $\im\Phi_{+}$, which count deformations of the Kähler potential $\omega$, and the NSNS fields $\phi$ and $B$. Notice that we have one extra, non-physical modulus here. Finally the real part of the final $H^{3,3}_{\bar{\partial}}$ gives deformations of $\tilde{B}\in \Gamma(\wedge^{6}T^{*})$, the six-form potential dual to $B$. Again we have an extra, non-physical modulus given by the imaginary part of $H^{3,3}_{\bar{\partial}}$. 

The two extra, non-physical moduli correspond to changing the $\mathcal{N}=1\subset \mathcal{N}=2$ that is picked out by our formalism. These moduli do not change the $\SU8$ structure (which gives us the physical fields in the theory), though they do rotate the $\SU7\subset \SU8$. Indeed, we note that choosing an $\mathcal{N}=1\subset \mathcal{N}=2$ is equivalent to choosing a $\Uni1\subset\SU2$. Hence there are 2 real or 1 complex parameters that encode this choice, precisely the counting we have. Note that these extra moduli appear only for Calabi--Yau compactifications as they are really $\mathcal{N}=2$ -- a generic GMPT solution is a genuine $\mathcal{N}=1$ solution and hence all the moduli are physical.

\subsection{Comments on the generic moduli problem}\label{sec: Generic moduli problem}

We would like to calculate the cohomology of the following complex for a generic integrable $L_{3}\subset E_{\mathbb{C}}$:
\begin{equation}
\Gamma(E_{\mathbb{C}})\xrightarrow{\: \dd_{1}\:} \Gamma(\mathfrak{Q}_{\UniR7}) \xrightarrow{\: \dd_{2}\:} \Gamma(\Wint_{\UniR7}).
\label{eq:complex 2}
\end{equation}
We can use the $\SU7$ structure to decompose the bundles as $J$ eigenspaces following~\eqref{eq:gen_vec-SU7}, \eqref{eq:ad-SU7} and~\eqref{eq:ru7_torsion}
\begin{equation}
\label{eigenspace param}
\begin{aligned}
E_{\mathbb{C}} 
&= \mathfrak{X}_{3}\oplus (\ext^{2}\mathfrak{X}^{*})_{1} \oplus (\ext^{5}\mathfrak{X}^{*})_{-1} \oplus \mathfrak{X}^{*}_{-3} \\
\ad \tilde{F}_\mathbb{C} 
&= \ad P_{\UniR7} \oplus (\ext^{3}\mathfrak{X})_{2}\oplus (\ext^{6}\mathfrak{X})_{4} \oplus (\ext^{3}\mathfrak{X}^{*})_{-2}\oplus (\ext^{6}\mathfrak{X}^{*})_{-4} \\
\Wint_{\UniR7} &= (\ext^{4}\mathfrak{X}^{*})_{-5} \oplus (\ext^{7}\mathfrak{X}^{*})_{-7}
\end{aligned}
\end{equation}
where $\mathfrak{X}$ transforms in the $\rep 7$ of $\SU7$. A natural parametrisation of embeddings is then
\begin{equation}\label{eigenspace param 2}
E_{\mathbb{C}}/L_{3} = (\ext^{5}\mathfrak{X}^{*})_{1} \oplus (\ext^{2}\mathfrak{X}^{*})_{-1} \oplus \mathfrak{X}^{*}_{-3}, \qquad 
\mathfrak{Q}_{\UniR7} = (\ext^{3}\mathfrak{X}^{*})_{-2}\oplus (\ext^{6}\mathfrak{X}^{*})_{-4}.
\end{equation}
As $L_{3}$ defines an integrable $\UniR7$ structure, we have a torsion-free compatible connection $\mathcal{D}$. Since $\dd_{1}$ and $\dd_{2}$ are defined in terms of the Dorfman derivative $\Dorf_{V}$ and $\mathcal{D}$ is torsion free, we can replace all Dorfman derivatives with $L^{\mathcal{D}}_{V}$, as in~\eqref{eq:Dorf-D}. This implies the maps $\dd_{1}$ and $\dd_{2}$ can be written in terms of $\mathcal{D}$. Moreover, viewing the derivative as a map $\mathcal{D}\colon R\to E^{*}\otimes R$, for any given generalised tensor bundle $R$, we can decompose $E^*$ and hence $\mathcal{D}$ into operators
\begin{equation}
\mathcal{D} = \mathcal{D}_{3}+\mathcal{D}_{-1}+\mathcal{D}_{1}+\mathcal{D}_{-3}.
\end{equation}
The compatibility of the generalised connection ensures that these operators map $\Uni7$ representations into $\Uni7$ representations in a way that will be clear in a moment. We can think of these operators as the generalisation of the Dolbeault operators to $\SU7$ structures.

Describing the operators $\dd_{1},\dd_{2}$ in this parametrisation, one finds that the complex \eqref{eq:complex 2} decomposes as
\begin{equation}
\begin{tikzcd}[column sep=huge, row sep = huge]
\Gamma(\ext^{2}\mathfrak{X}^{*})_{+1} \arrow[r, "\D_{-3}"] & \Gamma(\ext^{3}\mathfrak{X}^{*})_{-2} \arrow[r, "\D_{-3}"] & \Gamma(\ext^{4}\mathfrak{X}^{*})_{-5} \\
\Gamma(\ext^{5}\mathfrak{X}^{*})_{-1} \arrow[ru, "\D_{-1}"] \arrow{r}{\D_{-3}} & \Gamma(\ext^{6}\mathfrak{X}^{*})_{-4} \arrow[ru, "\D_{-1}"] \arrow[r, "\D_{-3}"] & \Gamma(\ext^{7}\mathfrak{X}^{*})_{-7}\\
\Gamma(\mathfrak{X}^{*}_{-3}) \arrow{ru}[near end]{\D_{-1}} \arrow[dotted]{ruu}[near start]{\D_{1}} &  & 
\end{tikzcd}
\end{equation}
Note that the involutivity of $L_{3}$ implies that $(\D_{-3})^{2}=0$. In fact $L_{3}$ defines a Lie algebroid and $\D_{-3}$ is the associated differential
\begin{equation}
    \D_{-3} \colon \ext^p \mathfrak{X}^* \rightarrow \ext^{p+1} \mathfrak{X}^* ,
\end{equation}
similarly to the situation for a Dirac structure in~\cite{Gualtieri04b}. It seems likely that under certain assumptions -- notably some generalised version of the $\del\delb$-lemma -- it is possible to write the cohomology of \eqref{eq:complex 2} in terms of the cohomology groups $H^\bullet_{\D_3}(M)$ of $\D_{-3}$. This would be in line with the theory of deformations of complex structures~\cite{Manetti04}, generalised complex structures~\cite{Gualtieri04b}, or more generally Dirac structures~\cite{Gualtieri:2017kdd}. However the existence of the $\D_1$ action between $\mathfrak{X}^*_{-3}$ and $\ext^3\mathfrak{X}^*_{-2}$ makes the analysis considerably more subtle than that for the $\Gx2$ and GMPT examples.

\section{Discussion}

In this paper we have rephrased generic $\mathcal{N}=1$, $D=4$ flux backgrounds in both M-theory and type II theories in terms of integrable $\SU7$-structures within $\ExR{7(7)}$ generalised geometry. The differential conditions on the $\SU7$ structure took the form of involutivity of a certain subbundle
defined by a $\Uni7\times\mathbb{R}^+\supset\SU7$ structure, and a moment map for the combined action of diffeomorphisms and gauge transformations. We showed how the examples of a conventional $\Gx2$ structure and the GMPT solutions can be understood as $\SU7$ structures, and discussed how our formalism allows a elegant derivation of the moduli of these solutions, extending previous results for the GMPT example. The space of involutive $\SU7$ structures admitted a natural pseudo-Kähler metric meaning the moment map condition could also be viewed as a complex quotient. This connects to the formalism of Geometrical Invariant Theory (GIT). We showed that the Kähler potential $\mathcal{K}$ on the space of structures plays the role of the norm functional, and can be viewed as a generalisation of Hitchin's $\Gx2$ functional. In particular, we showed that extremising $\mathcal{K}$ over the space of complex generalised diffeomorphisms reproduces Hitchin's extremisation procedure in the case of closed $\Gx2$ structures. Physically, the pseudo-Kähler metric is just a result of viewing the ten- or eleven-dimensional supergravity theory as a $D=4$, $\mathcal{N}=1$ with an infinite number of chiral fields parametrising the $\SU7$ structure. We derived the generic form of the superpotential for this reformulation and showed that it agreed with known examples.  

As for example recently emphasised in~\cite{Donaldson-g2}, despite significant progress in constructing examples, $\Gx2$ manifolds are far less well understood than, for example, their Calabi--Yau cousins. Hitchin's functional picture raises the hope that there might be a unique $\Gx2$ manifold (up to diffeomorphisms) for each stable closed three-form $\varphi$. However, there are examples where this does not hold~\cite{Fernandez87}. The moment map picture here suggests that there might be a notion of stability that picks out those closed structures that admit a solution. As we have stressed, a subtlety is that the Kähler metric on the space of structures is not positive definite. This means that stability may only be a sufficient condition for a solution. However we showed that, precisely in the $\Gx2$ case, the moment map is partially solved in a way that appears to remove the negative part of the metric, so that one is left again with a conventional GIT picture. The one-parameter subgroups relevant for stability would correspond to circle actions on the manifold, and in analogy to the Futaki invariant in the Kähler--Einstein case, there should be invariants associated to actions that leave the exceptional complex staction has a fixed point and the structure becomes type-changing.  

Another immediate area it would be interesting to explore is the explicit construction of flux backgrounds with potential phenomenological applications, and in particular identify their moduli. We note that the type-0 solutions are particularly simple extensions of $\Gx2$ holonomy, and notably give a case where the calculations here completely determine the moduli. The same is true for type IIB GMPT solutions where the underlying generalised complex structure is actually just a conventional complex structure. Important in both cases would be understanding the role of sources, particularly orientifold planes, necessary for the background to be compact. Closely related to this is the question of how calibrated cycles appear in our formalism. For $\Gx2$ compactifications, calibrated cycles also play an important role in non-perturbative physics. It would be interesting to understand how these cycles, as well as ``generalised calibrations'', can be encoded in our language. In Hitchin's generalised geometry, there are a large number of results relating the defining pure spinors to generalised calibrations~\cite{Koerber:2005qi,Martucci:2005ht,Koerber:2006hh,Koerber:2007xk} and we note that this was extended to generic $\mathcal{N}=2$ AdS$_5$ flux backgrounds in~\cite{deFelice:2017mhm}.

There are a number of other interesting directions for future study. An obvious generalisation is to backgrounds with an AdS factor instead of Minkowski. In some ways, this might be a richer problem to consider as one can have non-trivial fluxes without requiring the internal space to be non-compact. Roughly speaking, a consideration of the intrinsic torsion indicates that the involutivity condition will be deformed to include a non-vanishing singlet torsion (effectively the inverse AdS radius). We expect one will again have a moment map for the action of GDiff so that one can view the moduli space as a symplectic quotient. Unlike the Minkowski case, the moduli space is expected to be real, so we will not be able to reduce the moment map to modding out by the complexified symmetry group. These AdS backgrounds will be dual to 3d, $\mathcal{N}=1$ CFTs, and the moduli of these backgrounds will give marginal deformations of the CFTs. We hope to return to this fascinating topic in the near future.

Much of what we have discussed can be repeated for type I and heterotic theories, where the relevant generalised geometry is based on $\Orth{d,d+n}\times\mathbb{R}^+$~\cite{CSW11b,CMTW14}. Again, one finds that supersymmetry corresponds to the existence of an integrable $G$-structure, and the integrability conditions split into an involutivity condition and a moment map. One expects similar explicit expressions for the Kähler potential and the superpotential, as well as the moduli and the cohomologies that governs them, which all can be compared with previous results. There should again, formally, be a GIT picture of the symplectic quotient, and it would be interesting to compare, for instance, with the ``dilaton functional'' recently used by Garcia--Fernandez et al.~\cite{Garcia-Fernandez:2018emx} to argue for a Calabi--Yau type theorem for heterotic geometries. 

Topological string theories on backgrounds with $H$-flux are described by generalised complex geometry~\cite{Kapustin04,KL07,PW05,Pestun07}. The $\SU7$ structures we have described should give an extension of this to backgrounds with RR flux or to M-theory. Recall that there has been a proposal for topological M-theory~\cite{DGNV05,Nekrasov04} based on Hitchin's formulation of $\Gx2$ structures~\cite{Hitchin00a}. It would be particularly interesting to quantise these models following the prescription laid out in~\cite{PW05}. Note that some work in this direction has already been done in the case of $\Gx2$ and generalised $\Gx2$ structures~\cite{deBoer:2005pt,deBoer:2007zu}, although these results did not match with topological string calculations upon reduction. In light of our results, this is not so surprising. The natural generalisation of the Hitchin functional is the $\SU7$ Kähler potential. The moduli space calculation shows one should include fluctuations that are deformations by both exact three- and six-forms. Even though $H^{6}_{\dd}(M,\mathbb{C})$ is trivial, these fluctuations can still contribute to the one-loop calculation. 

Finally one might also use the formalism to address higher-derivative corrections to supergravity. These are essential for turning on fluxes on compact spaces: in M-theory, for example, eight-derivative $R^4$ corrections to the action of eleven-dimensional supergravity contribute to the stress-tensor and can be balanced against those of the four-form flux, permitting non-trivial fluxes on compact spaces. Even at this order, the full set of corrections is not known. Following recent work~\cite{Coimbra:2019dpt}, one might hope to extend the relevant generalised geometry to capture the higher-order corrections. One would no longer have a Leibniz algebroid, but a more general $L_\infty$ structure. One might hope that this structure is enough to constrain the form of the flux corrections, or even predict to higher orders in the derivative expansion. An important related question is whether a given supergravity background defines a good classical string background. That is, given a supergravity solution, can one correct it, order-by-order, so that it solves the full classical string equations of motion, including all higher-derivative corrections? The simplest case of a Calabi--Yau background without flux is known to provide such a good starting point~\cite{Alvarez-Gaume1986}. 
Understanding whether this also holds for generic flux backgrounds is a difficult and important problem~\cite{Sethi:2017phn}. A similar space-time analysis has also been performed for $\Gx2$ manifolds in M-theory~\cite{Becker:2014rea}, where it was found that $\Gx2$ holonomy is corrected order-by-order to a $\Gx2$ structure. There is no world-sheet calculation for this case, leaving one unsure of what space of structures the flow takes place in. Indeed, it was conjectured that the corrections are such that the defining three-form is always closed -- this might be akin to imposing our involutivity condition but relaxing the moment map condition. (Analogous deformations play a role in corrections to the notion of stability for D-branes~\cite{Douglas:2000ah}.) In both of these cases, it was useful to analyse the corrections using effective field theory and the space-time superpotential and K\"ahler potential. One could imagine using the formalism we have outlined to show that upon including higher-derivative corrections, the torsion-free $\SU7$ structure flows to an $\SU7$ structure with torsion, implying that supersymmetry is enough to guarantee the existence of the corrected classical string background.\footnote{Assuming the background also obeys flux quantisation.}

\acknowledgments

We thank Ruben Minasian and Edward Tasker for helpful discussions. AA is supported by DOE Grant No.~DESC0007901. DT is supported by an STFC PhD studentship. DW is supported in part by the EPSRC Programme Grant ``New Geometric Structures from String Theory'' EP/K034456/1, the EPSRC standard grant EP/N007158/1, and the STFC Consolidated Grant ST/L00044X/1. We acknowledge the Mainz Institute for Theoretical Physics (MITP) of the Cluster of Excellence PRISMA+ (Project ID 39083149) for hospitality and support during part of this work. 

\appendix

\section{Conventions}\label{Appendix:Conventions}

We use the musical isomorphism to denote raising or lowering an index with the conventional metric. For example, given a vector $v$ or a one-form $w$, we have a one-form $v^\flat = gv$ and a vector $w^\sharp = g^{-1}w$.

The Mukai pairing~\cite{Gualtieri04b} of two polyforms $\alpha$ and $\beta$ is
\begin{equation}\label{eq:mukai}
(\alpha,\beta)\coloneqq \alpha\wedge\lambda(\beta)\bigr|_\text{top},
\end{equation}
where $\lambda$ reverses the indices of the components of $\beta$, so that for a $p$-form $\lambda(\beta)_{m_1m_2\dots m_p}=\beta_{m_p \dots m_2m_1}$, and we project to the top-form.

\section{Embedding of \texorpdfstring{$\Orth{6,6}\subset\Ex{7(7)}\times\mathbb{R}^+$}{O(6,6) in E7(7) x R+} for type IIB}\label{sec:embedding}

We will follow the conventions and notation of~\cite{AW15} for $\ExR{(7(7)}$ generalised geometry applied to type IIB. Recall that the generalised tangent and adjoint spaces and their decompositions into $\Orth{6,6}$ generalised bundles take the form
\begin{equation}
\label{eq:Odd-decomp}
\begin{aligned}
    E &\simeq T \oplus 2T^* \oplus \ext^3T^* \oplus 2\ext^5T^* \oplus (T^*\otimes\ext^6T^*)  \\
    &\simeq E_{\Orth{6,6}} \oplus S^- \oplus (\ext^6T^*\otimes E_{\Orth{6,6}}) , \\
    \ad\tilde{F} &\simeq 4\bbR \oplus (T\otimes T^*) \oplus 2\ext^2T^* \oplus 2\ext^2T
        \oplus \ext^4T^* \oplus 2\ext^6T^* \oplus 2\ext^6T \\
    &\simeq 4\bbR \oplus \ad \tilde{F}_{\Orth{6,6}} \oplus S^+ 
        \oplus (\ext^6T\otimes S^+) \oplus \ext^6T^* \oplus \ext^6T .
\end{aligned}
\end{equation}
We use the following rules for embedding the $\Orth{6,6}$ structures into the $\Ex{7(7)}\times\mathbb{R}^+$ structures for type IIB.
\begin{align}
\begin{split}
E_{\Orth{6,6}} & \rightarrow E \\
v+\lambda & \mapsto v-s^{i}\lambda
\end{split} \\*[0.3em]
\begin{split}
\ad \tilde{F}_{\Orth{6,6}} & \rightarrow \ad \tilde{F} \\
r+\beta+B & \mapsto \tfrac{1}{8}\tr(r) + \bigl(r-\tfrac{1}{8}\mathbb{I}\,\tr(r)\bigr) + \tfrac{1}{4}\tr(r)(r^{i}\epsilon_{jk}s^{k} + s^{i}\epsilon_{jk}r^{k}) - s^{i}B + r^{i}\beta,
\end{split} \\*[0.3em]
\begin{split}
S^{+} &\rightarrow \ad \tilde{F} \\
\Sigma &\mapsto r^{i}\epsilon_{jk}r^{k}\Sigma_{(0)} + r^{i}\Sigma_{(2)} + \Sigma_{(4)} + s^{i}\Sigma_{(6)},
\end{split}\\*[0.3em]
\begin{split}
S^{-} & \rightarrow E \\
\Sigma & \mapsto r^{i}\Sigma_{(1)} + \Sigma_{(3)} + s^{i}\Sigma_{(5)},
\end{split}
\end{align}
where $\Sigma_{(k)}\in\Gamma(\ext^{k}T^{*})$ are the components of the polyform $\Lambda$ and $r^{i},s^{i}$ are real constant $\SL{2,\mathbb{R}}$ doublets such that $\epsilon_{ij}r^{i}s^{j}=1$.

\section{Detailed calculation of GMPT moduli}\label{Appendix: GMPT moduli}
We have the parametrisation
\begin{align}
E_\mathbb{C} / L_3&=\ee^{\Sigma}\bigl([L^{\mathcal{J}_{-}}_{-1} \oplus \bar{\mathcal{U}}_{\mathcal{J}_{-}}]\oplus S_{1}\oplus S_{-1}\oplus (\ext^{6}T^{*}\otimes [L^{\mathcal{J_{-}}}_{1}\oplus L^{\mathcal{J}_{-}}_{-1}])\bigr), \\
\ad(\mathcal{Q}_{\mathbb{R}^{+}\times \Uni7}) &= \ee^{\Sigma}\bigl(\ext^{2}(L^{\mathcal{J_{-}}}_{1})^{*}\oplus (S_{0}\oplus S_{-2}) \oplus r^{i}\ext^{6}T^{*}  \bigr)\ee^{-\Sigma},
\end{align}
where we have used $\ext^{5}T^{*}\simeq\ext^{6}T^{*}\otimes T$. We take $\chi=\chi_{0}+\chi_{-2}\in \Gamma(S_{0}\oplus S_{-2})$, $\varepsilon\in \Gamma(\ext^{2}(L^{\mathcal{J_{-}}}_{1})^{*})$ and $\Theta\in \Gamma(\ext^{6}T^{*})$ and consider the following generic deformation
\begin{equation}
L_{3}=\ee^{\Sigma}[L^{\mathcal{J_{-}}}_{1}\oplus \mathcal{U}_{\mathcal{J_{-}}}] \quad \rightarrow \quad L_{3}' = \ee^{\Sigma+\cliff{\varepsilon}\mu+\chi+r^{i}(\Theta+\frac{1}{2}(\Sigma,\cliff{\varepsilon}\mu-\chi))}[L^{\mathcal{J}_{-}^{\varepsilon}}_{1}\oplus \mathcal{U}_{\mathcal{J}_{-}^{\varepsilon}}], \label{GMPT deformation}
\end{equation}
where
\begin{equation}
\Sigma= C+8\,\ii\,\ee^{-3A}\im\Phi_{\mp},\qquad L^{\mathcal{J}_{-}^{\varepsilon}}_{1}=(1+\varepsilon)L^{\mathcal{J}_{-}}_{1}, \qquad \mathcal{U}_{\mathcal{J}_{-}^{\varepsilon}} = (1+\cliff{\varepsilon})\mathcal{U}_{\mathcal{J}_{-}}
\end{equation}
The latter two define a deformed generalised complex structure $\mathcal{J}_{-}^{\varepsilon}$. Note that $\Phi_{-}^{\varepsilon}=(1+\cliff{\varepsilon})\Phi_{-}$ is indeed the pure spinor associated to $L^{\mathcal{J}_{-}^{\varepsilon}}_{1}$. We define $\mu\in \Gamma(S_{2})$ in the following manner. Firstly, in what follows we will make the same simplification as in \cite{Tomasiello07} and assumed that the the generalised complex structure $\mathcal{J}_{-}$ satisfies the $\dd\dd^{\mathcal{J}_{-}}$-lemma. For us this will mean that
\begin{equation}
\image\partial\cap \ker\bar{\partial} = \image\bar{\partial}\cap \ker\partial = \image\bar{\partial}\partial, \label{ddJ-lemma}
\end{equation}
where $\partial$ and $\bar{\partial}$ are the generalised Dolbeault operators of $\mathcal{J}_{-}$.  With this we have
\begin{align}
\begin{array}{rcrcl}
(\dd\Sigma)_{-1}=0 \quad \Rightarrow& \quad & \bar{\partial}\Sigma_{0} &=& -\partial\Sigma_{-2} \\
\Rightarrow&& \partial\bar{\partial}\Sigma_{0} &=& 0 \\
\Rightarrow&&\partial\Sigma_{0} &=& \bar{\partial}\partial\alpha_{1}
\end{array}
\end{align}
for some $\alpha_{1}\in \Gamma(S_{1})$. We then define
\begin{equation}
\mu = \Sigma_{2}+\partial\alpha_{1}. \label{mu definition}
\end{equation}
Note that $\partial\alpha_{1}$ is not uniquely defined -- the ambiguity is some element of $\Gamma(S_{2})$ that is closed under $\bar{\partial}$. As we will see, this ambiguity can be absorbed in the definition of $\chi$. For definiteness, one can see that the deformation \eqref{GMPT deformation} to linear order is given by
\begin{equation}
\ee^{\Sigma}[\varepsilon+ (\cliff{\varepsilon}\mu+\chi) + r^{i}(\Theta - (\Sigma,\chi))]\ee^{-\Sigma} \in\Gamma( \ad (\mathcal{Q}_{\mathbb{R}^{+}\times \Uni7})).
\end{equation}

We now calculate the conditions for integrability of $L'_{3}$. Following the results of section \ref{sec:GMPT}, we find that we have integrability only if
\begin{equation}
\llbracket L^{\mathcal{J}_{-}^{\varepsilon}}_{1},L^{\mathcal{J}_{-}^{\varepsilon}}_{1}\rrbracket_{\Orth{6,6}} \subseteq L^{\mathcal{J}_{-}^{\varepsilon}}_{1} \label{GMPT def int 1}
\end{equation}
From \cite{Gualtieri04b} this implies
\begin{equation}
\dd_{L}\varepsilon = 0,
\end{equation}
where $\dd_{L}\colon\Gamma(\ext^{p}(L^{\mathcal{J}_{-}}_{1})^{*}) \rightarrow \Gamma(\ext^{p+1}(L^{\mathcal{J}_{-}}_{1})^{*})$ is the differential associated to the Lie algebroid structure $L^{\mathcal{J}_{-}}_{1}$. This means that $\cliff{\varepsilon}$ and $\bar\partial$ commute as operators on $S$:
\begin{equation}
\bar{\partial}\cliff{\varepsilon} = \cliff{\varepsilon} \bar{\partial}.
\end{equation}
Letting $S_{n},S_{n}^{\varepsilon}$ be the eigenspaces of $S$ with respect to $\mathcal{J}_{-},\mathcal{J}_{-}^{\varepsilon}$ respectively, we further require
\begin{equation}
[\dd(\Sigma+\cliff{\varepsilon}\mu+\chi+r^{i}\Theta)]_{S^{\varepsilon}_{-1}} = [\dd(\Sigma+\cliff{\varepsilon}\mu+\chi+r^{i}\Theta)]_{S^{\varepsilon}_{-3}} = 0,
\end{equation}
where the notation above means the projection of the polyform onto $S_{-1}^{\varepsilon}$ and $S_{-3}^{\varepsilon}$ respectively. We will still use subscript indices to denote projection onto $S_{n}$. Working to linear order in the deformation parameters and using the integrability of $L_{3}$, we find
\begin{align}
\begin{split}
0 &= (1+\cliff{\varepsilon}+\cliff{\bar{\varepsilon}})[\dd(\Sigma+\cliff{\varepsilon}\mu+\chi+r^{i}\Theta)]_{-1} - \cliff{\varepsilon}[\dd\Sigma]_{1} - \cliff{\bar{\varepsilon}}[\dd\Sigma]_{-3} \\
&= [\dd\cliff{\varepsilon}\mu]_{-1} + [\dd\chi]_{-1} - \cliff{\varepsilon}[\dd\Sigma]_{1} \\
&= \bar{\partial}\cliff{\varepsilon}\mu_{2} +\bar{\partial}\chi_{0} + \partial\chi_{-2} - \cliff{\varepsilon}\bar{\partial}\Sigma_{2} -\cliff{\varepsilon} \partial\Sigma_{0} \\
&= \cliff{\varepsilon}\bar{\partial}\Sigma_{2} + \cliff{\varepsilon}\bar{\partial}\partial\alpha_{1} - \cliff{\varepsilon}\bar{\partial}\Sigma_{2} - \cliff{\varepsilon}\partial\Sigma_{0} + \bar{\partial}\chi_{0} + \partial\chi_{-2} \\
&= \cliff{\varepsilon}\partial\Sigma_{0} - \cliff{\varepsilon}\partial\Sigma_{0} + \bar{\partial}\chi_{0} + \partial\chi_{-2} \\
&= \bar{\partial}\chi_{0}+\partial\chi_{-2}.
\end{split}
\end{align}
We also have
\begin{align}
\begin{split}
0&=(1+\cliff{\varepsilon}+\cliff{\bar{\varepsilon}})[\dd(\Sigma+\varepsilon\mu+\chi+r^{i}\Theta)]_{-3} -\cliff{\varepsilon}[\dd\Sigma]_{-1} \\
&= [\dd\chi]_{-3} \\
&= \bar{\partial}\chi_{-2}.
\end{split}
\end{align}

Taken together, we see that the integrability conditions are
\begin{equation}
\dd_{L}\varepsilon = 0, \qquad \bar{\partial}\chi_{0} + \partial\chi_{-2} = 0, \qquad \bar{\partial}\chi_{-2} = 0.
\end{equation}
We can simplify this further. Using the $\dd\dd^{\mathcal{J}_{-}}$-lemma we see that we can write $\partial\chi_{-2} = \bar{\partial}\partial\eta_{-1}$ for some $\eta_{-1}\in \Gamma(S_{-1})$. Then, defining $\tilde{\chi}_{0} = \chi_{0}+\partial\eta_{-1}$, we see that the integrability conditions become
\begin{equation}
\dd_{L}\varepsilon = 0, \qquad \bar{\partial}\tilde{\chi}_{0} = 0, \qquad \bar{\partial}\chi_{-2} = 0.
\end{equation}
Note again that $\partial\eta_{-1}$ is only defined up to a term that is $\bar{\partial}$-exact. We will see shortly that these terms correspond to trivial deformations.

To find the form of trivial deformations we take
\begin{equation}
V= \ee^{\Sigma}(W+ c\,\Phi_{-} + U+\nu+ r^{i}\sigma + \tau),
\end{equation}
where $W\in \Gamma(L^{\mathcal{J}_{-}}_{1})$, $U\in \Gamma(L^{\mathcal{J}_{-}}_{-1})$, $c\in C^{\infty}(M)$, $\nu=\nu_{1}+\nu_{-1}+\nu_{-3}\in \Gamma(S_{1}\oplus S_{-1}\oplus S_{-3})$, $\sigma\in \Gamma(\ext^{5}T^{*})$ and $\tau\in \Gamma(T^{*}\otimes \ext^{7}T^{*})$. Then we consider
\begin{equation}
L'_{3}=(1+L_{V})L_{3}.
\end{equation}
After a lengthy calculation we find that to linear order in $V$ this deformation is given by
\begin{equation}
\ee^{\Sigma}[ \dd_{L}U + (\dd_{L}U)\mu +(\dd\nu)_{0} + (\dd\nu)_{-2} + r^{i}(\dd\tilde{\sigma} -(\Sigma,(\dd\nu)_{0} +(\dd\nu)_{-2} ) ) ]\ee^{-\Sigma}.
\end{equation}
which is a section of $\Gamma(\ad(\mathcal{Q}_{\mathbb{R}^{+}\times \Uni7}))$. Here $\tilde{\sigma}$ is a 5-form that depends on $\sigma$, $\nu$ and $U$ and is of the form $\tilde{\sigma} = \sigma+f(\nu,U)$ where $f$ is some function whose form we do not need. A deformation is trivial if and only if
\begin{align}
\begin{split}
\varepsilon &= \dd_{L}U ,\\
\chi_{0} &= \bar{\partial}\nu_{1} + \partial\nu_{-1}, \\
\chi_{-2} &= \bar{\partial}\nu_{-1} + \partial\nu_{-3}, \\
\Theta &= \dd\tilde{\sigma}.
\end{split}
\end{align}
We can simplify this further using the $\dd\dd^{\mathcal{J}_{-}}$-lemma. Notice that we can write $\partial\nu_{-3}=\bar{\partial}\partial\eta_{-2}$ for some $\eta_{-2}\in \Gamma(S_{-2})$ and hence $\chi_{-2}$ is trivial if $\chi_{-2} = \bar{\partial}(\nu_{-1}+\partial\eta_{-2})=\bar{\partial}\tilde{\nu}_{-1}$. Moreover, if we calculate $\tilde{\chi}_{0}$ from these we find that $\tilde{\chi_{0}} = \bar{\partial}\tilde{\nu}_{1}$ for some $\tilde{\nu}_{1}\in \Gamma(S_{1})$. Hence trivial deformations are given by $\bar{\partial}$-exact $\tilde{\chi}_{0}$ and $\chi_{-2}$.

All of this shows that the inequivalent deformations are controlled by the following disjoint complex
\begin{equation}
\begin{tikzcd}[column sep=large]
(L^{\mathcal{J}_{-}}_{1})^{*} \arrow[r, "\dd_{L}"] & \ext^{2}(L^{\mathcal{J}_{-}}_{1})^{*}  \arrow[r, "\dd_{L}"] & \ext^{3}(L^{\mathcal{J}_{-}}_{1})^{*}  \\
S_{1} \arrow[r, "\bar{\partial}"] & S_{0} \arrow[r, "\bar{\partial}"] & S_{-1} \\
S_{-1} \arrow[r, "\bar{\partial}"] & S_{-2} \arrow[r, "\bar{\partial}"] & S_{-3} \\
\ext^{5}T^{*} \arrow[r, "\dd"] & \ext ^{6}T^{*}
\end{tikzcd}
\end{equation}
and so the deformations are counted by the cohomology
\begin{equation}
H^{2}_{\dd_{L
}}(M)\oplus H^{0}_{\bar{\partial}}(M)\oplus H^{-2}_{\bar{\partial}}(M) \oplus H^{6}_{\dd}(M,\mathbb{C}).
\end{equation}

\end{document}